\documentclass[superscriptaddress,preprintnumbers,amsmath,amssymb,prd,nofootinbib,preprint]{revtex4-1}
\pdfoutput=1
\usepackage{graphicx}
\usepackage{epstopdf}
\usepackage{dcolumn}
\usepackage{bm}
\usepackage{hyperref}
\usepackage{color}
\usepackage{amsmath}
\usepackage{cancel}
\usepackage{xpatch}

\makeatletter
\def\l@subsubsection#1#2{}
\patchcmd{\@ssect@ltx}
    {\addcontentsline{toc}{#1}{\protect\numberline{}#8}}
    {}
    {}
    {}
\makeatother
\begin{document}

\def\a{\alpha}
\def\b{\beta}
\def\c{\varepsilon}
\def\d{\delta}
\def\e{\epsilon}
\def\f{\phi}
\def\g{\gamma}
\def\h{\theta}
\def\k{\kappa}
\def\l{\lambda}
\def\m{\mu}
\def\n{\nu}
\def\p{\psi}
\def\q{\partial}
\def\r{\rho}
\def\s{\sigma}
\def\t{\tau}
\def\u{\upsilon}
\def\v{\varphi}
\def\w{\omega}
\def\x{\xi}
\def\y{\eta}
\def\z{\zeta}
\def\D{\Delta}
\def\G{\Gamma}
\def\H{\Theta}
\def\L{\Lambda}
\def\F{\Phi}
\def\P{\Psi}
\def\S{\Sigma}

\def\o{\over}
\def\beq{\begin{align}}
\def\eeq{\end{align}}
\newcommand{\gsim}{ \mathop{}_{\textstyle \sim}^{\textstyle >} }
\newcommand{\lsim}{ \mathop{}_{\textstyle \sim}^{\textstyle <} }
\newcommand{\vev}[1]{ \left\langle {#1} \right\rangle }
\newcommand{\bra}[1]{ \langle {#1} | }
\newcommand{\ket}[1]{ | {#1} \rangle }
\newcommand{\EV}{ {\rm eV} }
\newcommand{\KEV}{ {\rm keV} }
\newcommand{\MEV}{ {\rm MeV} }
\newcommand{\GEV}{ {\rm GeV} }
\newcommand{\TEV}{ {\rm TeV} }
\newcommand{\1}{\mbox{1}\hspace{-0.25em}\mbox{l}}
\newcommand{\headline}[1]{\noindent{\bf #1}}
\def\diag{\mathop{\rm diag}\nolimits}
\def\Spin{\mathop{\rm Spin}}
\def\SO{\mathop{\rm SO}}
\def\O{\mathop{\rm O}}
\def\SU{\mathop{\rm SU}}
\def\U{\mathop{\rm U}}
\def\Sp{\mathop{\rm Sp}}
\def\SL{\mathop{\rm SL}}
\def\tr{\mathop{\rm tr}}
\def\mpl{M_{\rm Pl}}

\def\IJMP{Int.~J.~Mod.~Phys. }
\def\MPL{Mod.~Phys.~Lett. }
\def\NP{Nucl.~Phys. }
\def\PL{Phys.~Lett. }
\def\PR{Phys.~Rev. }
\def\PRL{Phys.~Rev.~Lett. }
\def\PTP{Prog.~Theor.~Phys. }
\def\ZP{Z.~Phys. }

\def\dd{\mathrm{d}}
\def\ff{\mathrm{f}}
\def\BH{{\rm BH}}
\def\inf{{\rm inf}}
\def\ev{{\rm evap}}
\def\eq{{\rm eq}}
\def\SM{{\rm sm}}
\def\Mpl{M_{\rm Pl}}
\def\GeV{{\rm GeV}}
\newcommand{\Red}[1]{\textcolor{red}{#1}}
\newcommand{\TL}[1]{\textcolor{blue}{\bf TL: #1}}


\title{
Effective Theory of Flavor for Minimal Mirror Twin Higgs 
}

\author{Riccardo Barbieri}
\affiliation{Scuola Normale Superiore and INFN, Pisa, Italy}
\author{Lawrence J. Hall}
\affiliation{Department of Physics, University of California, Berkeley, California 94720, USA}
\affiliation{Theoretical Physics Group, Lawrence Berkeley National Laboratory, Berkeley, California 94720, USA}
\author{Keisuke Harigaya}
\affiliation{Department of Physics, University of California, Berkeley, California 94720, USA}
\affiliation{Theoretical Physics Group, Lawrence Berkeley National Laboratory, Berkeley, California 94720, USA}

\begin{abstract}
We consider two copies of the Standard Model, interchanged by an exact parity symmetry, $P$.   The observed fermion mass hierarchy is described by suppression factors $\epsilon^{n_i}$ for charged fermion $i$, as can arise in Froggatt-Nielsen and extra-dimensional theories of flavor.  
The corresponding flavor factors in the mirror sector are $\epsilon'^{n_i}$, so that spontaneous breaking of the parity $P$ arises from a single parameter $\epsilon'/\epsilon$,
yielding a tightly constrained version of Minimal Mirror Twin Higgs, introduced in
our previous paper.
Models are studied for simple values of $n_i$, including in particular one with SU(5)-compatibility, that describe the observed fermion mass hierarchy. The entire mirror quark and charged lepton spectrum is broadly predicted in terms of $\epsilon'/\epsilon$, as are the mirror QCD scale and the decoupling temperature between the two sectors.
Helium-, hydrogen- and neutron-like
mirror dark matter candidates are constrained by self-scattering and relic ionization.  In each case, the allowed parameter space can be fully probed by proposed direct detection experiments. Correlated predictions are made as well for the Higgs signal strength and the amount of dark radiation.
\end{abstract}

\date{\today}

\maketitle

\tableofcontents

\newpage

\section{Introduction}
A Mirror Sector, an identical copy of the Standard Model (SM)~\cite{Lee:1956qn,Kobzarev:1966qya}, is currently of considerable interest.  Two key results follow from introducing an approximate spacetime parity symmetry, $P$, that exchanges the two sectors.  First, dark matter may be mirror baryons~\cite{Goldberg:1986nk} with a density expected to be the same order as the baryon density. Second, the SM Higgs boson can be understood as a pseudo-Goldstone boson via the Twin Higgs mechanism~\cite{Chacko:2005pe}, even though it has order unity couplings, with a modest amount of fine-tuning. 

A key question is how $P$ is broken.  Simple schemes that have $P$ broken only via a Higgs mass term suffer from two key problems.  First, the theory is excluded from excessive dark radiation from the mirror sector.  Second, in such schemes mirror dark matter is in part hydrogen-like, with parameters that are excluded by self scattering.  Further there is the question of the origin of this $P$-breaking Higgs mass.

Recently we introduced Minimal Mirror Twin Higgs (MMTH) \cite{Barbieri:2016zxn}, where $P$ is broken only in the Yukawa couplings.  In the absence of an exotic cosmological history after the two sectors decouple (see~\cite{Craig:2016lyx,Chacko:2016hvu} for examples of such history), we showed that $P$-breaking in the Yukawa couplings is a necessity to solve the dark radiation problem, even if additional interactions allow the decoupling temperature to be arbitrary.
Also, a variety of candidates for mirror dark matter are possible that are not excluded and predict rich phenomenology\footnote{
The possibility to address both the dark matter and the dark radiation problems by Yukawa couplings of the light mirror fermions larger than the SM ones is proposed in \cite{Barbieri:2005ri}.}.  Furthermore, in MMTH a $P$-breaking Higgs mass term, necessary for the Twin Higgs mechanism, is generated by 1-loop radiative corrections.  We showed that MMTH has correlated signals in Higgs decays, direct detection of dark matter and dark radiation, over a region of parameter space where the fine-tuning for the electroweak scale is 10-50\%.

Nevertheless, MMTH itself leads to two questions: What is the origin of $P$ breaking in the Yukawa sector?  Given the large number of parameters in the Yukawa sector, how predictive can the theory be?  In Section~\ref{sec:flavor} we introduce a minimal flavor hierarchy for MMTH, defined in Eq.~(\ref{eq:Yukawa}), where the mirror fermion spectrum is predicted to leading order in terms of a single parameter $\epsilon'/\epsilon$. Such hierarchies arise in Froggatt-Nielsen theories~\cite{Froggatt:1978nt} with an Abelian flavor symmetry spontaneously broken by a small parameter $\epsilon$, as shown in Eq.~(\ref{eq:FN}), and they can also arise in extra-dimensional theories of flavor~\cite{Kaplan:2001ga}. 

In Section~\ref{sec:su5} we study in detail the resulting Higgs, dark radiation and dark matter signals in a particular model where the powers of $\epsilon$, the Froggatt-Nielsen charges, are compatible with SU(5) unification.  We give predictions for the Higgs signal strength and the amount of dark radiation, and focus on the nature and signals of mirror dark matter.  We show regions for hydrogen- and helium-like dark matter that are currently allowed by direct detection, self-scattering and relic ionization limits, and discover that there is a significantly larger parameter region for mirror neutron dark matter that is currently much less constrained.  We find that almost all regions for these dark matter candidates that are presently allowed can be probed by direct detection in experiments under way.

Variant models are briefly discussed in Section~\ref{sec:variant}. Although the predictions differ in detail, the broad picture is the same: all models with a single parameter describing charged fermion mass hierarchies are highly constrained by data.  Conclusions are drawn in Section~\ref{sec:concl} and several calculations and details are presented in Appendices \ref{sec:exdim} to \ref{sec:rc_EC}.

\section{Minimal Flavor Hierarchy}
\label{sec:flavor}

A key feature of the quark and charged lepton masses is their large hierarchies.  Any theory of flavor should incorporate a set of parameters $\epsilon_a \ll 1$ to describe these hierarchies.  Within the context of MMTH it is interesting to explore the possibility that the only breaking of $P$ arises spontaneously from a difference between these hierarchy parameters in the two sectors,  $\epsilon_a' \neq \epsilon_a$.  A general form for the $3 \times 3$ up, down and charged lepton Yukawa matrices in the two sectors in the effective theory below $\Lambda$ is
\begin{align}
y_{ij}(\epsilon_a) = \sum \lambda_{ij}^a \; \epsilon_a^{n_{ij}} \hspace{1in} y'_{ij}(\epsilon_a) = \sum \lambda_{ij}^a \; \epsilon'^{n_{ij}}_a
\label{eq:Yukawagen}
\end{align}
where $\lambda^a_{ij}$ are order unity and the same in each sector.  The powers $n^a_{ij}$ vary between theories, and the summation indicates that several such terms may be relevant for any $ij$.

In this paper we provide sharp predictions for MMTH by focussing on a simple scheme for flavor symmetry breaking in the effective theory below $\Lambda$, with a single hierarchy parameter in each sector so that the label $a$ may be dropped.  In this ``Minimal Flavor Hierarchy" each Yukawa matrix element is dominated by a single term of the form
\begin{align}
y_{ij} = \epsilon^{n_i} \, \lambda_{ij} \, \epsilon^{\bar{n}_j}  \hspace{1in}  y'_{ij} = \epsilon'^{n_i} \, \lambda_{ij} \, \epsilon'^{\bar{n}_j}.
\label{eq:Yukawa}
\end{align}
With this structure, the coupling to the $i$ ($j$) fermions on the left (right) receives a suppression of the hierarchy parameter to the $n_i$ ($\bar{n}_j$) power.  We stress that $P$ forces $n_i, \bar{n}_j$ and $\lambda_{ij}$ to be the same in the two sectors, while the spontaneous breaking of $P$ arises only via the single parameter $\epsilon' /\epsilon \neq 1$, which is constrained by data to typically be in the range of 2-3.

What is the UV completion of the theory that leads to the structure of (\ref{eq:Yukawa}) in the effective theory at the TeV scale?  Above $\Lambda$ the twin Higgs sector must be UV completed, for example in a composite Higgs~\cite{Batra:2008jy,Geller:2014kta,Barbieri:2015lqa,Low:2015nqa,Cheng:2015buv,Csaki:2015gfd,Cheng:2016uqk,Contino:2017moj} or supersymmetric theory~\cite{Falkowski:2006qq,Chang:2006ra,Craig:2013fga,Katz:2016wtw,Badziak:2017syq}.  Without addressing this completion, we can still discuss how the flavor breaking spurions $\epsilon^{n_i}, \epsilon^{\bar{n}_j} $ arise at high energies.  Possibilities include Frogatt-Nielsen (FN) \cite{Froggatt:1978nt} and extra-dimensional theories \cite{ArkaniHamed:1999dc,Kaplan:2001ga}.

We consider a FN theory with a U(1) flavor symmetry in each sector spontaneously broken by $\vev{\phi'} \neq \vev{\phi}$, which is the only breaking of $P$ in the theory.  The flavor structure of (\ref{eq:Yukawa}) results when the fermion charges $(Q_i, \bar{Q}_j)$ are chosen to be $(n_i, \bar{n}_j)$ and, for example, $\epsilon = \vev{\phi}/M$ and $\epsilon' = \vev{\phi'}/M$, where $M$ is the mass scale suppressing higher-dimensional operators which have order unity couplings $\lambda_{ij}$.  In summary
\begin{align}
\epsilon = \frac{\vev{\phi}}{M}, \hspace{0.5in} \epsilon' = \frac{\vev{\phi'}}{M},   \hspace{1in} (n_i, \bar{n}_j)=(Q_i, \bar{Q}_j).
\label{eq:FN}
\end{align}
The non-degeneracies between heavy FN fermions of the two sectors must not be so large that the Twin Higgs mechanism is upset. 
While there are many such models, they are greatly restricted since they must reproduce the known charged fermion masses. We find it convenient to take the charges to be integral and $\epsilon$ close to the Cabibbo angle, and study the predictions of three such models in detail. 

Small flavor parameters can arise from wavefunctions of zero-modes in extra dimensions~\cite{ArkaniHamed:1999dc}.  The analysis of this paper is based entirely on the Yukawa structure of (\ref{eq:Yukawa}) -- can it apply to extra-dimensional theories as well as 4D FN theories?  If the Higgs field is spread out in the bulk and fermion wavefunctions are Gaussian, as in \cite{ArkaniHamed:1999dc}, then the Yukawa matrix elements do not have the form of (\ref{eq:Yukawa}) as the overlap integral of the two fermion wavefunctions does not factor into a suppression factor for each fermion.  However, if the Higgs is localized in the bulk at $y_H$, the structure of $y_{ij}$ in (\ref{eq:Yukawa}) arises for any form of the wavefunctions of the fermions in the bulk, with $ \epsilon^{n_i} = \psi_i(y_H)$ and $ \epsilon^{\bar{n}_j} = \psi_j(y_H)$, and $\lambda_{ij}$ is the brane-localized coupling at $y_H$~\cite{Kaplan:2001ga}.  However, it is not clear what spontaneous breaking in the higher-dimensional set up would lead to $\epsilon' \neq \epsilon$ while leaving the powers $n_i, \bar{n}_j$ the same in both sectors. 

In Appendix~\ref{sec:exdim} we give two examples of how this could happen.   In one example, the fermions of the two sectors each live on orthogonal $S^1/Z_2$ spaces that intersect at the Higgs brane in a 2D bulk.  The parity $P$ interchanges these two spaces and is spontaneous broken by compactification to give different lengths, $L' \neq L$. We find the flavor structure of (\ref{eq:Yukawa}) is reproduced with
\begin{align}
\epsilon = e^{- \mu L}, \hspace{0.5in} \epsilon' = e^{- \mu L'},   \hspace{1in} (n_i, \bar{n}_j)=\left(\frac{M_i}{\mu}, \frac{\overline{M}_j}{\mu} \right).
\label{eq:6D}
\end{align}
where $M_i$ and $\overline{M}_j$ are bulk masses of the fermions and $\mu$ is an arbitrary scale which we choose to give $\epsilon$ close to the Cabibbo angle.

\section{$SU(5)$-Compatible Model}
\label{sec:su5}
In this section we investigate the prediction of a model with a U(1) flavor symmetry.
We consider a model consistent with the embedding of quarks and leptons into $SU(5)$ multiplets.
We discuss the mass spectrum of mirror fermions, its effect on the Higgs signal, dark matter phenomenology, and the amount of the dark radiation.
We expect the main features of the results to be similar for other $U(1)$ charge assignment as long as the observed fermions mass hierarchy is well reproduced, as in the two other models briefly discussed in Section~\ref{sec:variant}.

\subsection{Mass spectrum of  mirror fermions}

In this section we study U(1) flavor charges of fermions consistent with $SU(5)$~\cite{Babu:2003zz,Babu:2004th,Babu:2016aro}:
\begin{align}
\label{eq:charge su5}
Q,\bar{u},\bar{e}:~(4,2,0),~~
\bar{d},L:~(4,3,3).
\end{align}
The three numbers in each parenthesis denote charges of the first, second and third generation fermions, respectively.
Using this structure in Eqs.~(\ref{eq:Yukawa}) and (\ref{eq:FN}), the Yukawa couplings of the Standard Model (SM) fermions are given by
\begin{align}
y_t \sim 1 + O(\epsilon^4),~
y_c \sim \epsilon^4\left( 1 + O\left(\epsilon^4\right)\right),~~
y_u \sim \epsilon^8\left( 1 + O\left(\epsilon^4\right)\right)\nonumber \\
y_b \sim \epsilon^3\left( 1 + O\left(\epsilon^2\right)\right),~
y_s \sim \epsilon^5\left( 1 + O\left(\epsilon^2\right)\right),~~
y_d \sim \epsilon^8\left( 1 + O\left(\epsilon^2\right)\right)\nonumber\\
y_\tau \sim \epsilon^3\left( 1 + O\left(\epsilon^2\right)\right),~
y_\mu \sim \epsilon^5\left( 1 + O\left(\epsilon^2\right)\right),~~
y_e \sim \epsilon^8\left( 1 + O\left(\epsilon^2\right)\right),
\label{scaling}
\end{align}
where $\epsilon = \vev{\phi} / M$ and order unity coefficients from the $\lambda_{ij}$ are omitted.
Note that there is a correction of $O(\epsilon^2)$ or $O(\epsilon^4)$ to the leading order $\epsilon^n$ terms.
The derivation of the leading and correction terms are given in Appendix~\ref{sec:scaling} for down-type quarks.
The quality of the $SU(5)$ model as an explanation of the flavour hierarchy is exhibited in Appendix~\ref{sec:flavor_app}.

\begin{figure}[t]
\centering
\includegraphics[clip,width=.6\textwidth]{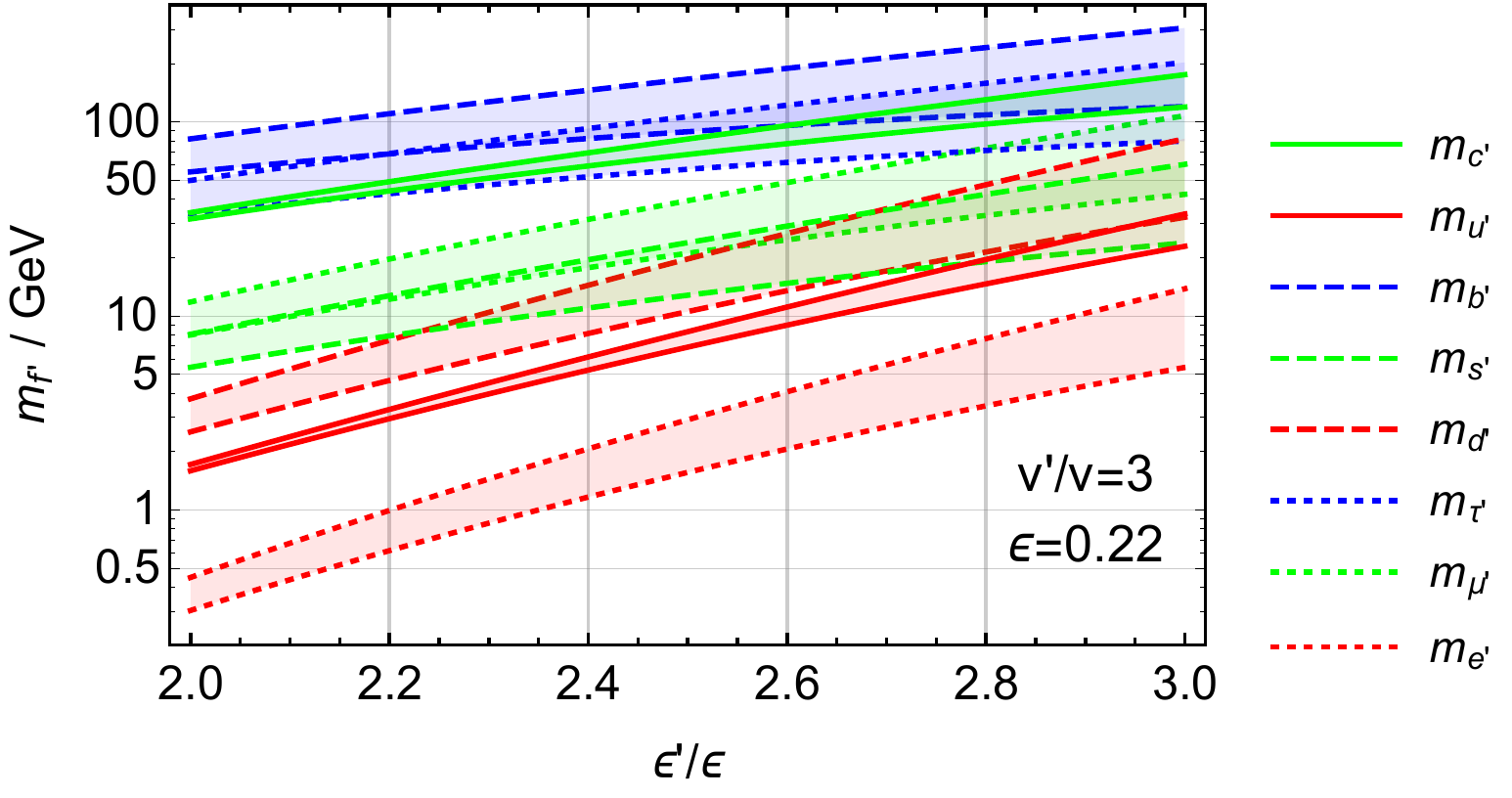}
\includegraphics[clip,width=.6\textwidth]{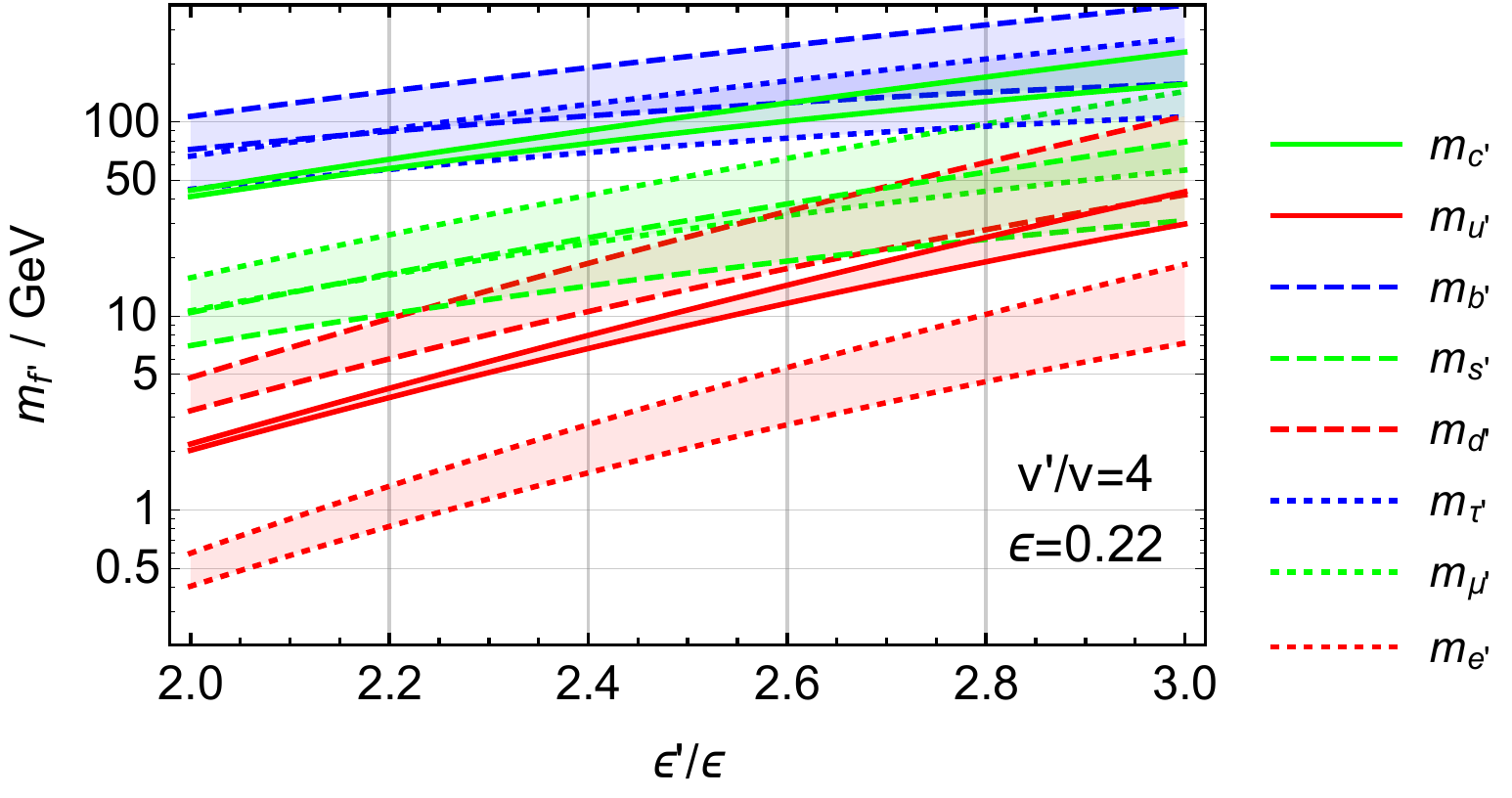}
\caption{
The mass spectrum of mirror fermions following from (\ref{scaling}) and (\ref{yukratio}).   The shaded bands, showing deviations from the simple scaling law, correspond to $|\delta_f| <1$.
Central values are taken for SM Yukawa couplings.
}
\label{fig:mf}
\end{figure}

For a fermion $f$ with a dependence $y_f \sim \epsilon^n \left(1 + O\left( \epsilon^m \right)   \right)$,
the ratio of the Yukawa couplings of the corresponding mirror fermion to that of the SM fermion, at the same scale above both masses, is given by
\begin{align}
\frac{y_{f'}}{y_f} =  \left( \frac{\epsilon'}{\epsilon}\right)^n \left( 1 + \delta_f\epsilon^{'m} - \delta_f\epsilon^{m} \right),
\label{yukratio}
\end{align}
where $\delta_f$ depend on the $\lambda_{ij}$ and hence are unknown $O(1)$ constants.
It should be noted that the top quark has $n=0$, and hence the SM and the mirror top yukawa couplings are  the same (up to small corrections of relative order $\epsilon^{'4},   \epsilon^4$)
which is required to suppress a too large correction to the Higgs mass term~\cite{Craig:2015pha}.
We use values of the SM Yukawa couplings shown in Table~\ref{tab:yukawa} at the renormalization scale $\mu=m_Z$~\cite{Antusch:2013jca}.
In Figure~\ref{fig:mf}, we show the masses of mirror fermions, including renormalization by the strong coupling.
The bands show the uncertainty due to the unknown constants $\delta_f$, and correspond to $|\delta_f|<1$.
The SM Yukawa couplings $y_u$ and $y_d$ suffer uncertainties of $30\%$ and $10\%$, but we assume central values in Figure~\ref{fig:mf}.

\begin{table}[tb]
\caption{Yukawa couplings of the SM fermions at the renormalization scale $\mu = m_Z$.}
\begin{center}
\begin{tabular}{|c|c|c|c|c|c|c|c|c|}
\hline
$y_e$ & $y_\mu$ & $y_\tau$ & $y_d$ & $y_u$ & $y_s$ & $y_c$ & $y_b$ & $y_t$ \\ \hline
$2.8\times 10^{-6}$ & $5.9\times 10^{-4}$ & $1.0\times 10^{-2}$   & $1.6\times 10^{-5}$ &  $7.4\times 10^{-6}$ &  $3.1\times 10^{-4}$ &  $3.6\times 10^{-3}$ &  $1.6\times 10^{-2}$ & $0.99$ \\ \hline

\end{tabular}
\end{center}
\label{tab:yukawa}
\end{table}%

The mass spectrum of the mirror particles also depends on the dynamical scale of  mirror QCD, $\Lambda_{QCD}'$.
 To estimate $\Lambda_{QCD}'$ and the mirror QCD phase transition temperature $T_c'$, we first take the mirror top quark mass to be $4 m_t$, corresponding to $v'/v=4$, and the other mirror quark masses to be $50$ GeV, and solve the renormalization group running of the mirror QCD coupling constant. We then find the renormalization scale such that $6/g_3^{'2} = 3.2$, we match the scale with the inverse of the lattice spacing and we estimate $T_c'$ based on the lattice calculation in~\cite{Okamoto:1999hi}. To estimate $T_c'$ for generic quark masses, we then use the scaling by the one-loop renormalization group equation.
The mirror QCD phase transition temperature is given by
\begin{align}
\label{eq:Tc}
T_c' \; \simeq \; & 2.3~{\rm GeV} \left( \frac{m_{t'}}{690~{\rm GeV}}\right)^{2/33} \prod_{q=d,s,b,u,c}\left( \frac{m_{q'}}{50~{\rm GeV}}\right)^{2/33}
\nonumber \\
\; \simeq \; &  2.1~{\rm GeV}\left( \frac{v'/v}{4}\right)^{4/11} \left( \frac{\epsilon'/ \epsilon}{ 2.5}\right)^{56/33}.
\end{align}
Note that the last expression does not depend on the $\delta_f$'s, as they should be cancelled with each other in the determinant of the mass matrix.

In the following sections we consider $\epsilon' / \epsilon$ in the range of 2-3, and find that experimental constraints will further reduce the allowed range. This range gives an origin for the needed breaking of Parity in the Higgs potential via the difference $y_f'\neq y_f$ in the Yukawa couplings of the light fermions~\cite{Barbieri:2016zxn} as well as the small difference between $y_t$ and $y_t'$.

We comment on the effect of the mass splitting between the heavy FN fields, which are introduced to generate the structure in  Eq.~(\ref{eq:FN}). We first consider the case where none of the masses of heavy fermions vanishes for $\epsilon=0$, which we assume in this paper.
Through the mixing between fermions, a small mass difference of $m'/m = 1 + O(\epsilon'^2)$ is expected, where $m$ and $m'$ are the mass scale of the heavy SM FN fermions and that of the heavy mirror FN fermions, respectively,
Although a difference between the gauge couplings $g_{3,2,1}$ and $g'_{3,2,1}$ is induced due to a threshold effect,
its effects on the breaking of the Parity in the Higgs potential is negligibly small.
The difference between $g_{3}$ and $g_3'$ does not affect the estimation of $\Lambda'_{QCD}$ and hence of $T_c'$ at the one-loop level, as the product of the fermion masses including light fermions are not affected by the mixing, and Eq.~(\ref{eq:Tc}) remains intact.

It is also possible that some of the heavy fermion masses vanishes for $\epsilon=0$.
In this case, $\epsilon\neq \epsilon'$ directly affects the mass splitting of those heavy fermions, and a mass splitting of $m'/m \sim \left(\epsilon'/\epsilon \right)^n$ is expected.
The Parity breaking threshold correction to the gauge coupling constant is given by
\begin{align}
\frac{\alpha'_i-\alpha_i}{\alpha_i} \simeq  \frac{\alpha_i}{2\pi}N {\rm ln}\left(\frac{\epsilon'}{\epsilon}\right)^n,
\end{align}
where $N$ is the multiplicity of the FN fermions with a large mass splitting.
As long as $\alpha'_{2(3)}-\alpha_{2(3)}/\alpha \lesssim 0.2 (0.5)$, the Parity breaking effect on the Higgs potential is small~\cite{Barbieri:2016zxn}, which requires $N n \lesssim 40$.
A displacement of $T_c'$ as well as the mirror electromagnetic gauge coupling is to be expected, which affect the amount of the dark radiation and the constraint on dark matter.
See Appendix A for analogous considerations when the Minimal Flavor Hierarchy arises from extra dimensions.

%
\subsection{Higgs signal}
\label{sec:hdecay}

\begin{figure}[t]
\centering
\includegraphics[clip,width=.48\textwidth]{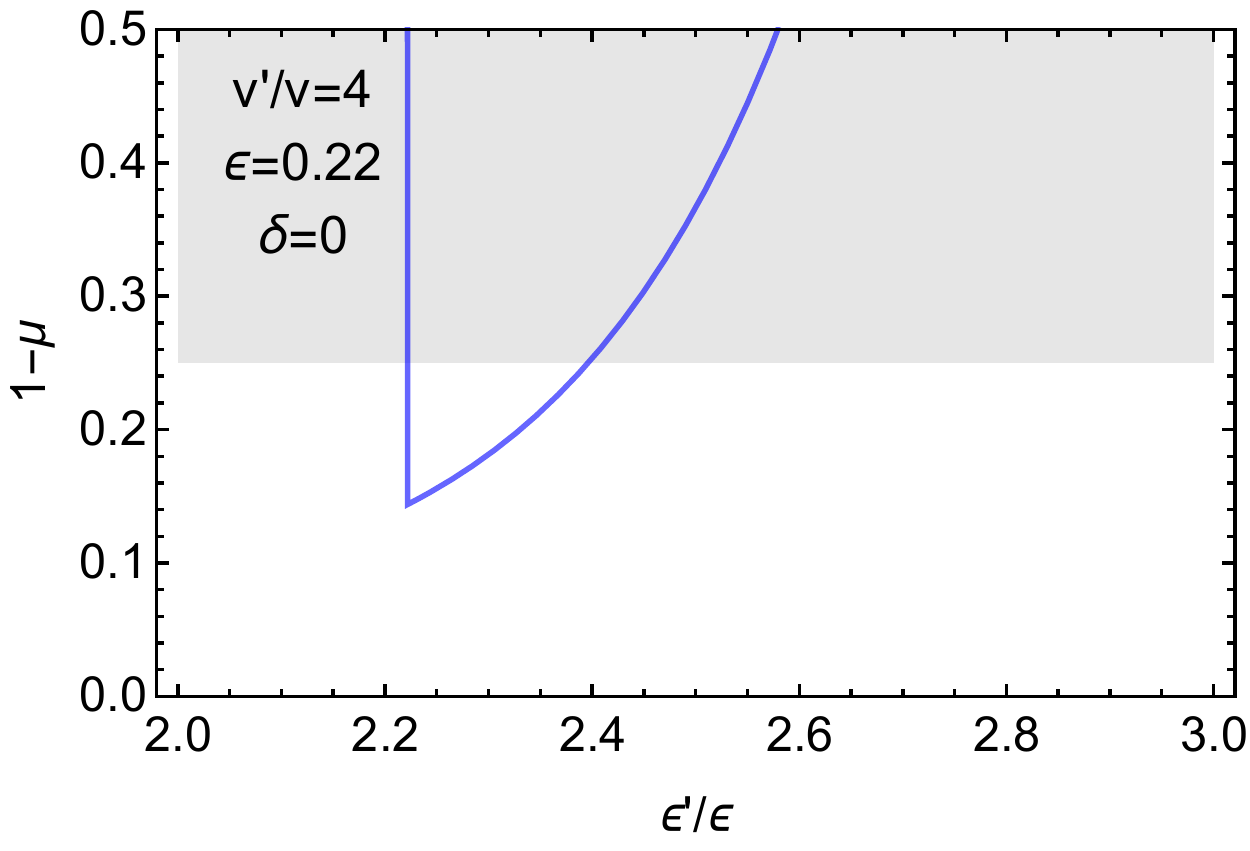}
\includegraphics[clip,width=.48\textwidth]{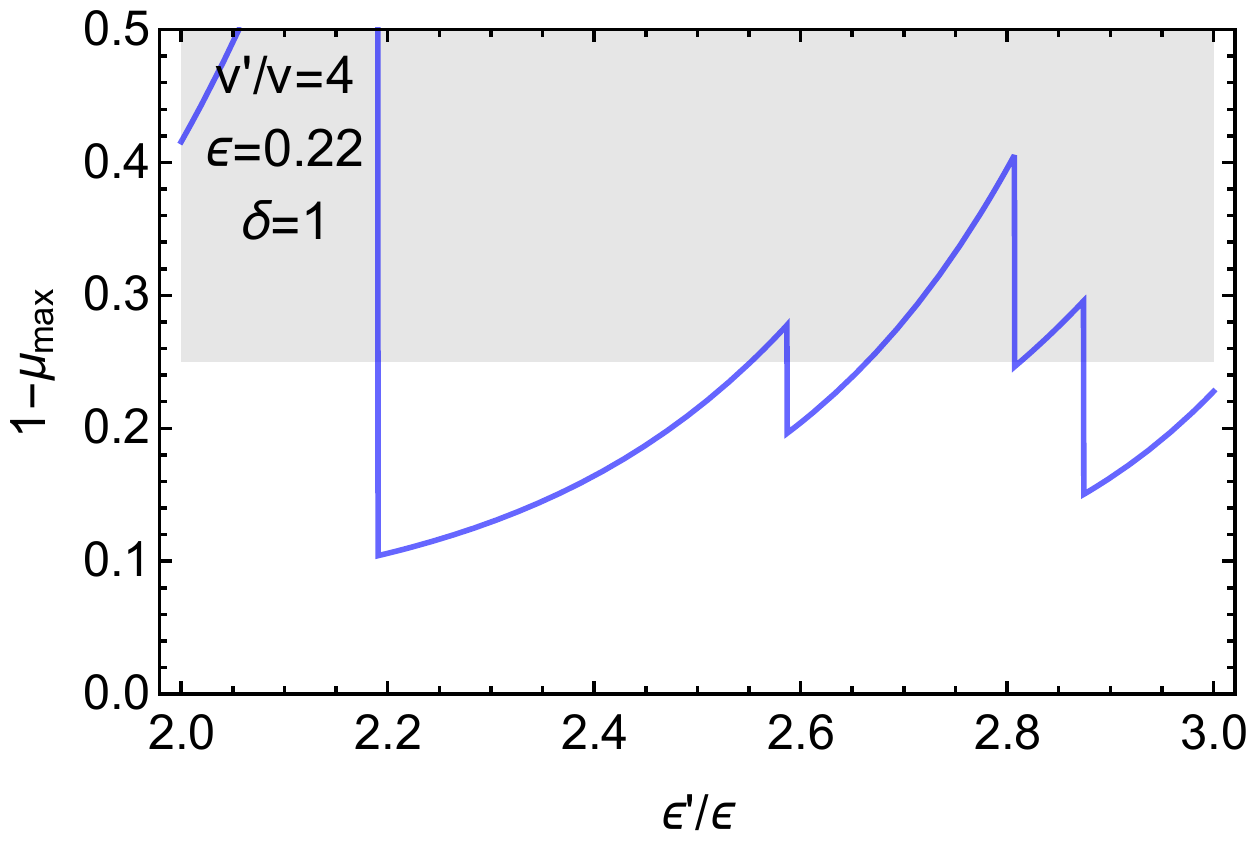}
\includegraphics[clip,width=.48\textwidth]{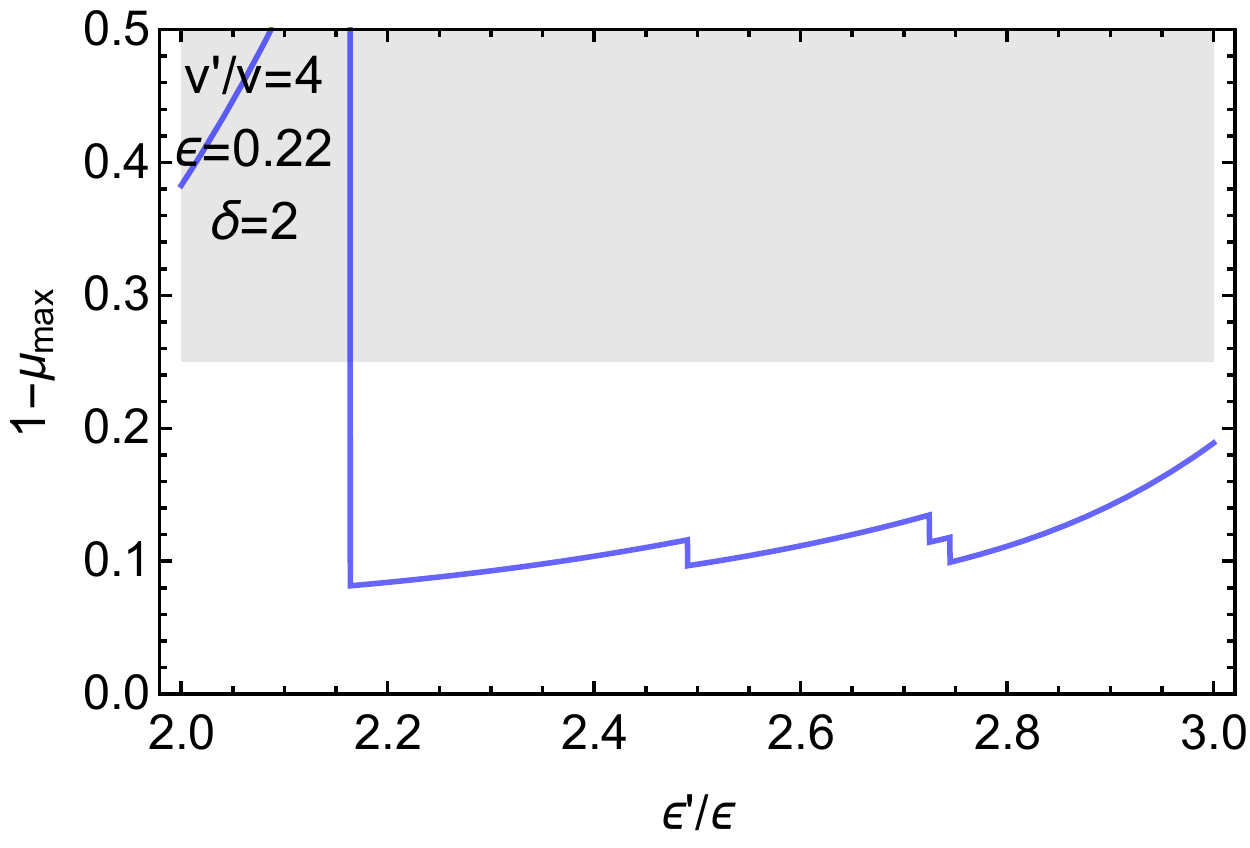}
\includegraphics[clip,width=.48\textwidth]{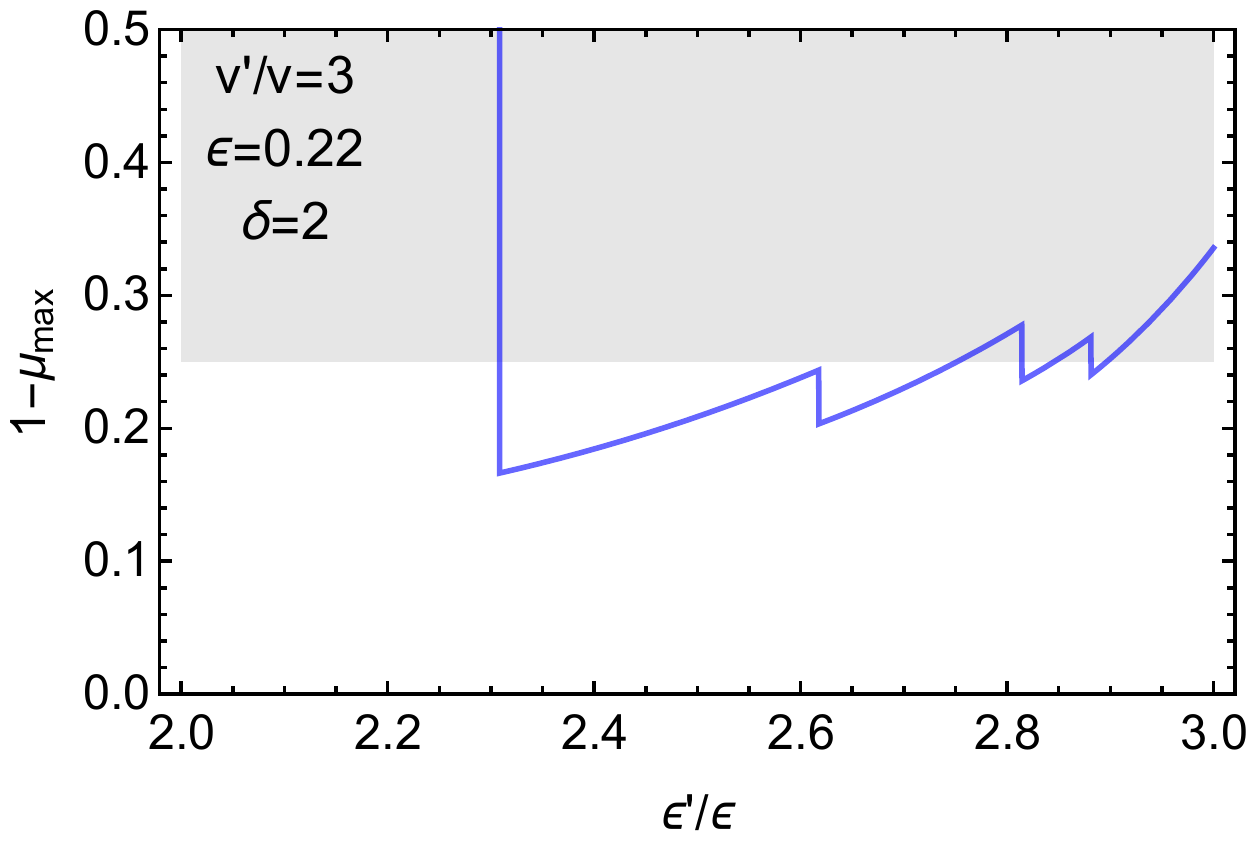}
\caption{
Prediction for the Higgs signal strength.
Panels with $\delta=1,2$ have the mass spectrum of mirror fermions chosen to minimize the invisible decay of the Higgs.
Decays to $c'$ exclude $\epsilon'/\epsilon$ less than about 2.2.  
}
\label{fig:hdec}
\end{figure}

In Twin Higgs models, the signal of the SM-like Higgs, $h$, is affected in two ways.
First, $h$ is an admixture of the two original doublets $H$ and $H'$, 
\begin{align}
h = c_\gamma H + s_\gamma H',~~~s_\gamma\equiv {\rm sin} \gamma \simeq v/v',
\end{align}
so that the couplings between $h$ and two SM particles are reduced by a factor of $c_\gamma$.
Second, $h$ also couples to a pair of mirror particles, so that it will decay to mirror fermions lighter than $m_h/2$ via the interaction
\begin{align}
{\cal L} \supset - y_{f'} \; H' f_L' \bar{f}_R'  \; \rightarrow \; - \frac{v}{\sqrt{2}v'} \, y_{f'} \; h f_L' \bar{f}_R' \; = \; -   \frac{v m_{f'}}{\sqrt{2}v^{'2} \delta_{f',m_h}} \; h f_L' \bar{f}_R'.
\end{align}
Here, $\delta_{f',\mu} \equiv y_{f'}(m_{f'}) / y_{f'}(\mu)$ encodes the effect of renormalization between a scale $\mu$ and $m_{f'}$.
These decays lead to an invisible branching ratio for $h$
\begin{align}
{\rm Br}_{\rm inv} = {\rm Br}(h\rightarrow f' \bar{f}') \simeq  0.1 \times \left( \frac{3}{v'/v} \right)^4 \sum_{f',2m_{f'}<m_h}  \frac{N_{f'}}{3}(\frac{m_{f'}}{10 {\rm GeV}})^2 
\delta_{f',m_h}^{-2}
\end{align}
where phase space has been neglected.
The invisible branching ratio, together with the reduction of the Higgs coupling to SM particles, results in a universal deviation from unity of the Higgs signal-strengths at colliders into any SM final state,
\begin{align}
1-\mu = 1- c_\gamma^2(1-{\rm Br}_{\rm inv})\simeq s_\gamma^2 +{\rm Br}_{\rm inv}.
\end{align}
In Figure~\ref{fig:hdec}, we show predictions on $1-\mu$ for $v'/v = 4$ and $3$.
The value of $\delta$ denotes the maximum absolute value of $\delta_f$ we allow.
We choose the sign and value of each $\delta_f$ so that $\mu$ becomes as large as possible.
Specifically, we first try $\delta_f=\delta$, and see if $m_{f'}>m_h/2$. If so, we choose $\delta_f$ to be $\delta$. If not, we choose $\delta_f =- \delta$.
The figure shows that $1-\mu$ can be smaller than the experimental bound, $\mu>0.75$~\cite{Khachatryan:2016vau} for ranges of $\epsilon'/\epsilon$ that depend on $v'/v$ and $\delta$.
Here we have adopted the constraint on the gluon fusion channel, as it has the smallest uncertainty.
$\epsilon'/\epsilon\lesssim 2.2$ is excluded because the mirror charm quark becomes  lighter than $m_h/2$.

%
\subsection{Mirror Dark Matter}

The lightest mirror baryon and the lightest mirror charged particle are stable, and may compose the dark matter of the universe.
We assume that the mirror sector also has non-zero matter asymmetry and that the asymmetric component of mirror matter explains the observed dark matter density.
Most of the discussion in this Section is applicable to generic mirror world scenarios.
Dark matter phenomenology in the mirror world scenario with $y=y'$ is discussed in~\cite{Berezhiani:2000gw,Ignatiev:2003js} and more recently in~\cite{Fukuda:2015ana,Fukuda:2017ywn}.

\subsubsection{Dark matter candidates}

\begin{figure}[t]
\centering
\includegraphics[clip,width=.48\textwidth]{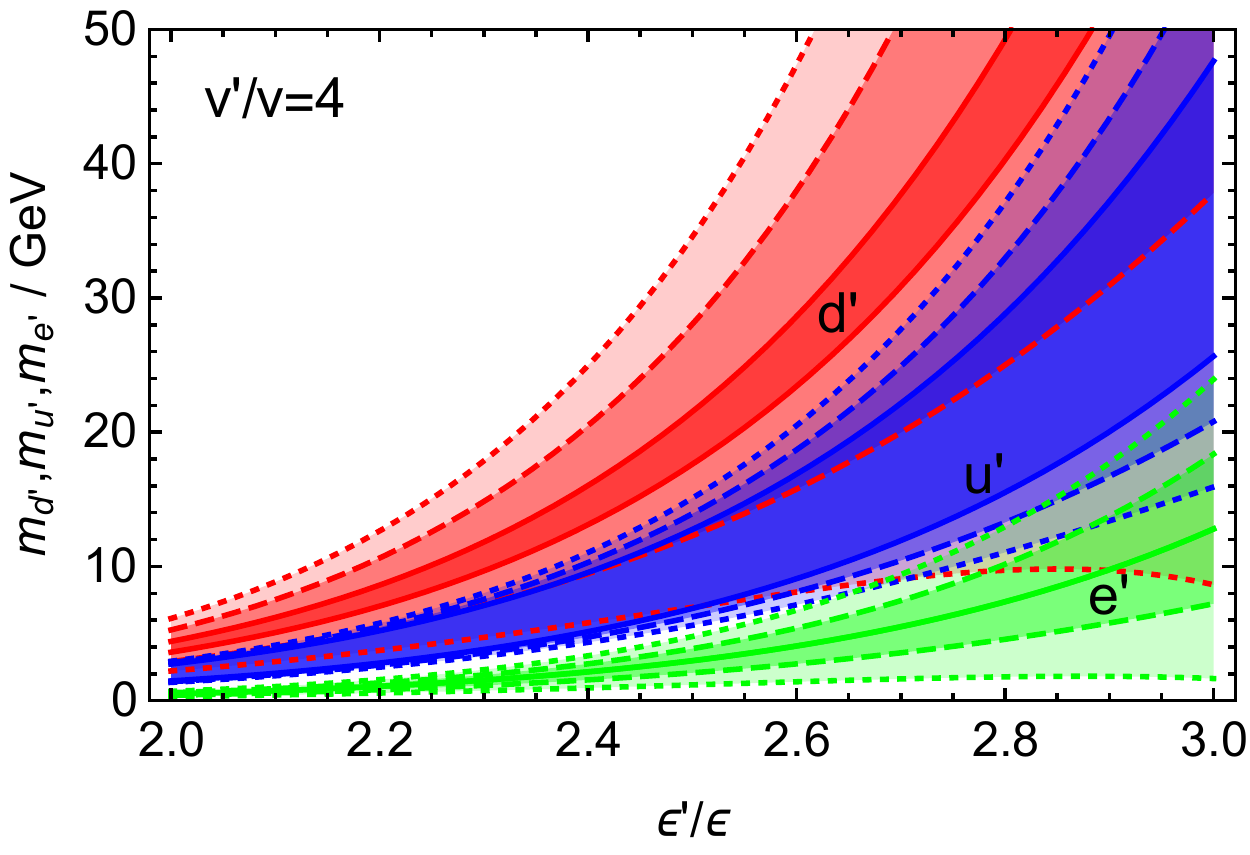}
\includegraphics[clip,width=.48\textwidth]{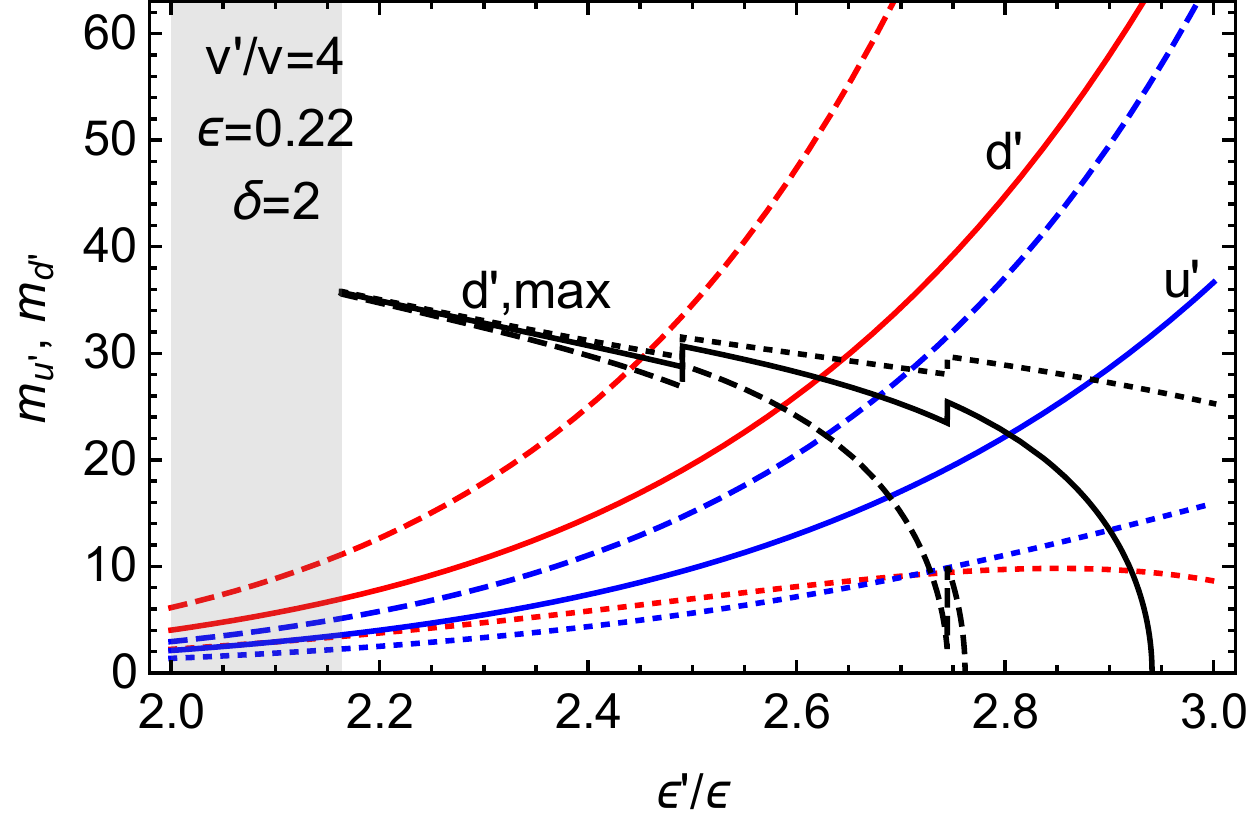}
\caption{
Left panel: the masses of $u'$, $d'$ and $e'$, including uncertainties from the SM up and down quark Yukawa couplings.
Solid, dashed and dotted lines show the cases with $\delta =$0, 1, 2 respectively. Right panel: red and blue lines show the central value and $\delta = 2$ ranges of the $d'$ and $u'$ masses, without any SM Yukawa uncertainties.  Black lines show the maximum $d'$ mass allowed from the Higgs signal strength, showing the central and $\delta = 2$ range as $u'$ and $d'$ masses are varied.
}
\label{fig:mude}
\end{figure}

The second and third generation mirror fermions decay into the first generation, so only the mirror up quark, down quark or electron may be stable.
The dark matter candidate depends on the mass relation between them.
In the left panel of Figure~\ref{fig:mude} we show the masses of $d'$, $u'$ and $e'$:
solid, dashed and dotted lines show ranges with $|\delta_f|\leq 0$, $1$ and $2$, respectively,
and uncertainties of the SM $u$ and $d$ Yukawa couplings, which we take to be $30\%$ and $10\%$, are included.  The right panel of Figure~\ref{fig:mude} shows the maximum value of $m_{d'}$ allowed by the Higgs signal strength, for values of $\delta$ described in the caption.

In most of the parameter space $m_{e'} < m_{u'} + m_{d'}$, so that the mirror electron is stable.  Depending on $m_{u'} , m_{d'}$ there are four candidates for the lightest baryon: $B'_{uuu}, B'_{uud},B'_{udd},B'_{ddd}$.  The $B'_{uuu},B'_{ddd}$ states are spin 3/2 and have an additional strong interaction contribution to their masses, $\Delta \sim T'_c$, compared to the spin 1/2 states $B'_{uud},B'_{udd}$.  From Figure~\ref{fig:mude} we see that there is a large region with $m_{d'} > m_{u'}$ and $m_{d'} - m_{u'} \gg m_{e'}$ so that the lightest baryon is $B'_{uuu}$ and $B'_{uud},B'_{udd},B'_{ddd}$ are unstable. The DM candidate is $(He)'_*$ composed of $(uuuee)$. (The star subscript indicates that the flavor structure of the nucleus differs from the corresponding SM case.)  The constraints on $(He)'_*$ dark matter are discussed later.

\begin{figure}[t]
\centering
\includegraphics[clip,width=.45\textwidth]{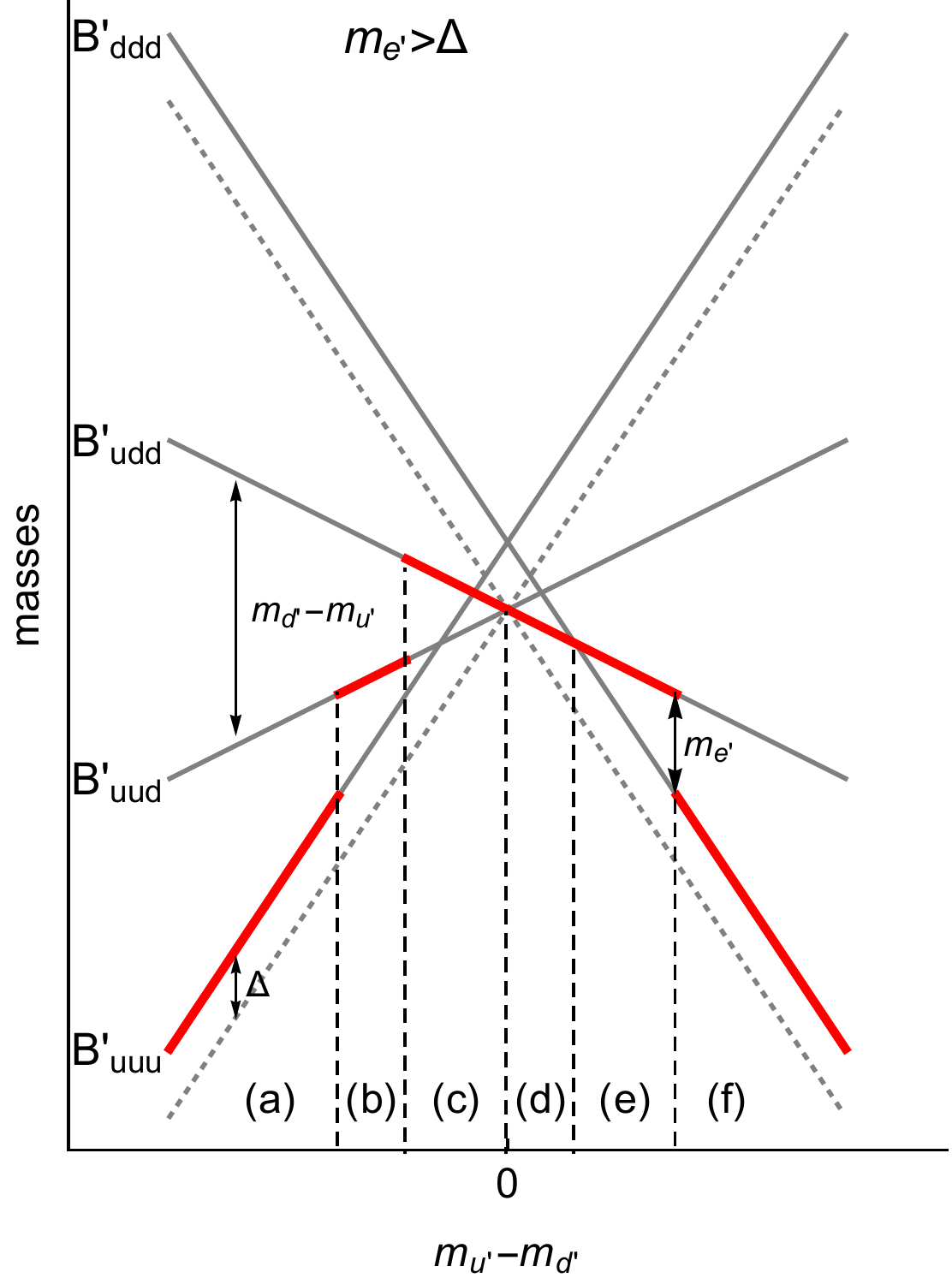}
\includegraphics[clip,width=.45\textwidth]{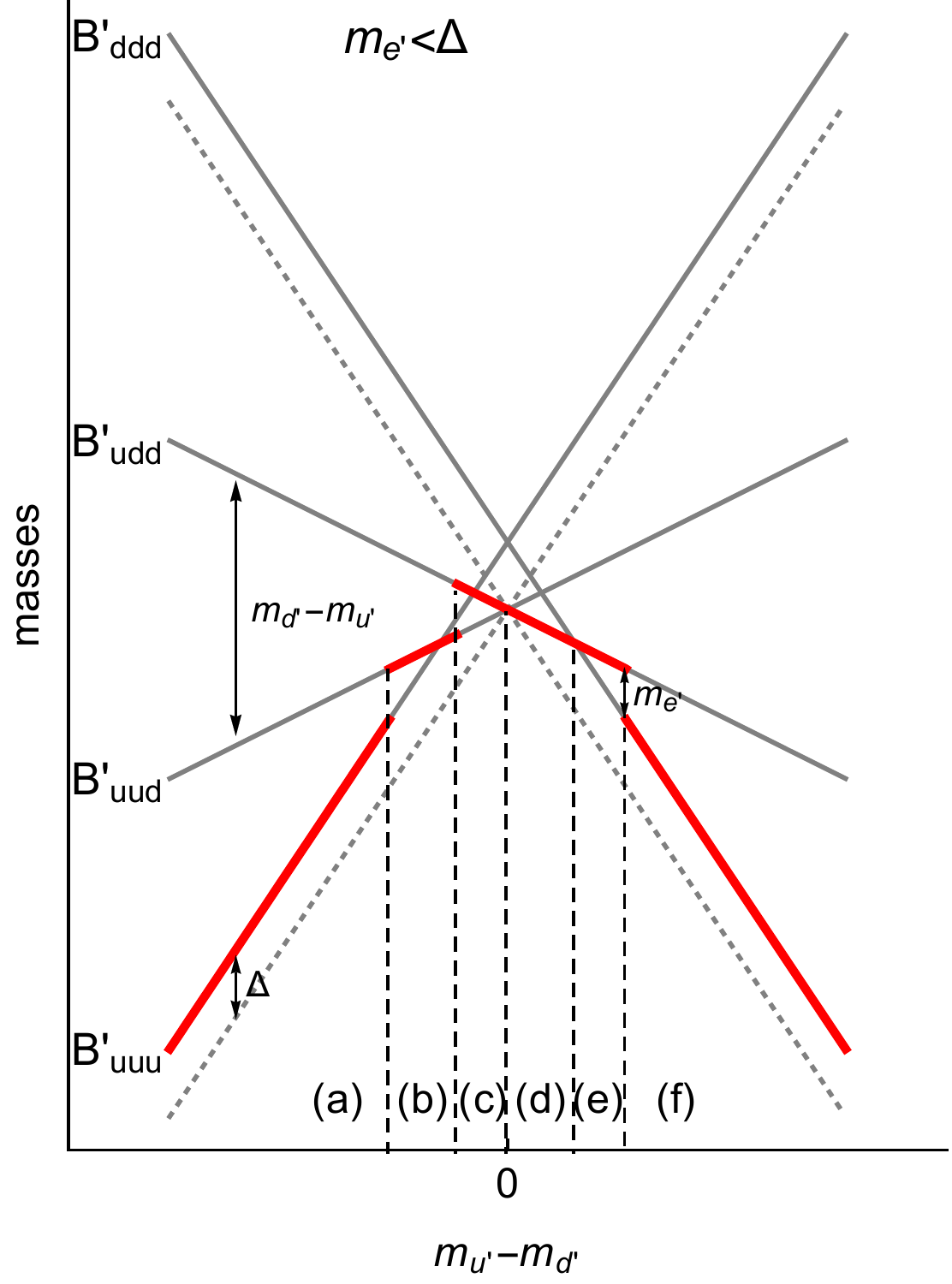}
\caption{
The mass spectrum of the mirror baryons as a function of $m_{u'}-m_{d'}$.
The dotted lines show the masses of $B'_{uuu}$ and $B'_{ddd}$ ignoring the contribution from the mirror QCD dynamics to the mass difference between the lightest spin-3/2 baryons and the lightest spin-1/2 baryons, $\Delta$.
The red lines show the mirror baryon of the dark matter candidate.
}
\label{fig:Bmass}
\end{figure}

In regions where $m_{d'} - m_{u'} \sim m_{e'}$, the other baryons, $B'_{uud},B'_{udd},B'_{ddd}$, could be the lightest baryon, and $B'_{uuu},~B'_{uud},~B'_{udd},~B'_{ddd},~e'$ may all be stable.  The spectrum of these baryons is sketched in Figure~\ref{fig:Bmass}, for $m_{e'} > \Delta$ ($m_{e'} < \Delta$) in the left (right) panel. In Appendix~\ref{sec:dueDM} we show that, after freeze-out of the mirror weak interactions at a temperature of about $ m_{e'}/18$, the baryon asymmetry is always carried by the lightest baryon, even if the heavier ones are stable. 

\begin{figure}[t]
\centering
\includegraphics[clip,width=.49\textwidth]{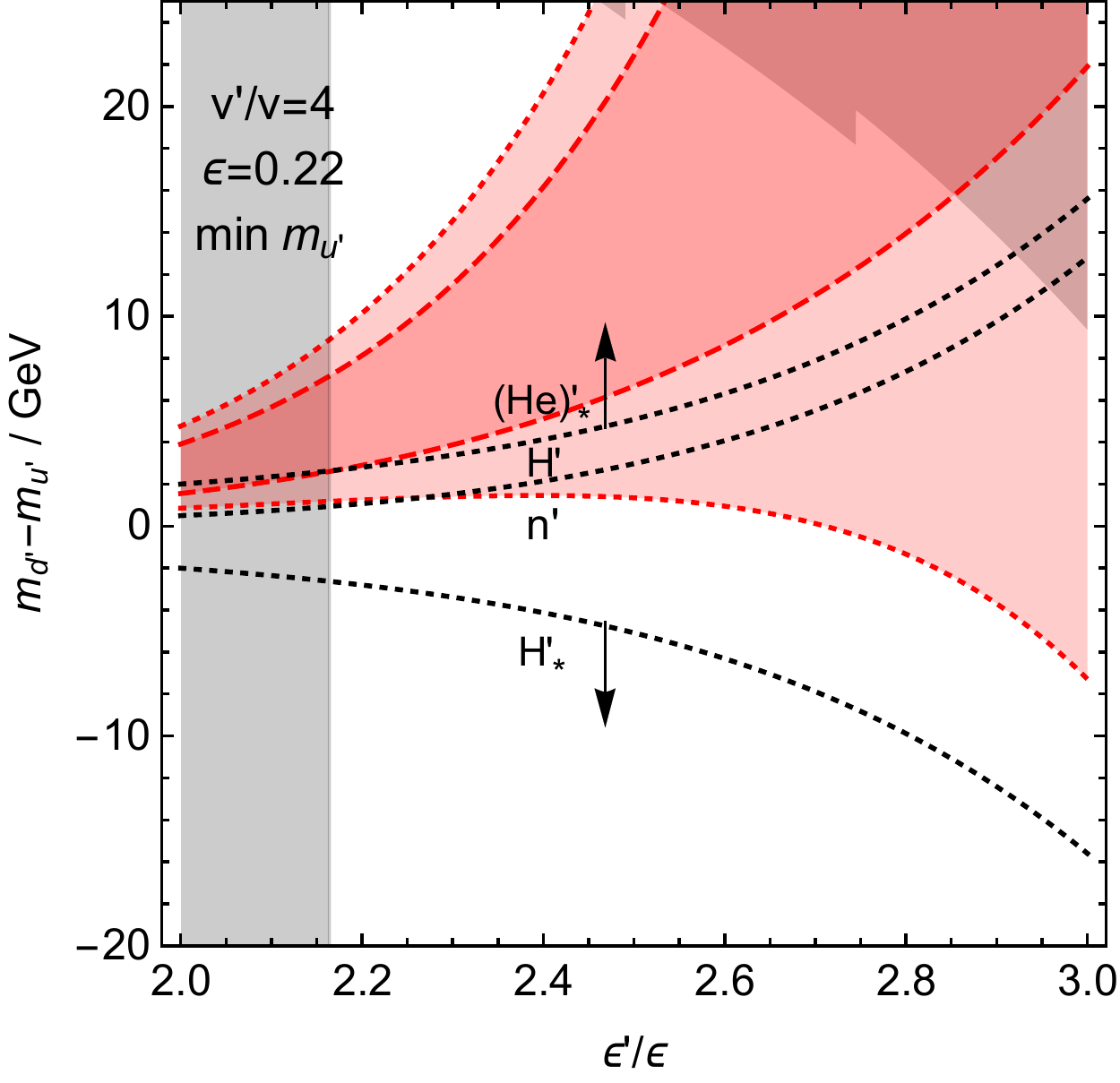}
\includegraphics[clip,width=.49\textwidth]{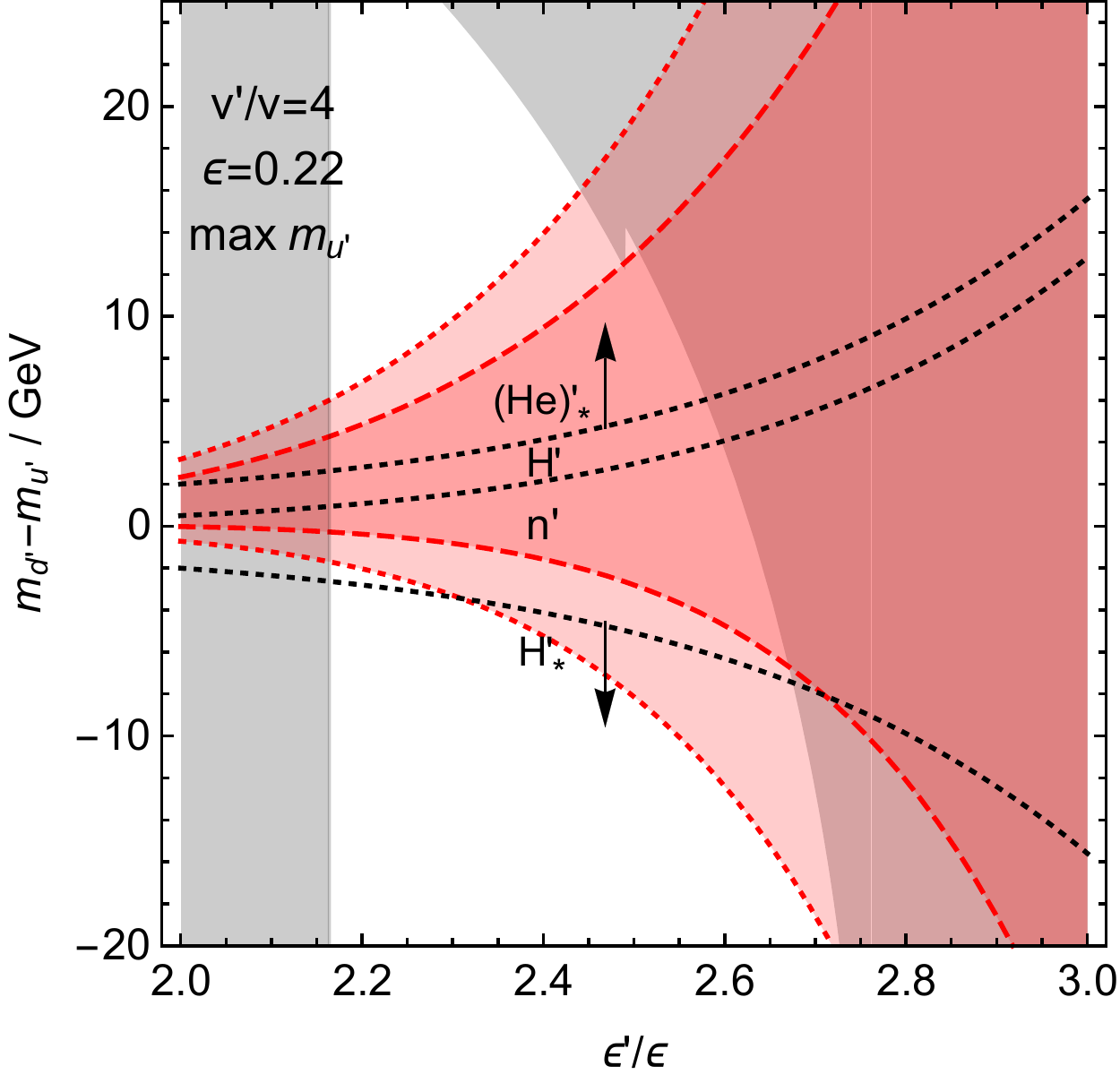}
\caption{
Dark (light) red shading gives the range of $m_{d'} - m_{u'}$  for $\delta =1 (2)$.  Black dashed lines separate regions where the DM candidate is $(He)'_*, H', n'$ and $H'_*$ with $\delta_e=0$.  The left (right) panel is for minimal (maximal) $m_{u'}$.  Gray shaded regions are excluded by the Higgs signal strength.   
The position of the upper and lower black dotted lines are uncertain and are shown for $\Delta = T'_c$.
}
\label{fig:mdiff}
\end{figure}

Hence there are four DM candidates 
\begin{align}
(He)'_*(uuuee), \hspace{0.3in} H'(uude), \hspace{0.3in} n'(udd), \hspace{0.3in} H'_*(ddd \bar{e})
\end{align}
Regions of parameter space leading to these four candidates are shown in Figure~\ref{fig:mdiff}, separated by black dashed lines, with the predicted regions in the $SU(5)$ model shown by dark (light) red shading for $\delta =1 (2)$, with $\delta_e=0$.  The $n'$ candidate is particularly important since the others are atoms and are significantly constrained by limits on self-scattering and relic ionization, as described below.  It is interesting and remarkable that the $n'$ region of Figure~\ref{fig:mdiff} is large, arising from a large region with $m_{d'} - m_{u'} \sim m_{e'}$, while the $H'$ and $H'_*$ regions are smaller.

While weak interaction freeze-out puts the baryon asymmetry into the lightest baryon, when atomic states form the electron capture process, if kinematically allowed, ensures that
\begin{align}
(He)'_*(uuuee) \rightarrow H'(uude), \hspace{0.2in}H'(uude) \rightarrow n'(udd) \hspace{0.2in} H'_*(ddd \bar{e}) \rightarrow n'(udd),
\end{align}
so that the DM candidate is the lightest of $(He)'_*, H', n'$ and $H'_*$.  It is the latter two processes that significantly enhance the $n'$ DM region. In Figure~\ref{fig:Bmass} the red line tracks the baryon of the DM candidate, and jumps where electron capture occurs, so that the DM candidate does not necessarily contain the lightest baryon.  The growth in the $n'$ DM region is particularly pronounced for large $m_{e'}$.  The resulting ranges of $m_{d'} - m_{u'}$ for each of the four DM candidates are shown in Table~\ref{tab:DM candidate}.

\begin{table}[t]
\caption{Ranges of $m_{d'} - m_{u'}\equiv \delta m_{d'u'}$ for the four Dark Matter candidates.}
\begin{center}
\begin{tabular}{|c||c|c|c|c|}
\hline
 & $m_{e'}+\Delta < \delta m_{d'u'} $ & $ m_{e'} <\delta m_{d'u'}< m_{e'}+\Delta  $ &  $ -\Delta-m_{e'} <\delta m_{d'u'}< m_{e'}  $ &  $ \delta m_{d'u'}< -m_{e'}-\Delta  $ \\ \hline
 DM & $B'_{uuu} + 2 e'$ & $B'_{uud} + e'$  & $B'_{udd} $ &  $B'_{ddd} + e' $ \\  \hline
\end{tabular}
\end{center}
\label{tab:DM candidate}
\end{table}%

\subsubsection{Direct detection via Higgs exchange}

\begin{figure}[t]
\centering
\includegraphics[clip,width=.7\textwidth]{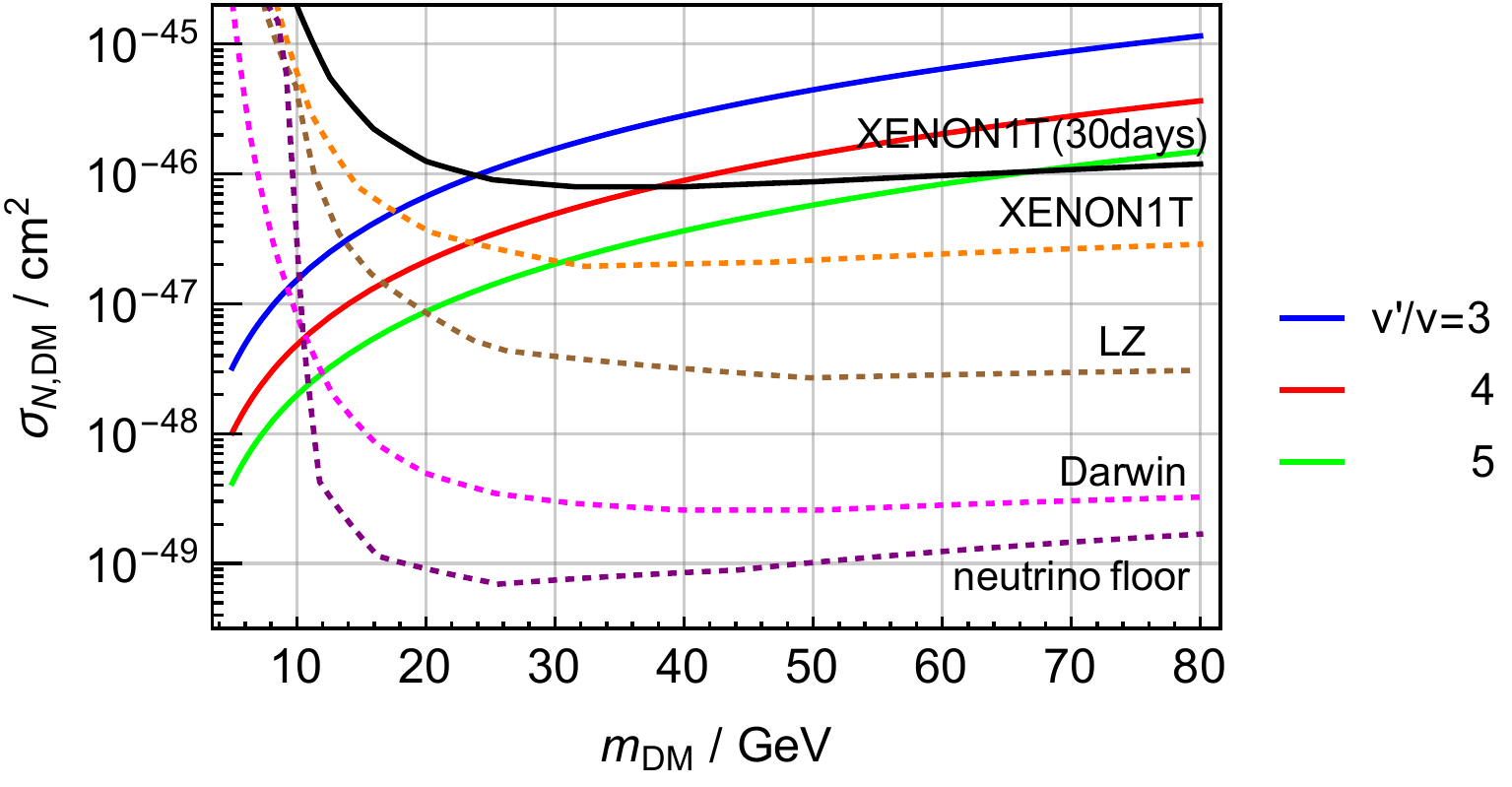}
\caption{
The scattering cross section between a dark matter particle and a SM nucleon as a function of the dark matter mass, which we assume is dominated by mirror fermion masses. The three full coloured 
lines correspond to $v'/v=3,4,5$.}
\label{fig:DD}
\end{figure}

Before investigating constraints and signals peculiar to each dark matter candidate, we discuss a signal universal to all the above candidates.
These dark matter particles interact with SM nucleons through the exchange of the SM-like Higgs, $h$, and can be observed in direct detection experiments~\cite{Barbieri:2016zxn,Craig:2015xla,Garcia:2015toa,Farina:2015uea}. The scattering cross section between a nucleon and a dark matter particle is given by~\cite{Barbieri:2016zxn}
\begin{align}
\sigma_{N,{\rm DM}} = \frac{0.028}{\pi} \frac{m_{\rm DM}^2 m_N^2}{v^{'4} m_h^4} \left( \frac{m_N m_{\rm DM}}{ m_N +m_{\rm DM} } \right)^2,
\end{align}
where $m_N$ and $m_{\rm DM}$ are the masses of the nucleon and the DM, respectively.
Here we assume that the mass of dark matter is dominated by mirror fermion masses.  This cross section is shown in Figure~\ref{fig:DD}.
We also show constraints from
the XENOT1T experiment (30days)~\cite{Aprile:2017iyp},
the expected sensitivities  of  XENON1T~\cite{Aprile:2015uzo}, LZ~\cite{Akerib:2015cja} and DARWIN~\cite{Aalbers:2016jon} experiments, as well as the neutrino floor~\cite{Billard:2013qya}.

%

\subsubsection{Constraint on $(He)'_*$ dark matter: Region (a) of Fig.~\ref{fig:Bmass}}

In Region (a) of Figure~\ref{fig:Bmass}, where $d'$ is sufficiently heavy, the lightest baryon is $B'_{uuu}$. Therefore the mirror matter asymmetry results in the asymmetric components of $B'_{uuu}$ and $e'$, which are stable cosmological relics. Once most of these combine into $(He)'_*$, they may explain the observed dark matter in the universe.

\begin{figure}[t]
\centering
\includegraphics[clip,width=.6\textwidth]{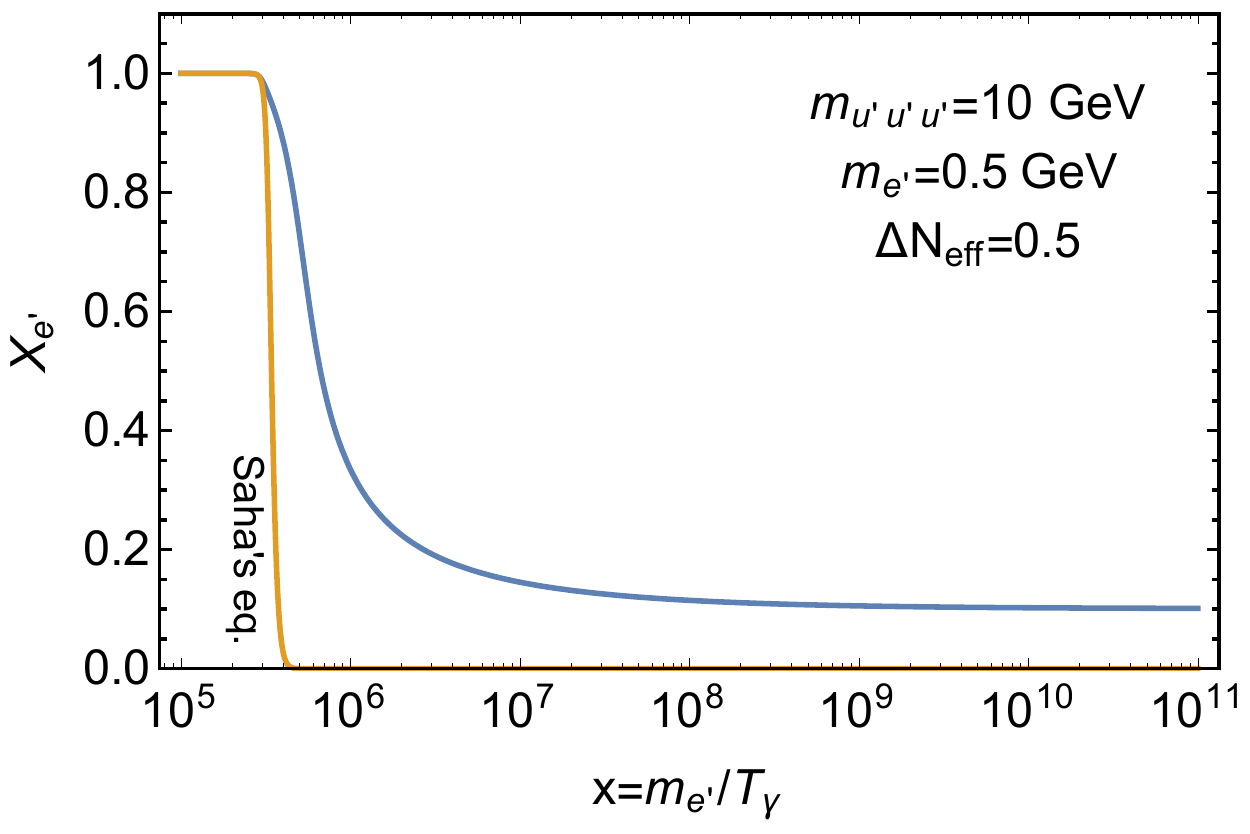}
\caption{
A sample evolution of the ionization fraction of $(He)'_*$.
}

\label{fig:He_rc}
\end{figure}

We calculate the recombination of $(He)'_*$, following the method described in~\cite{Seager:1999bc}, which calculates the recombination in the SM. We rescale recombination coefficients, etc, according to $m_{e'}/m_e$. This is applicable as long as $m_{u'u'u'} \gg m_{e'}$.
The temperature of mirror photons is determined via
\begin{align}
T_{\gamma'} / T_{\gamma} = \left( \frac{7}{29}  \Delta N_{\rm eff} \right)^{1/4} \left( \frac{4}{11}\right)^{1/3} \simeq 0.42 \left( \frac{\Delta N_{\rm eff}}{0.5}\right)^{1/4}.
\end{align}
A sample evolution of the ionization fraction of the mirror electron, $X_{e'}$, is shown in Figure~\ref{fig:He_rc}.
At low temperatures we find
\begin{align}
X_{e'} \simeq  0.05 \left(\frac{m_{u'u'u'}}{10~{\rm GeV}}\right)^{0.8} \left(\frac{m_{e'}}{0.23~{\rm GeV}}\right)^{0.8} \left( \frac{\Delta N_{\rm eff}}{0.5}  \right)^{1/4}
\end{align}
where we assume $m_{\rm u'u'u'}/m_e'  \gg 1$.
The sudden decoupling approximation from Saha's equation predicts $X_{e'}\propto m_{u'u'u'}m_{e'}$, but, as it can be seen in Figure~\ref{fig:He_rc}, the approximation is far from perfect.
Since the ionized components scatter with each other with a long-range force,  their fraction is constrained by the possible change of the mass-to-luminosity ratio in the Bullet Cluster~\cite{Markevitch:2003at,Randall:2007ph}, $X_{e'} \lsim 0.3$.

\begin{figure}[t]
\centering
\includegraphics[clip,width=.6\textwidth]{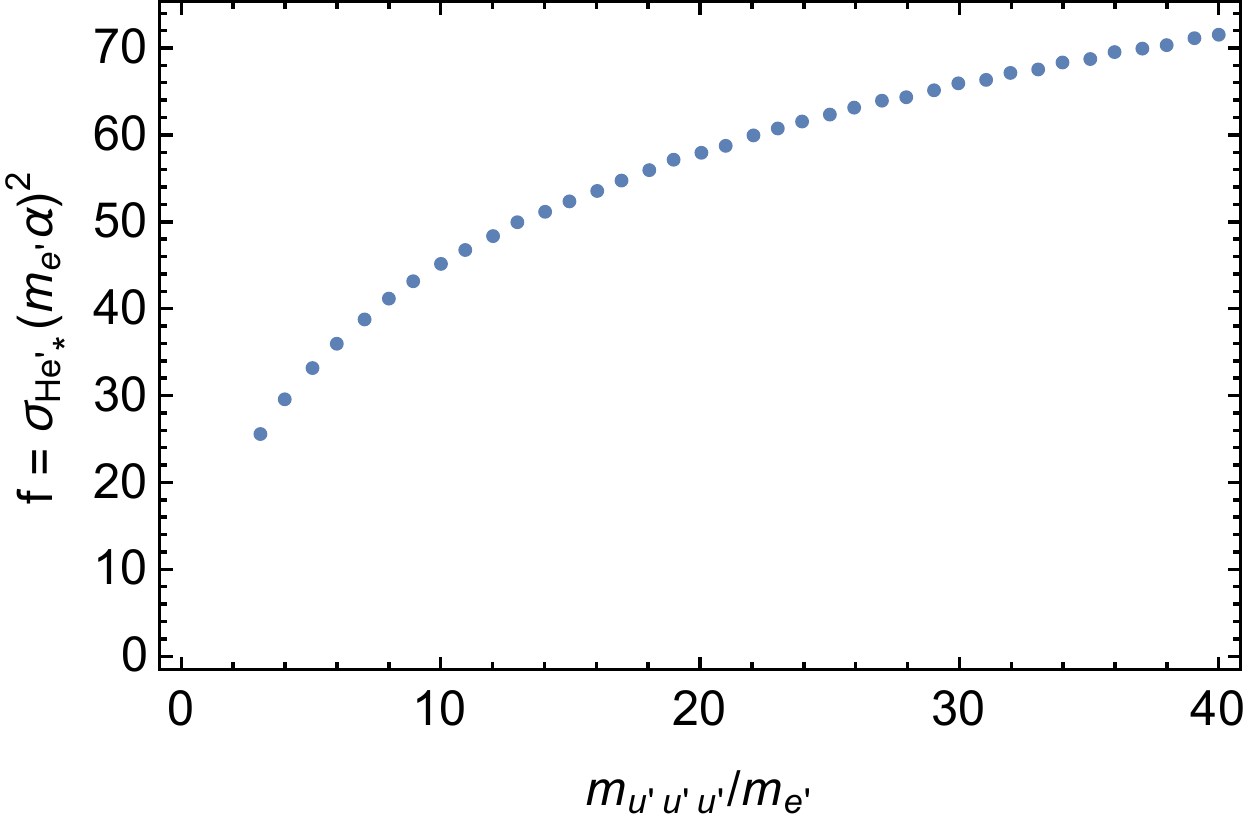}
\caption{
The normalized self-interaction cross section of $(He)_*'$, $f$, of Eq.~(\ref{eq:Xsec He}).
}

\label{fig:He_sca}
\end{figure}

The $(He)_*'$ self-interaction cross section  at low velocity is given by 
\begin{align}
\frac{\sigma}{m_{\rm DM}} \simeq \frac{f(m_{u'u'u'}/m_{e'})}{m_{e'}^2 \alpha^2} \frac{1}{m_{u'u'u'}} = 8.2~{\rm cm}^2/{\rm g} \times \frac{10~{\rm GeV}}{m_{u'u'u'}} \left( \frac{1~{\rm GeV}}{m_{e'}} \right)^2 \frac{f(m_{u'u'u'}/m_{e'})}{20}.
\label{eq:Xsec He}
\end{align}
We evaluate the function $f$ by calculating the s-wave scattering cross section using the HFDHE2 potential~\cite{HFDHE2}.
The numerical value of $f(m_{u'u'u'}/m_{e'})$ is given in Figure~\ref{fig:He_sca}.
We adopt the constraint ${\sigma}/{m_{\rm DM}} < 10 \; {\rm cm}^2/{\rm g}$~\cite{Kaplinghat:2015aga}.
The Bullet Cluster gives a stronger constraint on $\sigma / m_{\rm DM}$. However, the velocity of dark matter there is large, $v\sim 10^{-2} c$,
so that the typical momentum exchanged between dark matter exceeds the inverse of the Bohr radius of $(He)'_*$, giving a scattering cross section significantly suppressed relative to the low velocity one in  Eq.~(\ref{eq:Xsec He}).

\begin{figure}[t]
\centering
\includegraphics[clip,width=.45\textwidth]{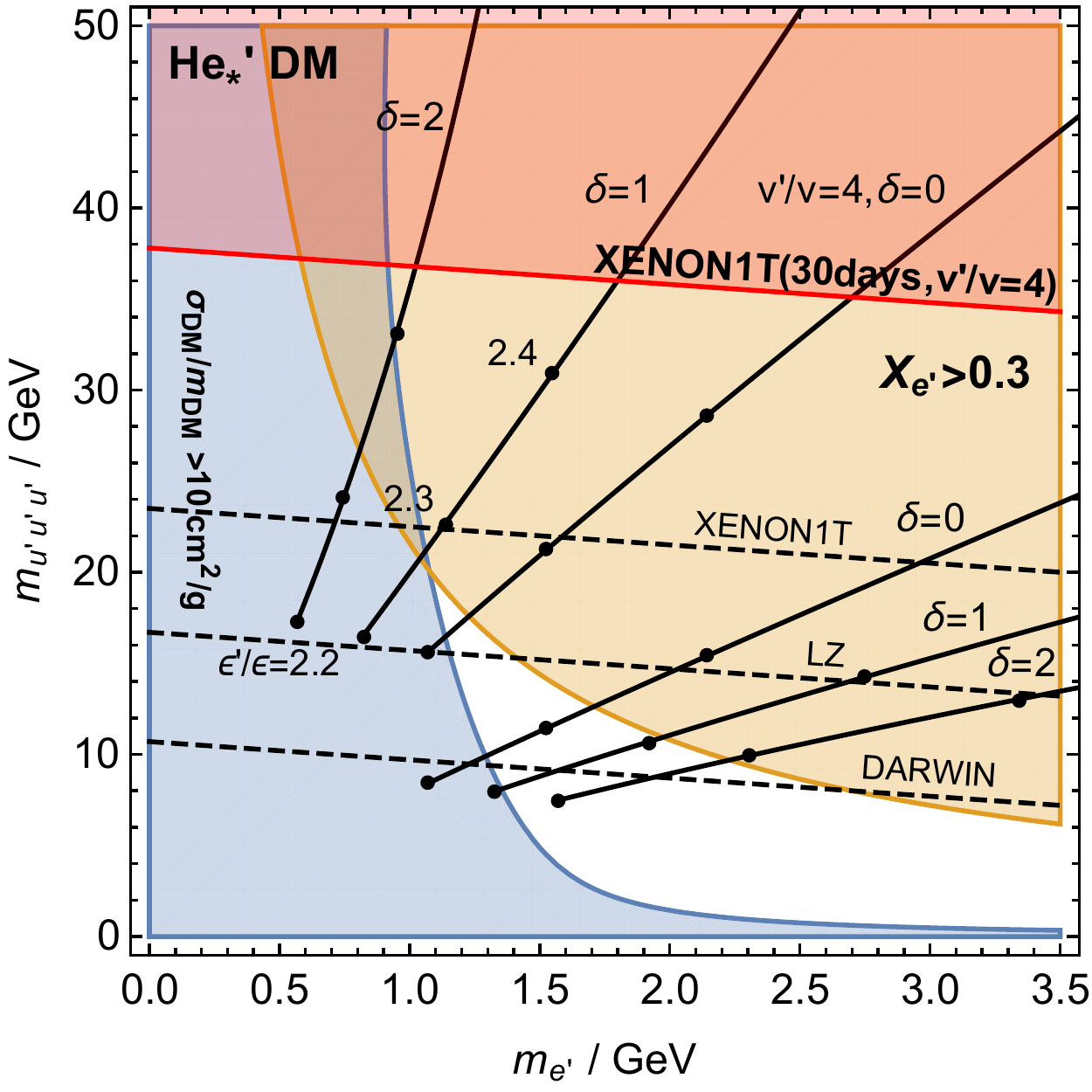}
\includegraphics[clip,width=.45\textwidth]{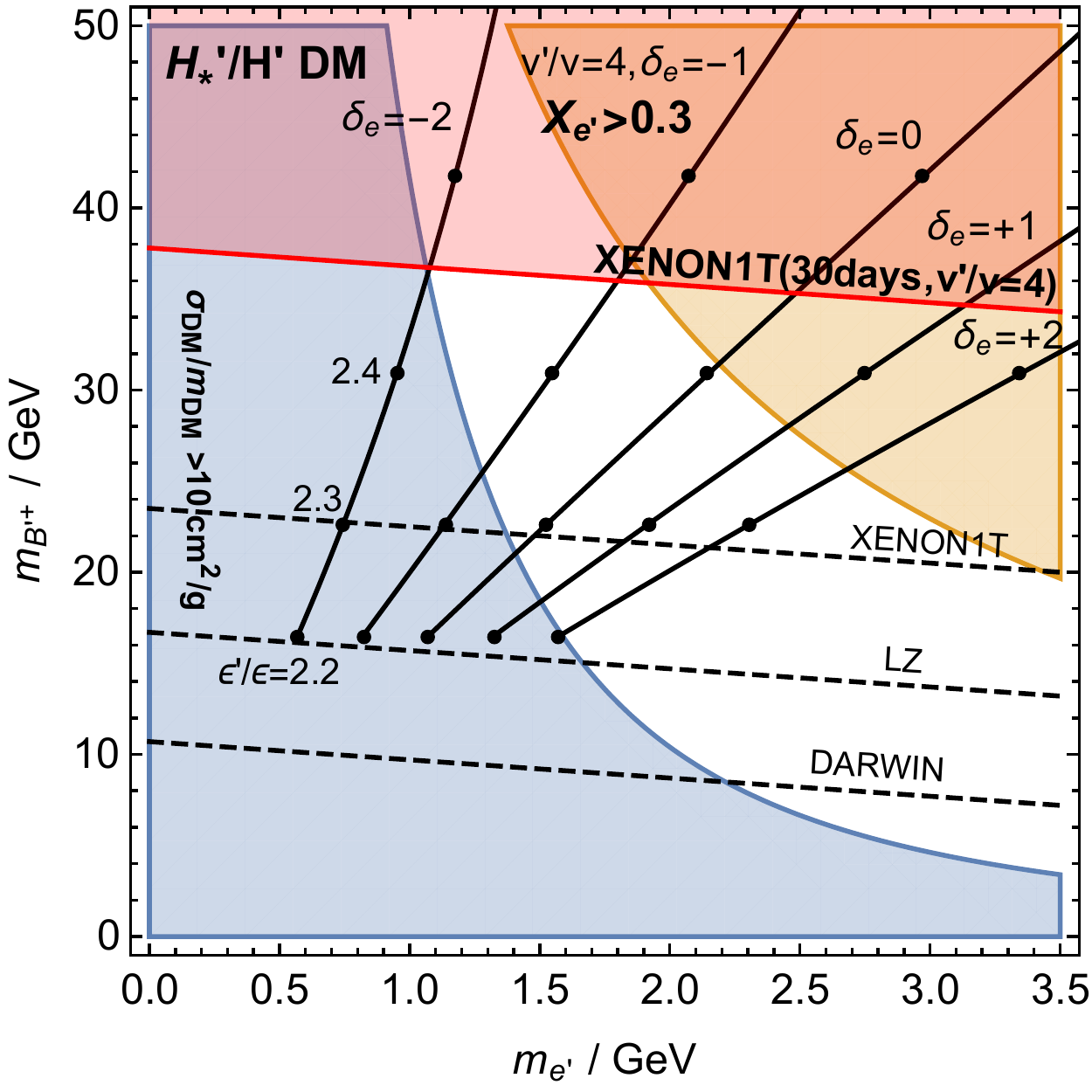}
\includegraphics[clip,width=.45\textwidth]{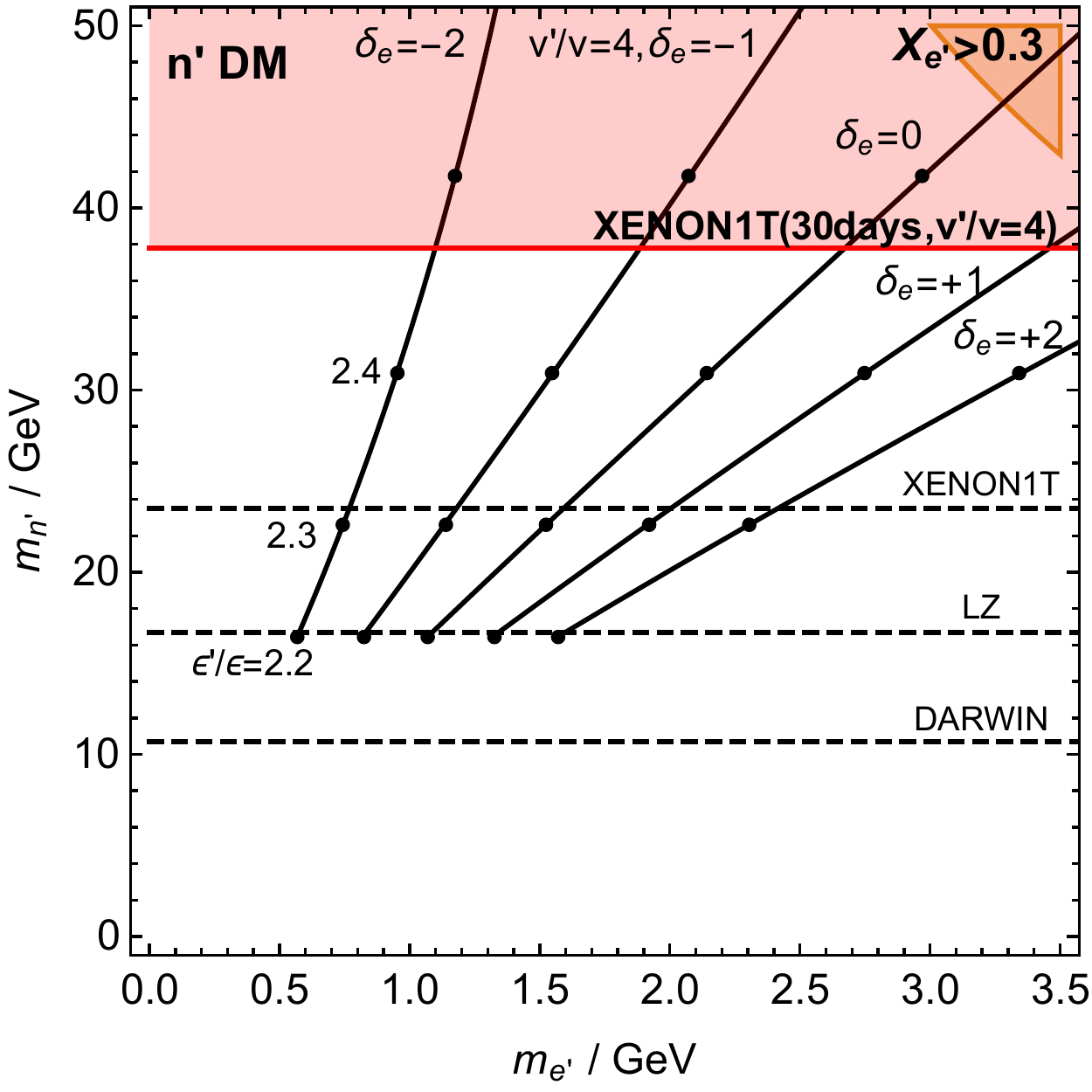}
\caption{
Constraints on the masses of $B_{u'u'u'}$ (top left panel) or $B_{d'd'd'}$/$B_{u'u'd'}$ (top right panel) and $e'$ from self-interactions of mirror atoms, the mirror ionization fraction and direct detection.
The bottom panel assumes that the mass of $B_{u'd'd'}=n'$ is below $m_{B'^+} + m_e'$ and  the mirror electron capture occurs inside the mirror atom.
Solid curves show predictions of the SU(5)-compatible model for a range of the uncertainties, as described in the text.  Dashed curves give expected reaches of future direct detection experiments.
}
\label{fig:selfscatt-ioniz}
\end{figure}

In the top left panel of Figure~\ref{fig:selfscatt-ioniz}, the shaded regions are excluded by the constraint on $(m_{e'}, m_{u'u'u'})$ from the ionization fraction and the self-interaction.
A portion of the parameter space is allowed.
Solid lines show the prediction of the $SU(5)$-consistent FN model for $(m_{e'}, m_{u'u'u'})$. The lines labeled ``$\delta=0,1,2$" show the range of the prediction with $|\delta_f|=0,1,2$, taking into account the 30\% uncertainty of the Yukawa coupling of the up quark.
We choose the signs of $\delta_{u,e}$ and the uncertainty of $y_u$ so that the upper (lower) two lines are located to the upper-left (lower-right).
Here we neglect the difference between $m(B_{u'u'u'})$ and  $m_{u'u'u'} = 3 m_{u'}$. For small $m_u'$ the contribution from the mirror  QCD dynamics is non-negligible, and the solid lines would slightly rise.
$\epsilon'/\epsilon \lesssim 2.2$ is excluded by the measurement of the Higgs signal strength.
It can be seen that $\epsilon'/\epsilon = 2.2-2.4$ predicts values of $(m_{e'}, m_{u'u'u'})$ consistent with the constraints, and the mass of  dark matter is in the range $(10-20)$ GeV.  All of this range is currently allowed by data from XENON1T, but much of the upper range will be probed by XENON 1T, LZ and DARWIN, as shown by the dashed lines.

%

\subsubsection{Constraints on $H' /H'_*$ dark matter: Regions (b) and (f) of Fig.~\ref{fig:Bmass}}

In Region (f) where $u'$ is sufficiently heavier than $d'$, the lightest baryon is $B'_{ddd}$, so that the mirror asymmetry is in the asymmetric components of $B'_{ddd}$ and $e'$. They may recombine into a neutral atom $H'_*$ and explain the observed dark matter.
The discussion here also applies to Region (b).
There the lightest baryon is  $B'_{uuu}$, but, once the recombination $B'_{uuu}+ e' \rightarrow He'_*$ happens, $He'_*$ decays into $B'_{uud}+\nu'$, and the recombination $B'_{uud}+ e' \rightarrow H'$ follows.
The first recombination process is more efficient than the second one due to the larger charge of the nucleon, so that we may approximate the whole recombination process as that of $B'_{uud}+ e'$.
We denote the mirror baryons of unit charge ($B'_{uud}$ or $B'_{ddd}$) as $B'^+$.

We calculate the recombination of the mirror baryon and mirror electron following~\cite{Seager:1999bc}. We find the ionization fraction,
\begin{align}
X_{e'} \simeq  0.05 \left(\frac{m_{B'^+} + m_{e'}}{10~{\rm GeV}}\right)^{0.9} \left(\frac{m_{{\rm red},e'B'}}{0.94~{\rm GeV}}\right)^{0.9} \left( \frac{\Delta N_{\rm eff}}{0.5}  \right)^{1/4},
\end{align}
where $m_{{\rm red},e'B'}$ is the reduced mass of the mirror electron-baryon system.
For fixed mirror baryon and electron masses, the ionization fraction of $H' /H'_*$ is smaller than that of $(He)'_*$, since the recombination cross section is larger for $H' /H'_*$.

For $m_{B'}/m_{e'}=O(1-10)$, the $H' /H'_*$ self-interaction cross section is given by~\cite{Cline:2013pca}
\begin{align}
\frac{\sigma}{m_{\rm DM}} \simeq \frac{100}{m_{e'}^2 \alpha^2} \frac{1}{m_{\rm DM}} \simeq 5.1~{\rm cm}^2/{\rm g} \times \frac{20~{\rm GeV}}{m_{B'}} \left( \frac{2~{\rm GeV}}{m_{e'}} \right)^2
\end{align}

In the top right panel of Figure~\ref{fig:selfscatt-ioniz}, we show the constraints on $(m_{e'}, m_{B'^+})$ from the ionization fraction and the self-interaction.
In order for $H' (H'_*)$ to be dark matter,  $m_{u'}$ must be similar to (larger than) $m_{d'}$.
The right bottom panel of Figure~\ref{fig:mdiff} shows that this is possible if $m_{u'}$ ($m_{d'}$) is larger (smaller) than its central value.
The figure also shows that for $|\delta_f|<1$, $m_{d'}$ is not much smaller than $m_{u'}$.
Based on these observations, in the right panel of Figure~\ref{fig:selfscatt-ioniz} we show predictions for $(m_{e'}, 3m_{d'})$ by solid lines, fixing $\delta_u = +1$ and the SM up Yukawa coupling larger than its central value by 30\%.   The various solid lines show that much of the allowed space is possible with $|\delta_e| < 2$.    Hence, with $\epsilon'/\epsilon = 2.2-2.5$ the prediction for $(m_{e'}, m_{B'^+})$ is consistent with the constraints, and the mass of dark matter is in the range $(20-50)$~GeV.  Xenon1T, LZ and DARWIN will probe all of this range.

\subsubsection{Constraints on $n'$ dark matter: Regions (c), (d) and (e) of Fig.~\ref{fig:Bmass}}

In Region (d) the lightest baryon is $B'_{udd}=n'$, so that the mirror asymmetry is in the  asymmetric component of $n'$.
There is no constraint from the ionization fraction or from the self-interaction cross section.

In Regions (c) and (e) the lightest baryon is not $n'$ but a charged mirror baryon. However, once  recombination happens, the mirror atom decays into $n'+\nu$, yielding $n'$ as a stable particle.
Still, the recombination may not be complete and there would be a constraint from the ionization fraction.
In Region (e) the recombination process is $B'_{ddd} + \bar{e}' \rightarrow H'_*$,
while in the right part of Region (c) it is $B'_{uud}+ e'\rightarrow H'$.
In the left part of Region (c) the recombination proceeds via $B'_{uuu} + e' \rightarrow He'_*$, $He'_*\rightarrow B'_{uud} + \nu'$, and $B'_{uud} + e' \rightarrow H'$.
The first and the second reaction is more efficient than the last one, so that we may approximate the whole process as $B'_{uud} + e \rightarrow H'$.
Thus in Regions (c) and (e) the recombination process is described as that of a mirror baryon with  unit charge ($B^{'+}$) and $e'$.

In Appendix~\ref{sec:rc_EC} we calculate  the ionization fraction with the inclusion of electron capture. We find that the ionization fraction is well-fitted by the following formula,
\begin{align}
X_{e'} \simeq  0.05 \left( \frac{m_{n'}}{10~{\rm GeV}} \frac{m_{{\rm red},e'B'}}{1.6~{\rm GeV}} \right)^{0.8} \left( \frac{\Delta N_{\rm eff}}{0.5}  \right)^{1/4}.
\end{align}
In the bottom panel of Figure~\ref{fig:selfscatt-ioniz},  the  corresponding constraint on $(m_{e'}, m_{n'})$ is shown.
The constraint is weaker than that on  $H' /H'_*$ dark matter, as the electron capture removes the mirror atom from the thermal bath, inhibiting  the inverse process $H' /H'_* + \gamma' \rightarrow B'^+ + e'$.
The solid lines are the same as those in the top right panel, and show that $\epsilon'/\epsilon < 2.4$ is allowed with $|\delta_e| < 2$.
The mass of dark matter is in the range $(20-60)$~GeV.
Part of the parameter region is excluded by XENON1T.
Xenon1T, LZ and DARWIN will probe all of this range.

\subsubsection{Mirror and SM matter asymmetries}
As we have seen, in viable parameter regions the mass of dark matter is $O(10)$ GeV.
Hence the observed dark matter abundance is explained by a mirror matter asymmetry of the same order as the SM matter asymmetry.
A difference of $O(1)$ in the asymmetries may arise in some scenarios of baryogenesis.
For example, if the baryon asymmetry is created by the Affleck-Dine mechanism~\cite{Affleck:1984fy,Dine:1995kz}, an $O(1)$ difference is expected from the difference of the initial mis-alignment in the angular direction of the scalar field responsible for  baryogenesis.

A dark matter mass of $m_p\cdot \Omega_{\rm DM}/\Omega_{\rm b}\approx 5$ GeV is close to being allowed for $He'_*$ dark matter. This would be consistent with equal matter asymmetries in the standard and in the mirror sectors.

\subsubsection{Possibility of mirror nucleosynthesis}
Mirror baryons collide with each other and may form bound states, namely mirror nuclei~\cite{Berezhiani:2000gw}.
Formation of nuclei of generic composite dark matter is discussed in~\cite{Krnjaic:2014xza,Hardy:2014mqa}.

In our case first we argue that mirror nuclei composed of more than two baryons are unlikely to be formed.
In  most of the parameter space the mass difference between the lightest mirror baryon and the next to lightest one is much larger than $m_{e'}/18$, so that  almost all of the mirror baryon number is stored in the lightest mirror baryons. Therefore, in order for the lightest mirror baryon to form a bound state with more than two baryons, a non-zero angular momentum is required due to  Fermi statistics. This leads to a positive energy of  order $1/ (m_{B'} r^2)$, where $r$ is the radius of the bound state. We expect that $r^{-1}$ is as large as the mass of the mediator of the mirror strong force.
In the parameter space of interest $m_{d',u'} > T_{c}'$, so $1/ (m_{B'} r^2)= O(m_{u',d'})$. On the other hand the possible binding energy would be at most $O(T_c') < O(m_{d'.u'})$. Thus we expect that  mirror nuclei composed of more than two lightest mirror baryons are unbound.

There could be a mirror nucleus composed of two lightest mirror baryons. A lattice QCD calculation with a  quark mass larger  than normal seems to make space for di-neutron and di-proton states~\cite{Beane:2012vq,Yamazaki:2012hi}.
Although it is not clear if a mirror di-neutron and di-proton exist for our mirror quark masses, or mirror di-$B_{uuu}'$ and di-$B_{ddd}'$ exist for any mirror quark mass, let us suppose that those states are stable and discuss the phenomenological consequence.
To verify this assumption, a dedicated lattice calculation is needed. 

Once the temperature drops below the binding energy, almost all of the lightest mirror baryons in Figure~\ref{fig:Bmass} are combined into di-baryon states.
In Region (a), a mirror baryon with  charge $4$ is formed. The recombination as well as the self-scattering cross section is affected, in a way that we do not pursue  further in this paper.
In Region (b), mirror di-protons are formed via the formation of di-baryons and the mirror electron capture.
The constraint on $He_*'$ is applicable but with  twice larger baryon mass. There is no viable parameter space for the $SU(5)$ model.
In Regions (c), (d) and (e), mirror di-neutrons are formed. The constraint on $n'$ is again applicable with twice larger baryon mass.
All parameter region of the $SU(5)$ model with $v'/v<4$ can be probed by the XENON1T.
In Region (f) mirror di-$B'_{ddd}$ are formed. The constraint on $He_*'$ is applicable but with  twice larger baryon mass.

\subsection{Dark Radiation}

In the early universe with a sufficiently large temperature the SM particles and their mirror partners interact with each other and have the same temperature. Below some temperature $T_d$ the interaction becomes inefficient and they evolve independently.
Mirror particles eventually decay/annihilate into mirror photons and neutrinos, which are observed as dark radiation.
The abundance of the dark radiation, traditionally  expressed as the excess of the effective number of neutrinos from the SM prediction, is
\begin{align}
\label{eq:DeltaNeff}
\Delta N_{\rm eff} = \frac{4}{7}g'_r \times \left( \frac{10.75}{g(T_d)} \right)^{4/3} \times \left(  \frac{g'(T_d)}{g'_r} \right)^{4/3},
\end{align}
where $g(T)$ and $g'(T)$ are the effective entropy degrees of freedom (d.o.f) of the SM particles and the mirror particles at temperature $T$, respectively.
The second factor in the r.h.s. of eq.~\ref{eq:DeltaNeff}   expresses the heating of the SM neutrinos, whereas the third factor expresses the heating of the dark radiation.
$g'_r$ is the d.o.f.~of the radiation component of the mirror sector.
In the minimal model where the mirror neutrinos are nearly massless, $g'_r = 29/4$.
We extract the d.o.f.~of the SM particles $g(T)$ from~\cite{Borsanyi:2016ksw}.

\subsubsection{Generic decoupling temperature}

In this Subsection we treat $T_d$ as a free parameter.
If $T_d > T_c'$, the mirror gluons give a large contribution to $g'(T_d)$, and $\Delta N_{\rm eff}$ is larger than the constraint from the Planck satellite, $\Delta N_{\rm eff}<0.65$ ($2\sigma$). We only consider the case with $T_d < T_c'$, and neglect the contribution from the mirror gluons to $g'(T_d)$.

The contributions of the mirror photons, neutrinos and leptons to $g'(T_d)$ are readily estimated using the ideal gas approximation.
The mirror quarks, on the other hand, cannot be treated as an ideal gas, especially for $T_d < T_c'$, where the dynamics of the mirror quarks is better described as a gas of mirror hadrons.
Figure~\ref{fig:mf} shows that among mirror hadrons, the ones composed of mirror up quarks are the most important ones.
We estimate the contribution from the mirror QCD sector, treating the hadron gas as an ideal gas composed of mirror $\sigma$ ($J=0,CP=++$), $\eta'$ ($J=0,CP=+-$) and $\omega$ ($J=1, CP=--$), with their masses given by
\begin{align}
m_{\sigma}'^2 =& \left(2 m_{u'}\right)^2 + \left(1.5 T_c' \right)^2,\\ \nonumber
m_{\eta'}'^2 =& \left(2 m_{u'}\right)^2 + \left(3 T_c' \right)^2,\\ \nonumber
m_{\omega'}'^2 =&\left(2 m_{u'}\right)^2 + \left(4 T_c' \right)^2.\\
\end{align}
The contribution proportional to $T_c'^2$ is inferred from the Standard Model QCD spectrum.

\begin{figure}[t]
\centering
\includegraphics[clip,width=.48\textwidth]{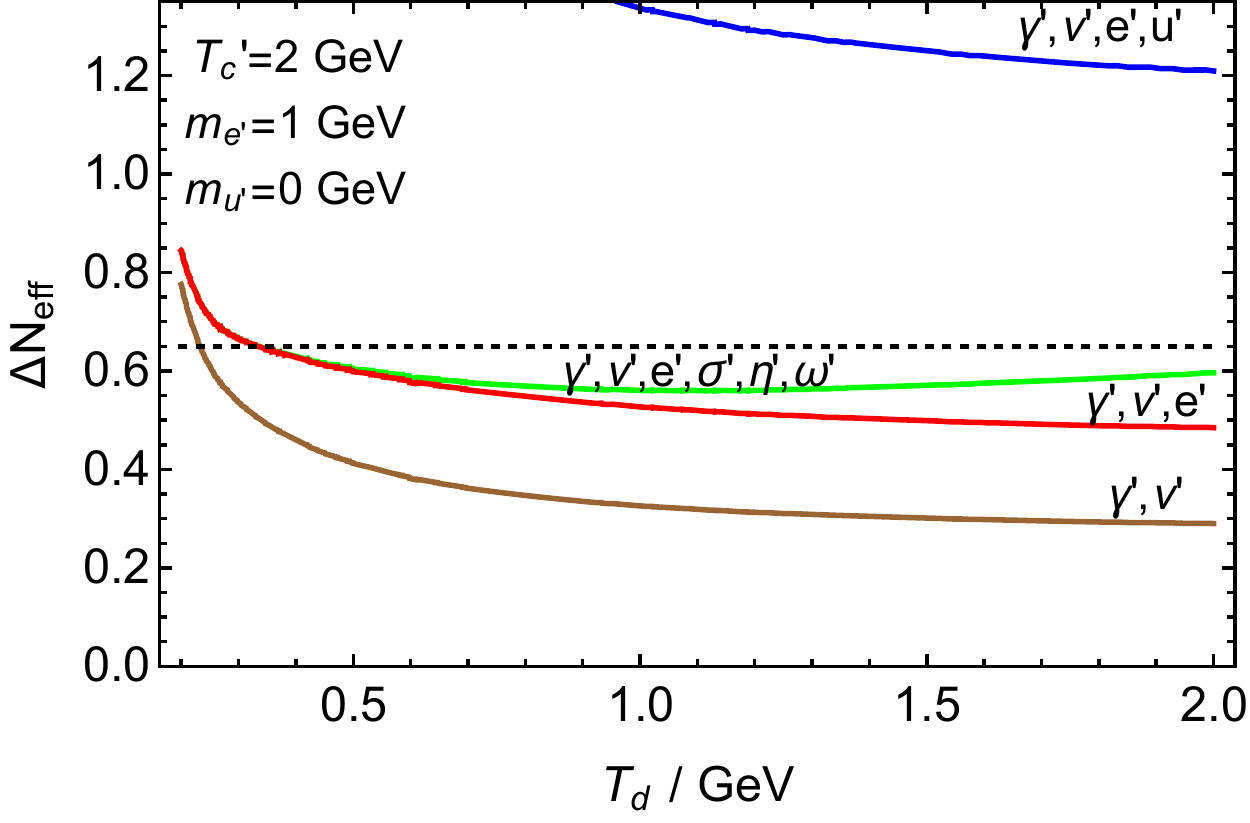}
\includegraphics[clip,width=.48\textwidth]{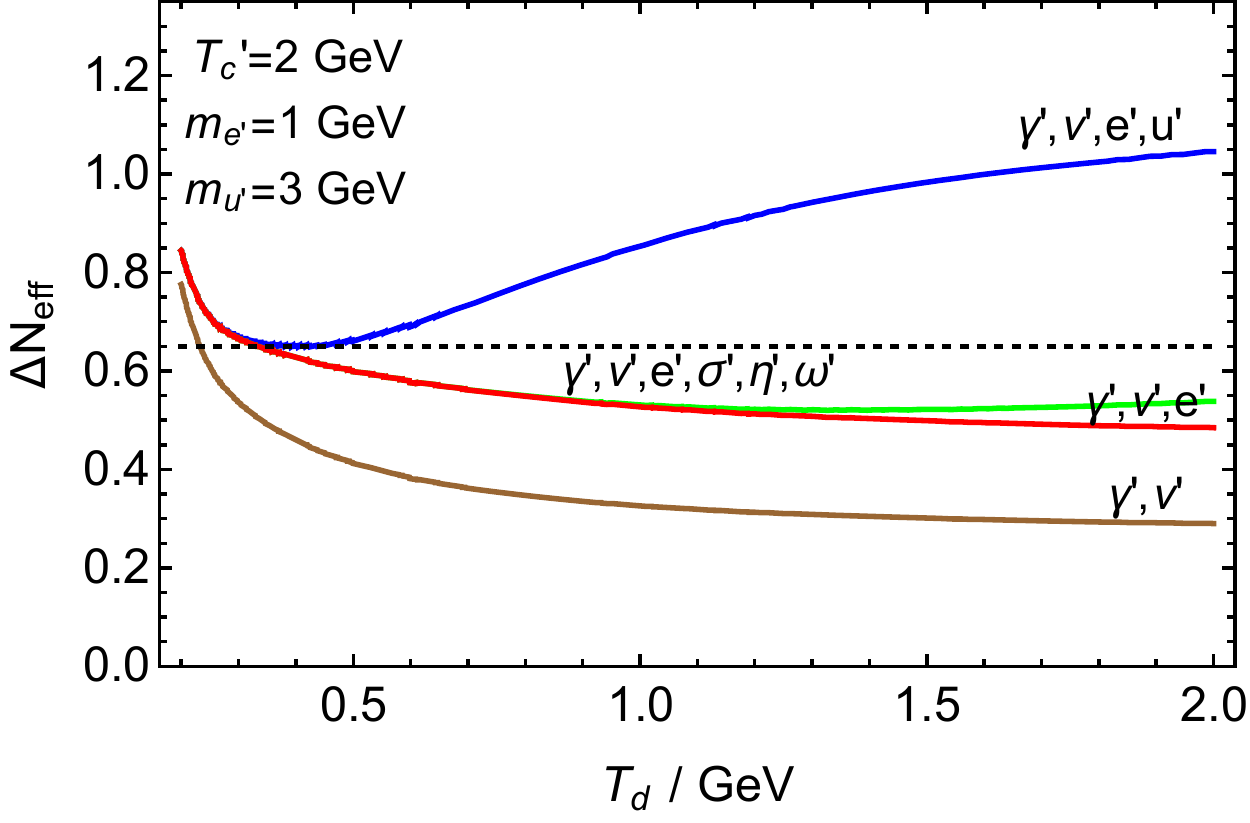}
\caption{
The dark radiation abundance predicted as a function of the decoupling temperature.
}
\label{fig:DNeff_ue}
\end{figure}

In Figure~\ref{fig:DNeff_ue}, we show the prediction of $\Delta N_{\rm eff}$ as a function of $T_d$ with  fixed $m_{e'}$, $m_{u'}$ and $T_{c'}$, neglecting the contributions from the other mirror fermions.
The brown, red and green lines show the contribution from $\gamma'\nu'$, $\gamma'\nu'e'$ and $\gamma'\nu'e'\sigma'\eta''\omega'$, respectively.
These figures show that $\Delta N_{\rm eff}$ is dominated by the contribution from $\gamma'$, $\nu'$ and $e'$.
For comparison, we also show $\Delta N_{\rm eff}$ calculated using the quark picture with the ideal gas approximation by a blue line:  confinement suppresses the abundance of dark radiation.

\begin{figure}[t]
\centering
\includegraphics[clip,width=.4\textwidth]{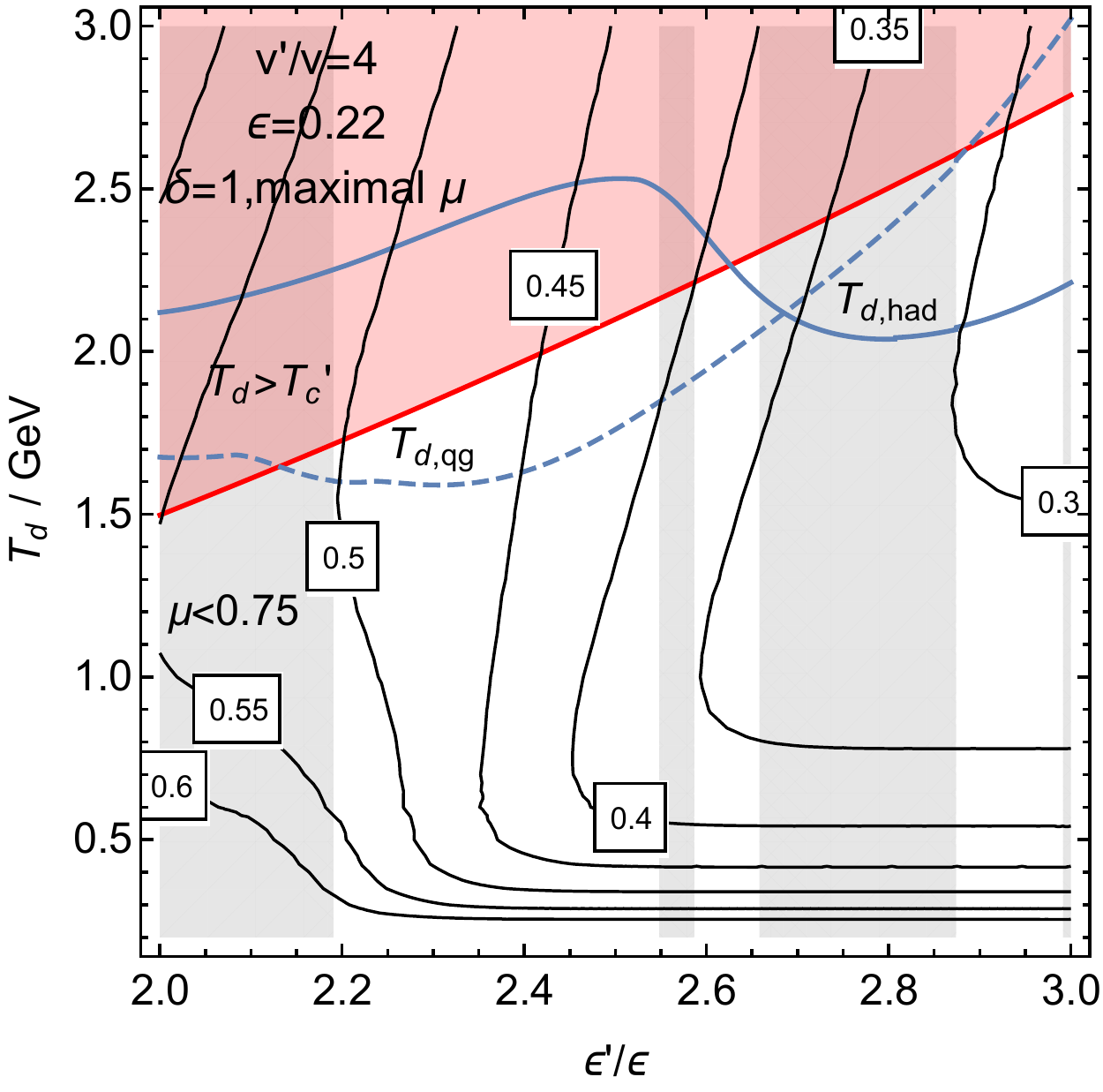}
\includegraphics[clip,width=.4\textwidth]{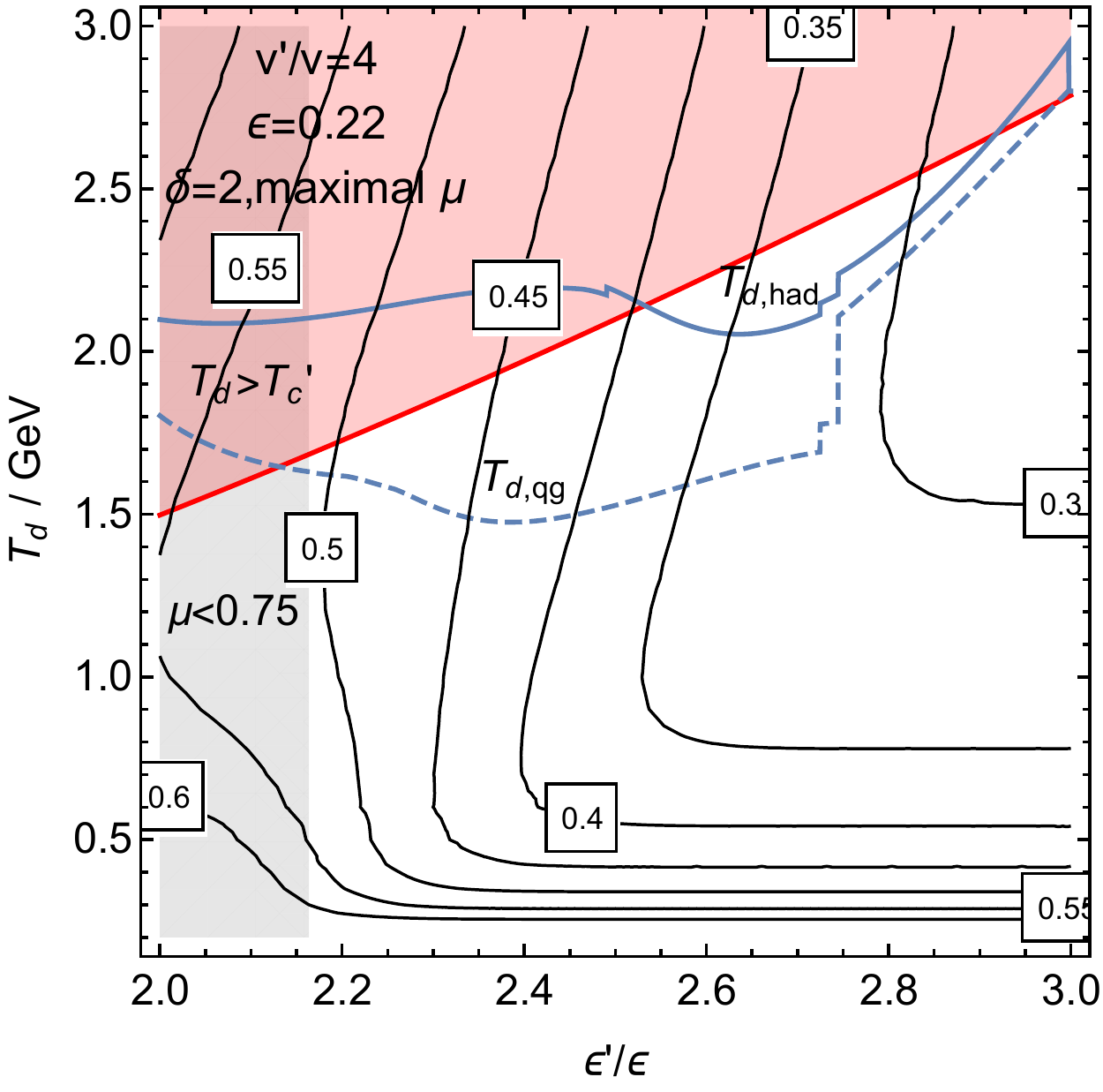}
\includegraphics[clip,width=.4\textwidth]{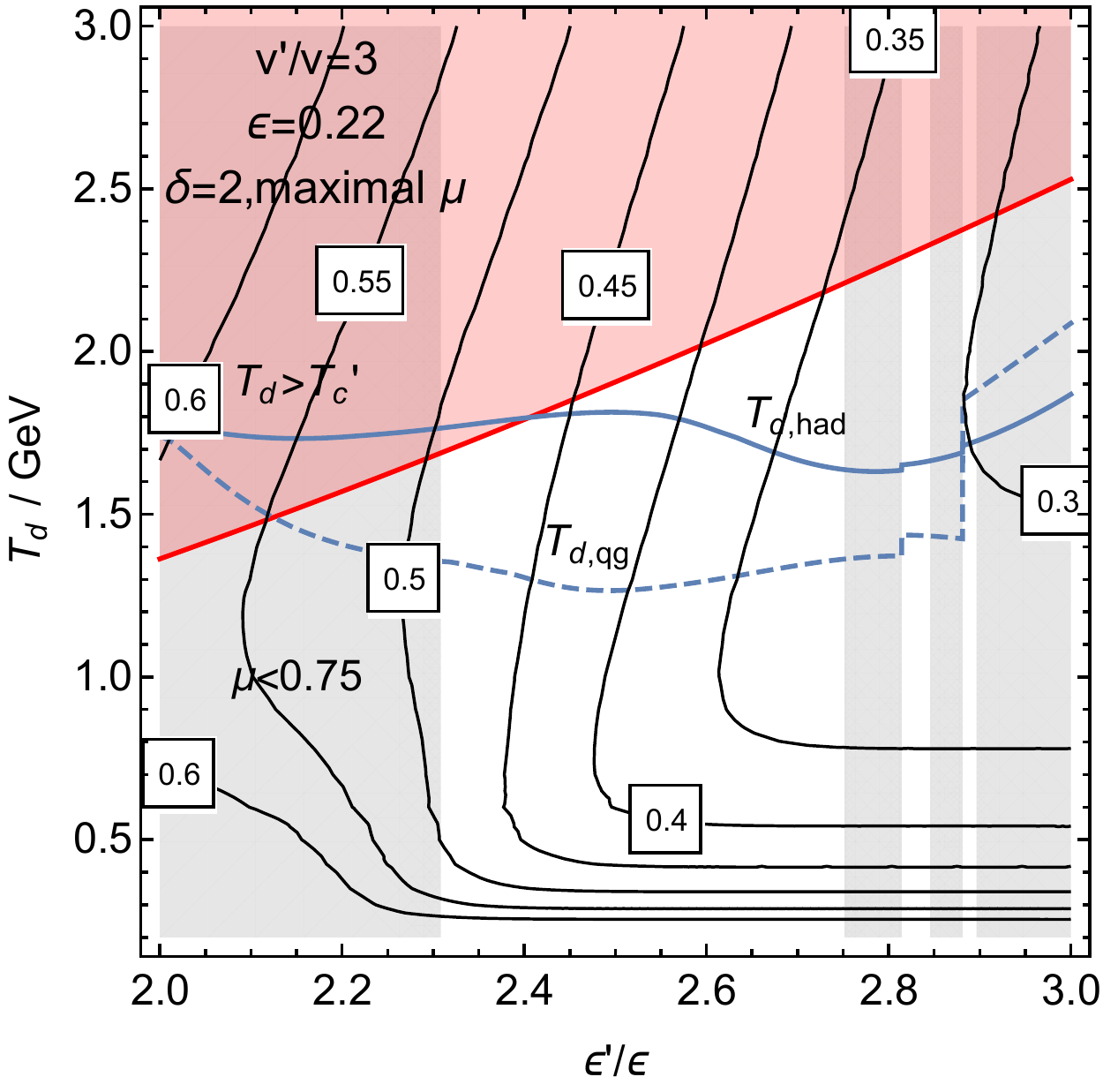}
\caption{
The predicted dark radiation abundance expressed as $\Delta N_{eff}$.
We choose the mass spectrum of mirror fermions to minimize the invisible decay of the Higgs except for the mirror electron.
In the red shaded region $T_d > T_c'$ and the abundance of the dark radiation is too large.
The solid and dashed blue lines show the decoupling temperature via the Higgs exchange in the hadron ($T_{d, had}$) and the quark-gluon picture ($T_{d, qg}$), respectively.
For the quark-gluon picture the decoupling temperature is mainly determined by the annihilation of mirror quarks, while for the hadron picture it is mainly determined by the decay of mirror glueballs.
}
\label{fig:DNeff}
\end{figure}

In Figure~\ref{fig:DNeff}, we show the prediction of $\Delta N_{\rm eff}$ as a function of $\epsilon'/\epsilon$ and $T_d$.
Here we choose the sign and the value of each $\delta_f$ so the $\mu$ becomes as large as possible, expect for $\delta_{e}$, for which we take $\delta_e=\delta$ to suppress $\Delta N_{\rm eff}$.
The red line shows the mirror QCD phase transition temperature $T_c'$. Above this line the contribution from mirror gluons makes $\Delta N_{\rm eff}$ unacceptably large.
Blue shaded regions are excluded due to too small $\mu$, as  discussed in Section~\ref{sec:hdecay}.
The amount of the dark radiation is typically $\Delta N_{\rm eff}= 0.3-0.6$.

%

\subsubsection{Decoupling temperature from Higgs exchange}

In this Subsection  we estimate the decoupling temperature determined by the Higgs exchange between the SM particles and the mirror partners.
The interaction rate between the mirror leptons and the Standard fermions is readily estimated using the ideal gas picture, following~\cite{Barbieri:2016zxn}.
The scattering cross section between a mirror fermion $f'$ and a SM fermion $f$ is given by
\begin{align}
\label{eq:scatter}
\sigma v_{\rm rel}(ff'\rightarrow ff') = \frac{1}{8\pi}\left(\frac{m_f}{v}\right)^2 \left( \frac{v m_{f'}}{v^{'2}} \right)^2 \frac{m_f m_{f'}}{m_f + m_{f'}} \frac{p_{\rm cm}}{m_h^4},
\end{align}
where we take a non-relativistic limit.
Here $p_{\rm cm}$ is the momentum of the fermion in the center of mass frame. In the thermal bath, it has a typical size
\begin{align}
p_{\rm cm}^2 = \frac{4T (m_f + m_{f'}+ \sqrt{m_f m_{f'}})}{3 \left( 2 + m_f/m_{f'} + m_{f'}/m_f\right)}.
\end{align}
The annihilation cross section of a pair of $f'$ into a pair of $f$ is given by
\begin{align}
\sigma(f'\bar{f}' \rightarrow f \bar{f})v_{\rm rel} = \frac{N_f}{4\pi} \left(\frac{m_f}{v}\right)^2 \left( \frac{v m_{f'}}{v^{'2}} \right)^2 \frac{ (m_{f'}^2 - m_f^2)^{3/2}}{m_{f'}^3m_h^4} p_{f'}^2.
\end{align}
Here $p_{f'}$ is the momentum of $f'$ in the center of mass frame. In the thermal bath, it is as large as $p_{f'}^2 \simeq 3m_{f'} T/2 $.
$N_f$ is the multiplicity of the Dirac fermion $f$: for one lepton (quark) $N_f = 1(3)$.
The transfer rate of the energy density of mirror particles into SM particles is then given by
\begin{align}
\label{eq:transfer fermion}
\frac{\rm d}{{\rm d}t}\rho'|_{f'} =& \sum_f  \left(4 N_f n_{F}\left(m_f,T\right)\right)  \left( 4 N_{f'}n_{F}\left(m_{f'},T\right)\right)   \sigma v_{\rm rel}(ff'\rightarrow ff') \times \Delta E \nonumber \\
&+ \sum_f N_{f'}4 n_F(m_{f'},T)^2 \sigma v_{\rm rel}(f'\bar{f}'\rightarrow f\bar{f}) \times 2 m_{f'},
\end{align}
where $n_F(m,T)$ is the number density of a fermion of mass $m$ in the thermal bath at temperature $T$,
and $\Delta E \simeq T$ is a typical energy transfer by the scattering $f f' \rightarrow f f'$.

The scattering with mirror QCD charged particles requires a dedicated treatment.
We use in succession quark and hadron pictures with an ideal gas approximation to calculate the energy transfer rate.

Let us first treat mirror QCD charged particles as an ideal gas of mirror quarks and gluons.
The scattering cross section between a mirror fermion $f'$ and a SM fermion $f$ is given by Eq.~(\ref{eq:scatter}).
The annihilation cross section of a pair of mirror quarks $q'$ into a pair of $f$ is given by
\begin{align}
\sigma(q'\bar{q}' \rightarrow f \bar{f})v_{\rm rel} \simeq \frac{N_f}{4\pi} \left(\frac{m_f}{v}\right)^2 \left( \frac{v m_{f'}}{v^{'2}} \right)^2 \frac{ (m_{f'}^2 - m_f^2)^{3/2}}{m_{f'}^3m_h^4} p_{f'}^2 \times \frac{2\pi x}{ 1- e^{-2\pi x}} \left( 1 + x^2 \right),~~x = \frac{4}{3} \frac{\alpha_s'}{v_{\rm rel}},
\end{align}
where we have included the Sommerfeld effect~\cite{Sommerfeld} for a $p-$wave annihilation~\cite{Cassel:2009wt}. The fine structure constant should be evaluated at the scale $\mu \simeq 4/3 m_{q'} \alpha_s'/2$~\cite{Nagano:1999nw}, so we solve the consistency condition
\begin{align}
\frac{2}{3}m_{q'} \alpha_s'(\mu) = \mu
\end{align}
to determine the appropriate scale. We put $v_{\rm rel} = \sqrt{T/m_{q'}}$ to estimate the Sommerfeld enhancement factor.
The contribution of a mirror quark $q'$ to the energy transfer rate is given by Eq.~(\ref{eq:transfer fermion}).

The mirror gluons also couple to the SM Higgs,
\begin{align}
{\cal L} = \frac{v}{v'}\frac{h}{\sqrt{2} v'} \frac{\alpha_s'}{12\pi} \sum_{q'} \left( 1 + \frac{11}{4\pi} \alpha_s'(m_{q'})\right) G_{\mu\nu}^{a'} G^{\mu \nu a'} \simeq \frac{v}{v'}\frac{h}{\sqrt{2} v'} \frac{\alpha_s'}{2\pi} G_{\mu\nu}^{a'} G^{\mu \nu a'}.
\end{align}
The annihilation cross section of a pair of mirror gluons into a pair of SM fermions  $f$ is
\begin{align}
\sigma (g'g' \rightarrow f \bar{f}) v_{\rm rel} = \frac{2 N_f}{\pi} \left( \frac{v}{v'}\right)^2 \left( \frac{\alpha_s' }{2\pi} \right)^2 \left(\frac{m_f}{v}\right)^2 \frac{1}{v'^2} \left( \frac{p_{\rm cm}}{m_h} \right)^4 \left(1 - \frac{m_f^2}{p_{\rm cm}^2}\right)^{3/2},
\end{align}
while the scattering cross section is
\begin{align}
\sigma (g' f \rightarrow g'f) = \frac{4}{3\pi} \left( \frac{v}{v'}\right)^2 \left( \frac{\alpha_s' }{2\pi} \right)^2 \left(\frac{m_f}{v}\right)^2 \frac{1}{v'^2} \left( \frac{p_{\rm cm}}{m_h} \right)^4,
\end{align}
where we take the non-relativistic limit for $f$.
Due to the absence of the Sommerfeld effect, however, the energy transfer from mirror gluons is negligible in comparison with that from mirror quarks.

We define the decoupling temperature by $({\rm d}\rho'/{\rm d}t)/\rho' = H$, where $H$ is the $T$-dependent expansion rate of the universe.
In Figure~\ref{fig:DNeff}, we show the decoupling temperature $T_{d,qg}$ determined by the Higgs exchange with the quark picture by dotted lines.
We find that $T_{d,qg}$ can be lower than $T_c'$. The decoupling temperature is dominantly determined by the annihilation of mirror quarks.
We note, however, that this does not mean that the actual decoupling temperature $T_d$ can be below $T_c'$.
As the temperature drops and becomes close to $T_c'$, the ideal gas approximation of mirror quarks is not straightforwardly  applicable, and is expected to break down for $T_d<T_c'$. Our estimate at least shows, however, that the decoupling temperature is close to $T_c'$.

Let us next treat the mirror QCD charged particles as an ideal gas of mirror hadrons.
We include the scattering and the annihilation of mirror glueballs.
A spin-0 glueball with $CP=++$ mixes with the SM Higgs and decays  into SM fermions.
A result of a lattice calculation is available for the lightest one, $S_{0^{++}}'$.
Using the lattice calculation for the relevant matrix element and for the glueball mass~\cite{Meyer:2008tr},
\begin{align}
<0|g_s^{'2} G_{\mu\nu}^{a'} G^{\mu \nu a'} |S_{0^{++}}>\simeq 2.7 m_{S_{0^{++}}'}^3,~~m_{S_{0^{++}}'}\simeq 5.3 T_c',
\end{align}
the decay rate of $S_{0^{++}}'$ into a pair of Standard Model fermion $f$ is given by
\begin{align}
\Gamma (S_{0^{++}}' \rightarrow  f \bar{f}) = \frac{1}{32\pi}\left( \frac{v}{v'}\right)^2 \left( \frac{1 }{8\pi^2} \right)^2 \left(\frac{m_f}{v}\right)^2 \frac{1}{v'^2} \frac{2.7^2 m_{S_{0^{++}}'}^7}{m_h^4}  \left(1- \frac{4 m_f^2 }{m_{S_{0^{++}}'}^2} \right)^{3/2}.
\end{align}
The scattering cross section of a mirror glueball $S_i'$ can be estimated by the trace anomaly,
\begin{align}
<S_i'| \frac{11}{32\pi^2} g_s^{2'} G_{\mu\nu}^{a'} G^{\mu \nu a'} |S_i'> = 2m_{S_i'}^2,
\end{align}
where we assume that the mass of the mirror glueball is not affected by the masses of mirror fermions, which is the case for  sufficiently large mirror fermion masses and/or large $N_c$.
The scattering cross section between a mirror glueball $S_i'$ and $f$ is given by
\begin{align}
\sigma v_{\rm rel}(fS_i'\rightarrow fS_i') = \frac{1}{8\pi} \left( \frac{4}{11} \right)^2\left(\frac{m_f}{v}\right)^2 \left( \frac{v m_{S_i'}}{v^{'2}} \right)^2 \frac{m_f m_{S_i'}}{m_f + m_{S_i'}} \frac{p_{\rm cm}}{m_h^4}.
\end{align}
We take into account the scattering with mirror glueballs of spin $S_{S_i}=0,1,2$ and $CP=++,+-,-+,--$ , whose masses are estimated in~\cite{Chen:2005mg}.
The contribution of the mirror glueballs to the energy transfer rate is given by
\begin{align}
\frac{\rm d}{{\rm d}t}\rho'|_{S} =& \sum_{f,i}  \left(4 N_f n_{F}\left(m_f,T\right)\right)  \left( \left( 2 S_{S_i} +1 \right)n_{B}\left(m_{S_i'},T\right)\right)   \sigma v_{\rm rel}(fS_i'\rightarrow fS_i') \times \Delta E \nonumber \\
&+ \sum_f n_B(m_{S_{0^{++}}},T) \Gamma (S_{0^{++}} \rightarrow  f \bar{f}) \times m_{S_{0^{++}}},
\end{align}
where $n_B(m,T)$ is the number density of a boson of mass $m$ in the thermal bath at temperature $T$.

We also include the annihilation and the scattering of mirror quarkonia.
The decay rate of a mirror quarkonium with spin-0 and $CP=++$, $\chi_{q'}$, into a pair of SM fermions is approximately given by
\begin{align}
\Gamma(\chi_{q'}\rightarrow f\bar{f}) \simeq \sigma(q'\bar{q}' \rightarrow f \bar{f})v_{\rm rel}|_{p_{q'} = m_{q'} \alpha_s'} \frac{1}{8\pi} \left( m_{q'} \alpha_s' \right)^3.
\end{align}
The scattering cross section between a mirror quarkonium $\chi_{i}'$ and $f$ is given by
\begin{align}
\sigma v_{\rm rel}(f\chi_i'\rightarrow f\chi_i') = \frac{1}{8\pi} \left( \frac{4}{11} \right)^2\left(\frac{m_f}{v}\right)^2 \left( \frac{v m_{\chi_i'}}{v^{'2}} \right)^2 \frac{m_f m_{\chi_i'}}{m_f + m_{\chi_i'}} \frac{p_{\rm cm}}{m_h^4}.
\end{align}
Here we assume that the mass of the quarkonium is dominated by the mirror quark mass.
We take into account the scattering of all the quarkonia composed of $d'$, $s'$, $b'$, $u'$, $c'$ with spin-CP $0^{+-}$ ($\eta$-like) and $1^{--}$ ($J/\psi$-like).

In Figure~\ref{fig:DNeff} we show   by solid lines the decoupling temperature $T_{d,had}$ determined by the Higgs exchange in the hadron picture.
In some of the parameter space $T_{d,had}$ is lower than $T_c'$. The decoupling temperature is dominantly determined by the decay of glueballs.
The estimated $T_{d,had}$ is however close to $T_c'$, and the thermal effect may be important (e.g.~that on the glueball mass).
The raise of $T_{d,had}$ when lowering  $\epsilon'/\epsilon$ below about $2.6$
 is due to the kinematic suppression of the decay of the lightest glueball into $b\bar{b}$.
Inclusion of higher resonances might make $T_{d,had}$ smaller than $T_c'$ also for $\epsilon'/\epsilon \lsim 2.6$.

\subsubsection{Decoupling temperature from kinetic mixing}

The kinetic mixing between the hypercharge gauge fields,
\begin{align}
 \frac{1}{2} \frac{\epsilon_{\rm kin}}{{\rm cos}\theta_W^2} B^{\mu \nu} B'_{\mu \nu},
\end{align}
can maintain thermal equilibrium between the SM and mirror sectors through the scattering between a mirror charged fermion and the SM photon.  The mirror electron is the lightest mirror charged fermion and decoupling does not occur until the temperature drops below its mass.  For $T\ll m_{e'}$, the scattering cross section for the process $e' \gamma' \leftrightarrow e'\gamma$ is given by
\begin{align}
\sigma (f' \gamma' \leftrightarrow f' \gamma)v = \frac{8\pi}{3} \epsilon_{\rm kin}^2 \alpha^2  \frac{1}{m_{e'}^2}.
\end{align}
The scattering rate becomes smaller than the expansion rate of the universe below a temperature $T_{d,kin}$,
\begin{align}
T_{d,kin} \simeq \frac{m_{e'}}{4 + 2{\rm ln}\frac{\epsilon_{\rm kin}}{10^{-6}}}.
\end{align}
Sufficient suppression of $\Delta N_{\rm eff}$ requires $0.2$ GeV $<T_{d,kin}<T'_{c}$.
For the mirror electron mass we are interested in, this is achieved for $\epsilon_{\rm kin}\sim 10^{-7}-10^{-6}$.
Kinetic mixing of this size is excluded if dark matter is mirror atoms, but is allowed if dark matter  is composed of mirror neutrons~\cite{Barbieri:2016zxn}.

\section{Variant Models}
\label{sec:variant}

While in principle there are many models based on Eqs. (\ref{eq:Yukawa},\ref{eq:FN}), they are greatly restricted by the need to account for the known fermion masses and quark mixings.  To illustrate the broad persistence, given this constraint, of the mirror fermion spectrum obtained in Section IIIA, we briefly consider in this Section two variants of the SU(5)-compatible model examined there.
In both the new models~\cite{Feruglio:2015jfa} we take the FN charge of the $\mathcal{Q}_1$ multiplet to deviate by one unit from the 
charge of  $\bar{u}_1, \bar{e}_1$ in order to get  the same scaling law in terms of $\epsilon$ as  in the Volfenstein parameterization of  the CKM angles,  $V_{us} \approx \lambda_c, V_{cb} \approx \lambda_c^2, V_{ub} \approx \lambda_c^3$, in terms of $\lambda_c=0.22$. 

The FN charges of the two models and the corresponding scaling law of the masses are:

\begin{itemize}
\item Model B1
\begin{align}
Q:(3,2,0),~
\bar{u}:(4,2,0),~
\bar{e}:(4,2,0),~
\bar{d},L:~(4,3,3)
\end{align}
\begin{align}
m_t \sim 1 + O(\epsilon^4),~
m_c \sim \epsilon^4\left( 1 + O\left(\epsilon^4\right)\right),~~
m_u \sim \epsilon^7\left( 1 + O\left(\epsilon^4\right)\right)\nonumber\\
m_b \sim \epsilon^3\left( 1 + O\left(\epsilon^2\right)\right),~
m_s \sim \epsilon^5\left( 1 + O\left(\epsilon^2\right)\right),~~
m_d \sim \epsilon^7\left( 1 + O\left(\epsilon^2\right)\right)\nonumber\\
m_\tau \sim \epsilon^3\left( 1 + O\left(\epsilon^2\right)\right),~
m_\mu \sim \epsilon^5\left( 1 + O\left(\epsilon^2\right)\right),~~
m_e \sim \epsilon^8\left( 1 + O\left(\epsilon^2\right)\right)
\end{align}

\item Model B2
\begin{align}
Q:(3,2,0),~
\bar{u}:(4,2,0),~
\bar{e}:(4,2,0),~
\bar{d},L:~(3,2,2)
\end{align}
\begin{align}
m_t \sim 1 + O(\epsilon^4),~
m_c \sim \epsilon^4\left( 1 + O\left(\epsilon^4\right)\right),~~
m_u \sim \epsilon^7\left( 1 + O\left(\epsilon^4\right)\right)\nonumber\\
m_b \sim \epsilon^2\left( 1 + O\left(\epsilon^2\right)\right),~
m_s \sim \epsilon^4\left( 1 + O\left(\epsilon^2\right)\right),~~
m_d \sim \epsilon^6\left( 1 + O\left(\epsilon^2\right)\right)\nonumber\\
m_\tau \sim \epsilon^2\left( 1 + O\left(\epsilon^2\right)\right),~
m_\mu \sim \epsilon^4\left( 1 + O\left(\epsilon^2\right)\right),~~
m_e \sim \epsilon^7\left( 1 + O\left(\epsilon^2\right)\right)
\end{align}
\end{itemize}
How well these models account for the known masses and mixings is illustrated in Appendix~\ref{sec:flavor_app}, where they are also compared with the SU(5)-compatible model of Section IIIA. 

\begin{figure}[t]
\centering
\includegraphics[clip,width=.6\textwidth]{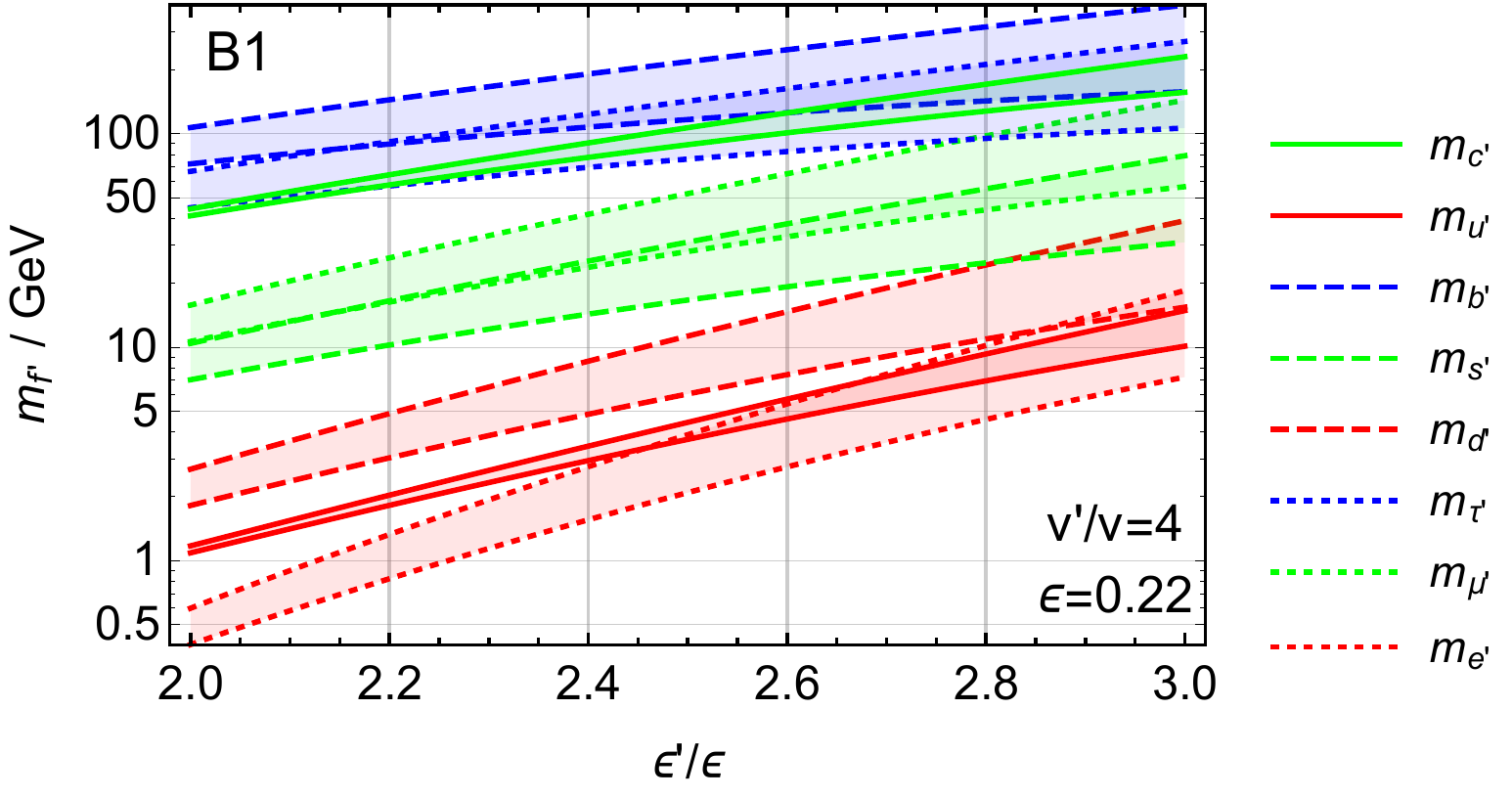}
\includegraphics[clip,width=.6\textwidth]{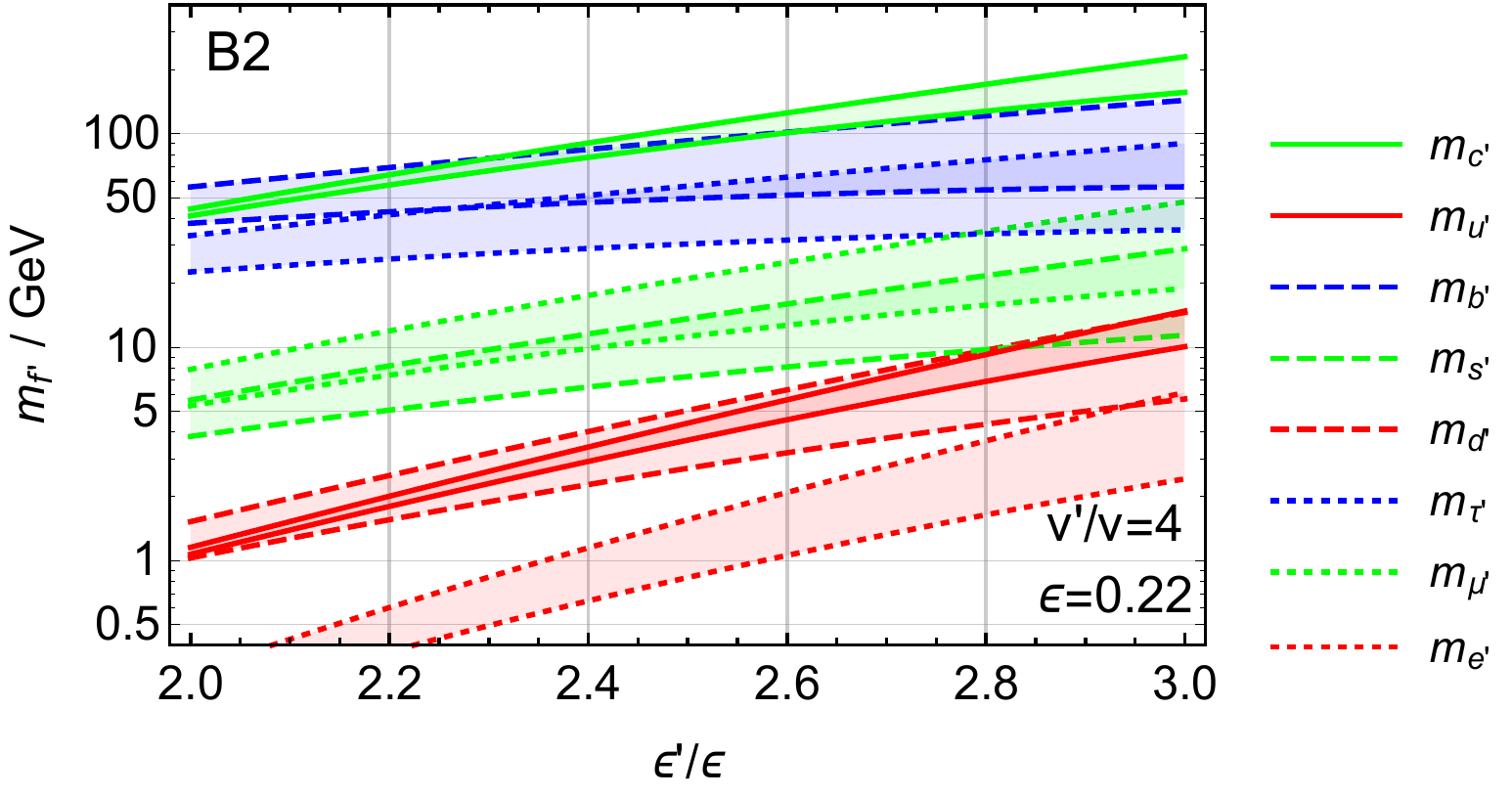}
\caption{
The mass spectrum of the mirror fermions in models B1, B2. The bands show a possible deviation from the simple scaling law with $|\delta| <1$.
Here we have taken the central value for the Yukawa couplings of the SM fermions.
}
\label{fig:mf_B}
\end{figure}

\begin{figure}[t]
\centering
\includegraphics[clip,width=.48\textwidth]{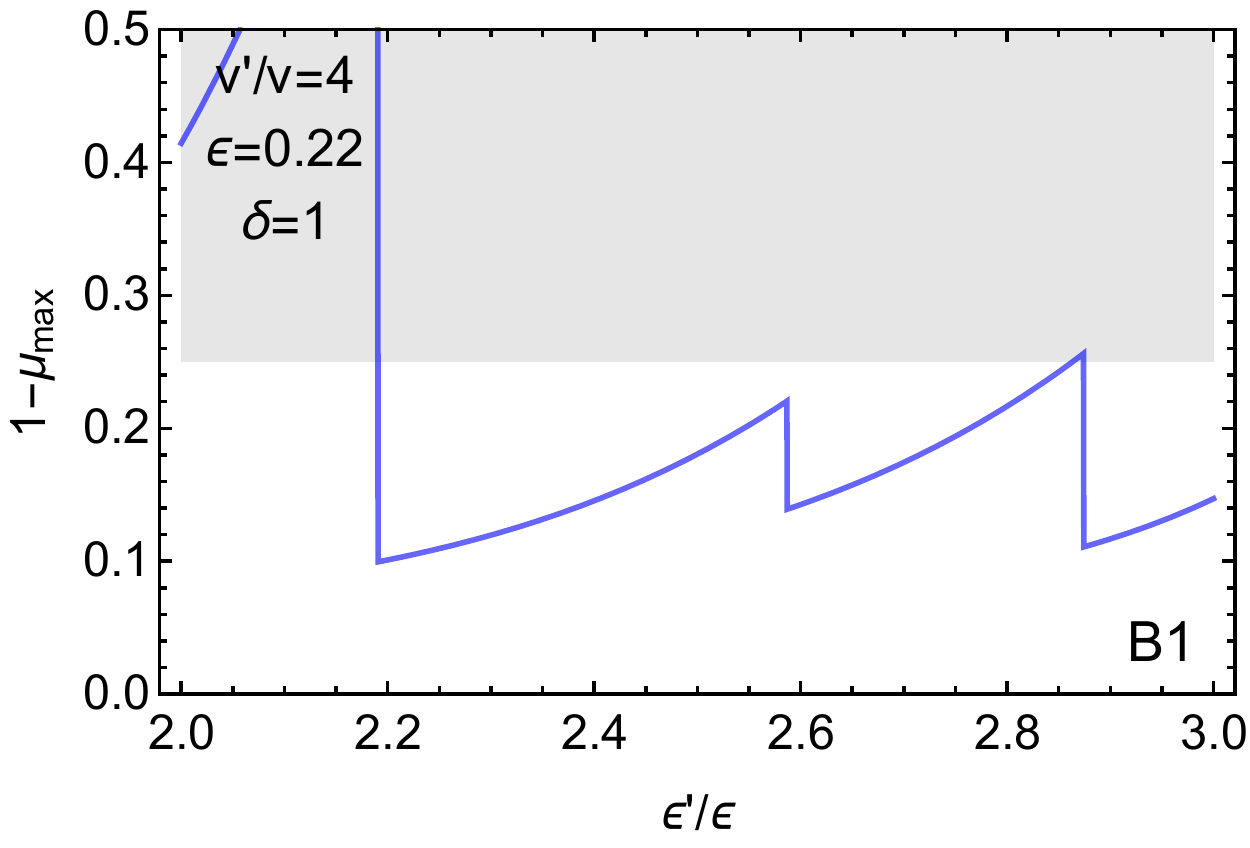}
\includegraphics[clip,width=.48\textwidth]{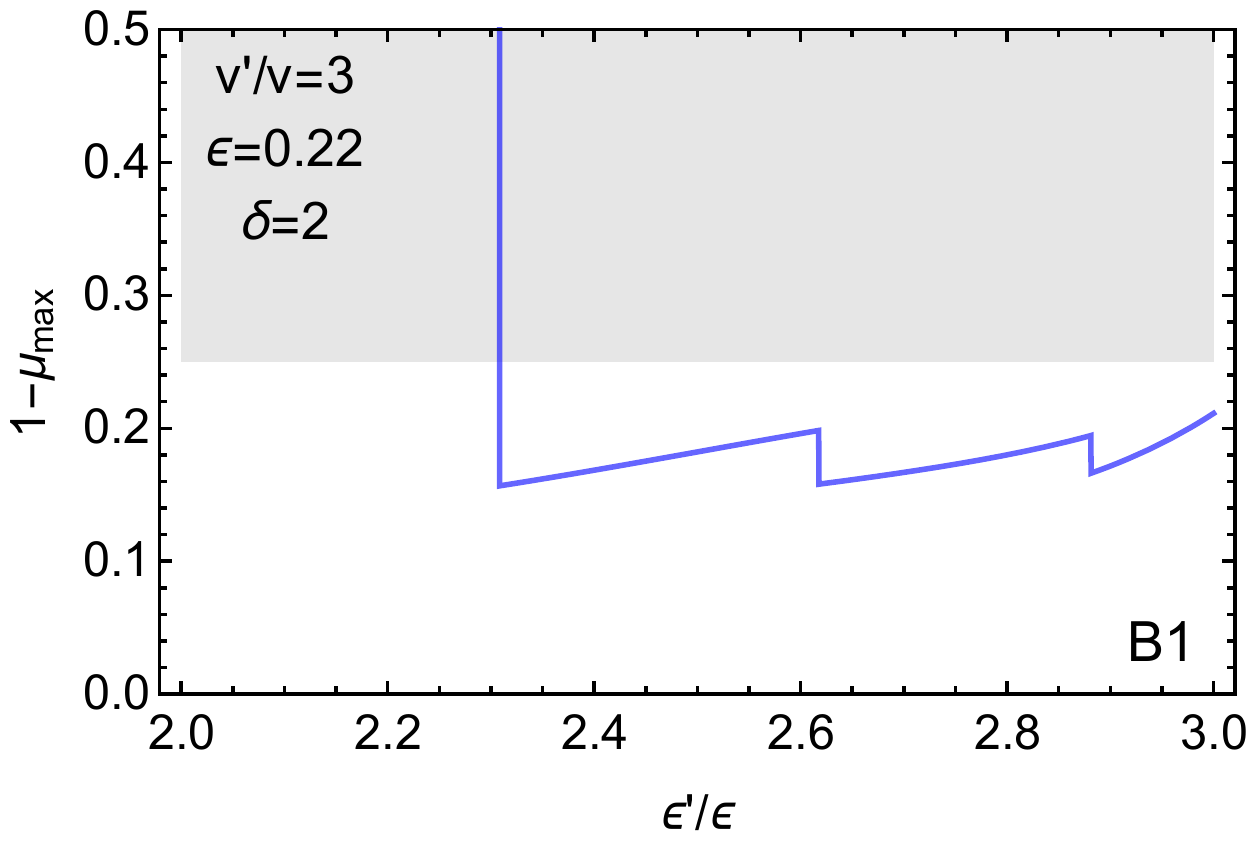}
\includegraphics[clip,width=.48\textwidth]{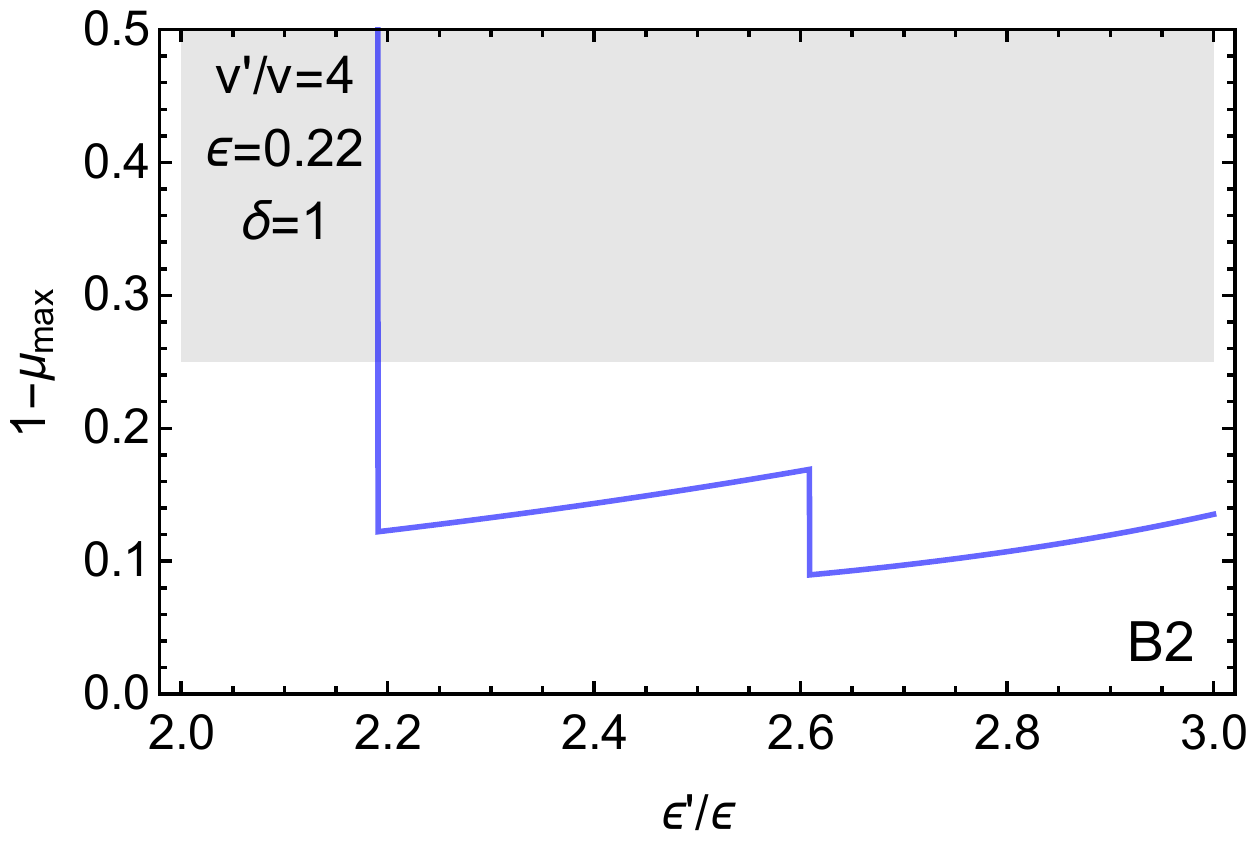}
\includegraphics[clip,width=.48\textwidth]{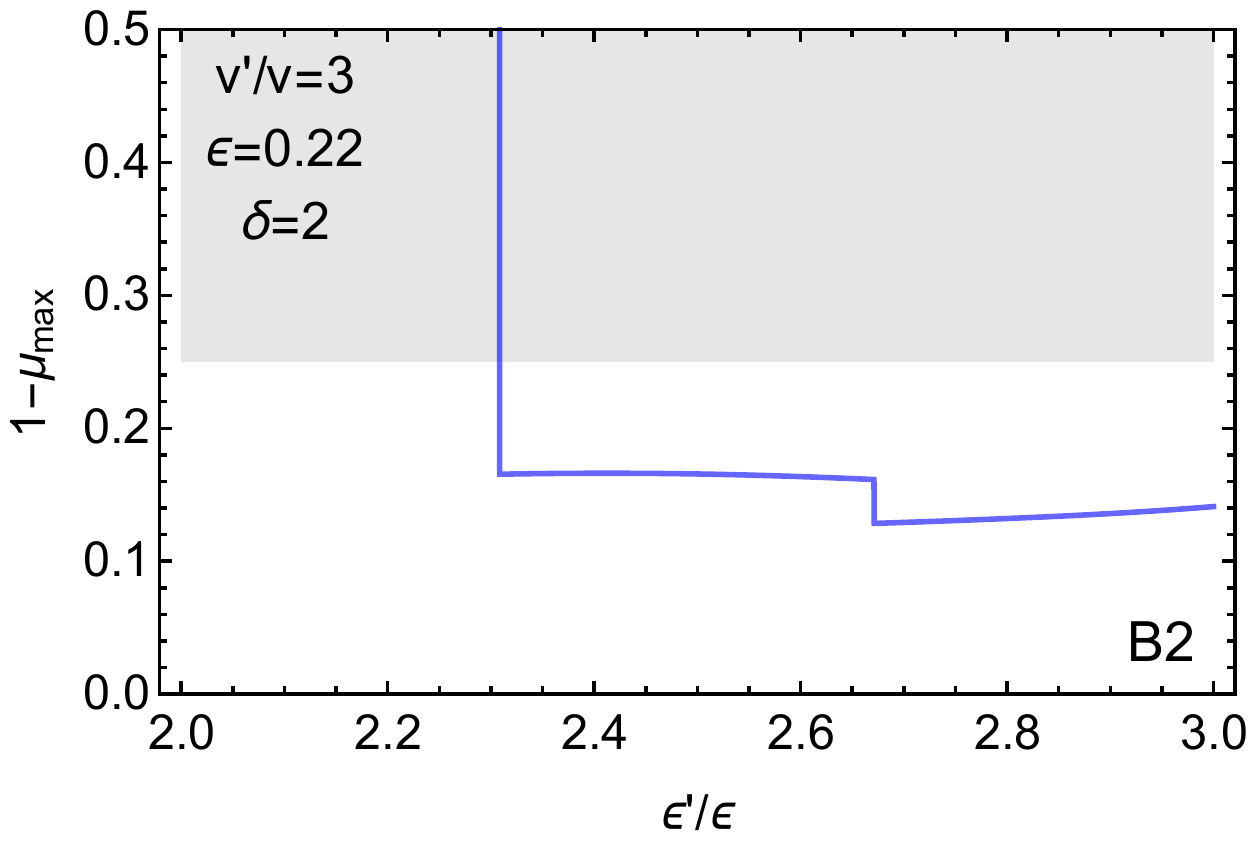}
\caption{
Prediction of the Higgs signal strength in models B1, B2. Panels with $\delta=1,2$ have the mass spectrum of mirror fermions chosen to minimize the invisible decay of the Higgs.
Decays to $c'$ exclude $\epsilon'/\epsilon$ less than about 2.2.}
\label{fig:hdec_4}
\end{figure}

Based on Eq.~(\ref{yukratio}), similarly to Figure~\ref{fig:mf}, we show in Figure~\ref{fig:mf_B}
the masses of the mirror fermions. The consistency of these models with the constraints from Higgs decays is shown in Figure~\ref{fig:hdec_4}. Concerning Dark Matter, the overlap of the masses of $u', d'$ in Figure~\ref{fig:mf_B} for the model B2 makes it relatively more likely that in this case  $B_{udd}'$ be the lightest stable mirror baryon.
%

\section{Conclusions}
\label{sec:concl}

Can Minimal Mirror Twin Higgs be the reason why LHC has not found, so far, any signal of New Physics and, at the same time, explain the surprising similar size of Dark Matter and baryon densities? In~\cite{Barbieri:2016zxn} we have argued in favour of this possibility, attributing the needed breaking of parity only to a difference in the Yukawa couplings between standard and mirror fermions, except the top. We were led to this hypothesis by the need to keep under control, in the absence of an exotic cosmological history, the amount of mirror radiation. 

In this paper we have made the further step of identifying the source of the difference in the standard and mirror Yukawa couplings: a different single scaling parameter, $\epsilon$ versus $\epsilon'$, that is at the origin of the hierarchy in the masses of the charged fermions. In this way the masses of the light mirror fermions are raised, while the top Yukawa couplings remain similar, and the separation between the heaviest and the lightest is reduced, with respect to the masses of the standard fermions, by almost two orders of magnitude. This can be done in a general scheme that we call ``Minimal Flavor Hierarchy". While there can be many such models, different in the physical origin and in the detailed parameters, the range of the predicted signals is greatly reduced by the need to reproduce the known charged fermion masses. Therefore, although we have based our detailed predictions on a specific Froggatt-Nielsen model with SU(5)-compatible U(1) charges, we believe that their main features have a broader validity.

From a phenomenological point of view the new main achievement in the present paper is contained in the part of Section~\ref{sec:su5} where we discuss the various DM configurations, which can be in the form of mirror atoms, Hydrogen-like or Helium-like, or of mirror neutrons. A special summary of the overall situation is in Figure~\ref{fig:selfscatt-ioniz}.  It is remarkable that one can give a detailed prediction of the possible DM configurations and that the entire allowed regions, mostly controlled by the single parameter $\epsilon'/\epsilon$, are within reach of foreseen direct detection experiments for a wide range of the uncertainties. As already pointed out in~\cite{Barbieri:2016zxn}  we expect other correlated signals in Higgs decays and in the amount of dark radiation.  In theories with Minimal Flavor Hierarchies these predictions are sharpened, as shown in Figures~\ref{fig:hdec},\ref{fig:hdec_4}  and Figure~\ref{fig:DNeff} respectively.

\section*{Acknowledgement}
The work of L.H.~and K.H.~was supported in part by the Director, Office of Science, Office of High Energy and Nuclear Physics, of the US Department of Energy under Contract DE-AC02-05CH11231 and by the National Science Foundation under grants PHY-1316783 and PHY-1521446.

\appendix

\section{Minimal Flavor Hierarchy from Extra Dimensions }
\label{sec:exdim}

We first review the model of the flavor hierarchy introduced in~\cite{Kaplan:2001ga}.
We consider a flat extra dimension compactified to an orbifold $S_1/Z_2$, with fixed points $y=0, \pm L/2$.
For a fermion $\psi$ the following boundary condition is imposed to obtain a chiral fermion in the low energy 4D theory,
\begin{align}
\psi(x,-y) = i\gamma_5 \psi(x,y),~~\psi(x, \frac{L}{2} + y) = i\gamma_5 \psi(x, \frac{L}{2} - y),~~
\gamma_5 = - i  \begin{pmatrix} 1 & \\ & -1\end{pmatrix}.
\end{align}
The fermion $\psi$ has a mass term with a non-trivial profile in the extra dimension,
\begin{align}
{\cal L}_{5D} = \bar{\psi} \left( i \gamma^N \partial_N - m(y)  \right) \psi,~~
m(y) = \left\{
\begin{array}{ll}
M &: 0 < y < L/2 \\
-M &: -L/2 < y <0.
\end{array}
\right.
\end{align}
The profile is consistent with the boundary condition as well as with the $Z_2$ symmetry, and may be dynamically generated with a thin domain wall of a scalar field.
The equation of motion of the wave function of the zero-mode of  $\psi$ is given by
\begin{align}
\partial_y \psi_{0,\pm} = \pm M \psi_{0,\pm}.
\end{align}
The solution for this equation is symmetric for $y\leftrightarrow -y$ due to the profile of $m(y)$, and only $\psi_{+,0}$ is consistent with the boundary condition.
The normalized zero mode wave function is given by
\begin{align}
\psi_0(y) = \sqrt{\frac{2M}{e^{ M L}-1 }} e^{M y}.
\end{align}
The zero mode is localized around $y=0$ for $M<0$, and around $y=L/2$ for $M>0$.

The structure of the Yukawa couplings in Eq.~(\ref{eq:Yukawa}) arises when the SM fermions, with bulk mass $M_i$ different from each other, are localized around $y=L/2$, while the Higgs field is confined to the brane at $y=0$.
From the 5D brane couplings
\begin{align}
\label{eq:5dyukawa}
{\cal L}_{5D} = - \delta(y) \frac{\lambda_{ij}}{M_*} H f_{L,i}\bar{f}_{R,j},
\end{align}
we obtain the 4D Yukawa couplings
\begin{align}
{\cal L}_{4D} = - y_{ij} \, H f_{L,i} \bar{f}_{R,j},~~y_{ij} = \frac{\lambda_{ij}}{M_*} \; \psi_{f_{L,ii},0} (0) \psi_{\bar{f}_{R,j},0} (0) \propto \; e^{-M_i/L} \lambda_{ij} e^{-\bar{M}_j/L}
\end{align}
The $O(1)$ top yukawa coupling is obtained by localizing $Q_3$ and $\bar{u}_3$ at $y=0$.

\begin{figure}[t]
\centering
\includegraphics[clip,width=.6\textwidth]{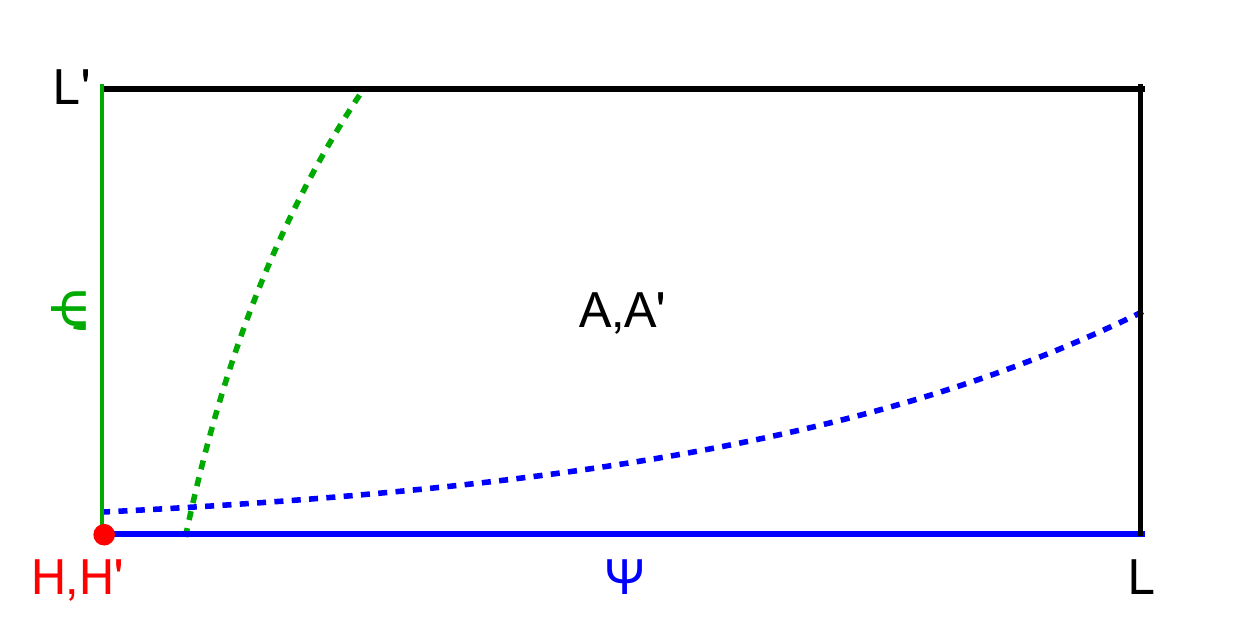}
\caption{
Sketch of a 6D theory that leads to the MMTH scenario.
}
\label{fig:exdim}
\end{figure}

To obtain the minimal flavour hierarchy of MMTH, as described in Section~\ref{sec:flavor}, we consider the 6D configuration depicted in Figure~\ref{fig:exdim}.
The extra dimensions are compactified to $T/(Z_2\times Z_2)$, with fixed points at $(y_5,y_6) =(0,0)$, $(L/2,0)$, $(0,L'/2)$ and $(L/2,L'/2)$.
The SM and mirror fermions are confined to the 5D brane $y_6=0$ and $y_5=0$ respectively.
Those fermions have exponential profiles in each 5D brane via the mechanism shown above.
The Higgs sector is confined to the 4D brane at $(y_5,y_6) = (0,0)$.
The $Z_2$ symmetry, which is now understood as the symmetry $y_5 \leftrightarrow y_6$, is spontaneously broken by $L'<L$, which gives $\epsilon' > \epsilon$.

We assume that the gauge fields live in the 6D bulk which ensures the identity of the gauge couplings from the 6D bulk, $g= g'$, at the tree level.
A difference between them could arise from the quantum correction from KK modes and the 5D bulk gauge couplings.
The former is loop suppressed and is much smaller than the tree level one unless the cut off scale is much larger than the KK scale.
The latter is also suppressed if $L,L' \gsim M_*^{-1}$, where $M_*$ is the cut off scale, due to the volume factor.
It is also possible to obtain non-Minimal Mirror Twin Higgs with $g \neq g'$ with the above two corrections, or confining gauge fields to the 5D bulks. This might be beneficial for two reasons. First, $g_3'>g_3$ raises the mirror QCD phase transition temperature $T_c'$, which helps suppressing the abundance of dark radiation.
Second, $\alpha' > \alpha$  makes recombination for the mirror atomic dark matter more efficient. It also suppresses the self-interaction of atomic dark matter, widening the allowed parameter range.
We do not pursue this possibility in the present paper.

So far we have treated the Higgs field as a fundamental field.
In some UV completions of MMTH the Higgs could be composite.
Then the above derivation of the suppression factor $e^{-ML}$ should be applied to the operators which eventually lead to the SM Yukawa couplings.
For example, if the Yukawa couplings originate from mixing between fundamental SM fermions and composite fermions, we may apply the above discussion to the mixing instead of the Yukawa couplings in Eq.~(\ref{eq:5dyukawa}).

\section{Scaling Law}
\label{sec:scaling}

The down Yukawa matrix for the SU(5) compatible model in Eq.~(\ref{eq:charge su5}) is of the form
\begin{align}
{\cal L} = H^* Q_i  Y_{d,ij} \bar{d}_j,~~
Y_{d,ij} = \epsilon^3
\begin{pmatrix}
a\epsilon^5 & b\epsilon^4 & c\epsilon^4 \\
d\epsilon^3 & e\epsilon^2 & f\epsilon^2 \\
g\epsilon^1 & h & i
\end{pmatrix}.
\end{align}
The square of the matrix is
\begin{align}
Y_d Y_d^\dag = \epsilon^6
\begin{pmatrix}
\epsilon^{8} \left( b^2 + c^2 + a^2 \epsilon^2\right) & \epsilon^{6} \left( be + cf + a d \epsilon^2\right) & \epsilon^{4} \left( bh + ci + a g \epsilon^2\right) \\
 \epsilon^{6} \left( be + cf + a d \epsilon^2\right) & \epsilon^{4} \left( e^2 + f^2 + d^2  \epsilon^2\right) & \epsilon^{2} \left( eh + fi + d g \epsilon^2\right) \\
 \epsilon^{4} \left( bh + ci + a g \epsilon^2\right) & \epsilon^{2} \left( eh + fi + d g \epsilon^2\right) & h^2 + i^2 + g^2 \epsilon^2  \\
\end{pmatrix}.
\end{align}
From this we obtain the bottom Yukawa coupling,
\begin{align}
y_b^2 = \left(h^2 + i^2\right)\epsilon^6 \left( 1 + O(\epsilon^2)\right).
\end{align}
By integrating out the bottom quark, the $2\times 2$ squared Yukawa matrix of the first two generations is given by 
\begin{align}
 \left(Y_d Y_d^\dag\right)_{ds,11}\simeq & \frac{ (c h-b i)^2}{h^2+i^2} \epsilon^{14} \left( 1 + \frac{\epsilon ^2 \left(-a \left(h^2+i^2\right)+b g h+c g i\right)^2}{\left(h^2+i^2\right) (c h-b i)^2}\right), \nonumber \\
 \left(Y_d Y_d^\dag\right)_{ds,12} \simeq & \frac{(c h-b i) (f h-e i)}{h^2+i^2} \epsilon ^{12}\left( 1 + \frac{ \left(-a \left(h^2+i^2\right)+b g h+c g i\right) \left(-d \left(h^2+i^2\right)+e g h+f g
   i\right)}{\left(h^2+i^2\right) (c h-b i) (f h-e i)} \epsilon ^2\right),\nonumber \\
 \left(Y_d Y_d^\dag\right)_{ds,22}   \simeq &    \frac{\left(fh-ei\right)^2}{h^2 + i^2}\epsilon^{10} \left( 1 +  \frac{\left(-d \left(h^2+i^2\right)+e g h+f g i\right)^2}{\left(h^2+i^2\right) (f h-e i)^2} \epsilon ^2 \right).
\end{align}
Therefore the Yukawa coupling of the strange quark is 
\begin{align}
y_s^2 = \frac{\left(fh-ei\right)^2}{h^2 + i^2}\epsilon^{10}\left( 1 + O(\epsilon^2)\right),
\end{align}
and, by integrating $s$ out, we obtain
\begin{align}
y_d^2 = \frac{(-a e i+a f h+b d i-b f g-c d h+c e g)^2}{(f h-e i)^2}\epsilon^{16}\left( 1 + O(\epsilon^2)\right).
\end{align}

\section{Evidence for the Minimal Flavor Hierarchy}
\label{sec:flavor_app}

How well does the flavor structure of (\ref{eq:Yukawa}) account for the observed hierarchies of quark and charged lepton masses in the three FN models considered in this paper?
With $\epsilon$ of about 0.2 and relative corrections of order $\epsilon^2$ or smaller, the leading scaling terms in Eq.~(\ref{scaling}) give a quite accurate approximation for the charged fermion masses and  quark mixing angles in the SM. 
These leading terms are shown in Table \ref{tab:masses} and \ref{tab:angles} for the models considered in the text by fitting the experimental numbers without subleading corrections. In the SU(5) model the coefficients of the leading terms shown in the Tables are determined by a single scaling variable, taken to be $\epsilon=0.22$, and five integers. In model B1 we take $\epsilon= 0.22$ and in model B2 $\epsilon = 0.18$.

The closeness to unity of the coefficients of the leading scaling terms shown in Table \ref{tab:masses} and \ref{tab:angles} represents evidence for the FN picture of the flavour parameters. The neutrino masses and the PMNS angles can also be described by extending  the models discussed in the text with right handed neutrinos~\cite{Babu:2016aro,Feruglio:2015jfa}.

\begin{table}
\renewcommand{\arraystretch}{1.15}
\centering
\begin{tabular}{l|c|c|c|c|c|c|c|c|}

model &$\frac{m_b}{m_t}$ & $\frac{m_\tau}{m_t}$ &  $\frac{m_c}{m_t}$ & $\frac{m_s}{m_t}$ &$\frac{m_\mu}{m_t}$ & $\frac{m_u}{m_t}$& $\frac{m_d}{m_t}$& $\frac{m_e}{m_t}$\\
\hline
$SU(5)$  & $1.6 \epsilon^3$ & $1.1 \epsilon^3$ & $1.8 \epsilon^4$ & $1.0\epsilon^5$  &$ 1.25 \epsilon^5  $&$2.5 \epsilon^8  $&$ 4.5 \epsilon^8  $&$ 0.6 \epsilon^8$  \\
\hline
B1  &  $1.6 \epsilon^3 $&$ 1.1 \epsilon^3 $&$ 1.8 \epsilon^4 $&$1.0  \epsilon^5  $&$1.25 \epsilon^5  $&$0.55 \epsilon^7  $&$ 1.0 \epsilon^7  $&$ 0.6 \epsilon^8$   \\
\hline
B2  &  $0.5 \epsilon^2 $&$ 0.4 \epsilon^2 $&$ 4.0 \epsilon^4 $&$0.45 \epsilon^4  $&$0.6 \epsilon^4  $&$2.2 \epsilon^7  $&$ 0.7 \epsilon^6  $&$ 0.5 \epsilon^7$   \\
\hline

\end{tabular}
\caption{Leading scaling terms for the charged fermion masses in: i) SU(5), with $\epsilon = 0.22$; ii) model B1, with $\epsilon = 0.22$; iii) model B2, with $\epsilon = 0.18$.}
\label{tab:masses}
\end{table}

\begin{table}
\renewcommand{\arraystretch}{1.15}
\centering
\begin{tabular}{l|c|c|c|}
model &$V_{us} $&  $V_{cb} $& $V_{ub}$\\
\hline
SU(5)  &  $4.5 \epsilon^2 $& $1.0  \epsilon^2$ & $2.3 \epsilon^4$ \\
\hline
B1  & $1.0 \epsilon $& $1.0  \epsilon^2$ & $0.5 \epsilon^3 $ \\
\hline
B2  & $1.2 \epsilon$  & $1.5 \epsilon^2$  & $1.8 \epsilon^3$  \\
\hline
\end{tabular}
\caption{Leading scaling terms for the CKM mixings in: i) SU(5), with $\epsilon = 0.22$; ii) model B1, with $\epsilon = 0.22$; iii) model B2, with $\epsilon = 0.18$. }
\label{tab:angles}
\end{table}

\section{Mirror matter asymmetry for $m_{u'}\sim m_{d'}$}
\label{sec:dueDM}

As commented in Section IIIC6, it is natural to assume that the mirror sector has non-zero baryon and lepton asymmetries similar to the SM ones.
As the universe cools, the symmetric components annihilate and almost disappear, and only the asymmetric components remain.
The dark matter component is determined by the scattering of the following particles,
\begin{align}
B'_{uuu},~B'_{uud},~B'_{udd},~B'_{ddd},~e',~\nu'.
\end{align}

Let us first consider the $B'_{uud}\equiv p'$, $B'_{udd}\equiv n'$, $e'$ and $\nu'$ system.
For simplicity we drop the superscript $'$ from now on.
The corresponding number densities are given by
\begin{align}
n_p = 2 \left( \frac{m_p T}{2\pi}\right)^{3/2}e^{- m_p/T + \mu_p / T},~
n_{\bar{p}} = 2 \left( \frac{m_p T}{2\pi}\right)^{3/2}e^{- m_p/T - \mu_p / T},\\
n_n = 2 \left( \frac{m_n T}{2\pi}\right)^{3/2}e^{- m_n/T + \mu_n / T},~
n_{\bar{n}} = 2 \left( \frac{m_n T}{2\pi}\right)^{3/2}e^{- m_n/T - \mu_n / T},\\
n_e = 2 \left( \frac{m_e T}{2\pi}\right)^{3/2}e^{- m_e/T + \mu_e / T},~
n_{\bar{e}} = 2 \left( \frac{m_e T}{2\pi}\right)^{3/2}e^{- m_e/T - \mu_e / T},\\
n_\nu \simeq \frac{3 \zeta(3)}{4\pi^2} T^3 + \frac{1}{12} T^3 \frac{\mu_\nu}{T},~
n_{\bar{\nu}} \simeq \frac{3 \zeta(3)}{4\pi^2} T^3 - \frac{1}{12} T^3 \frac{\mu_\nu}{T}.
\end{align}
The asymmetries are given by
\begin{align}
\label{eq:Dp}
\Delta_p \equiv \frac{n_p- n_{\bar{p}}}{T^3} = 4 \left( \frac{m_p }{2\pi T }\right)^{3/2}e^{- m_p/T} {\rm sinh}\frac{\mu_p}{T},\\
\label{eq:Dn}
\Delta_n \equiv \frac{n_n- n_{\bar{n}}}{T^3} = 4 \left( \frac{m_n }{2\pi T}\right)^{3/2}e^{- m_n/T} {\rm sinh}\frac{\mu_n}{T},\\
\label{eq:De}
\Delta_e \equiv \frac{n_e- n_{\bar{e}}}{T^3} = 4 \left( \frac{m_e }{2\pi T}\right)^{3/2}e^{- m_e/T} {\rm sinh} \frac{\mu_e}{T},\\
\Delta_\nu \equiv \frac{n_\nu- n_{\bar{\nu}}}{T^3} = \frac{1}{6} \frac{\mu_\nu}{T}.
\end{align}
The charge neutrality condition, the conservation of the baryon asymmetry $B\equiv (n_B-n_{\bar{B}})/T^3$, and that of the lepton asymmetry $L\equiv (n_L-n_{\bar{L}})/T^3$ require that
\begin{align}
\Delta_e = \Delta_p, \\
\Delta_n = B-\Delta_p, \\
\label{eq:L}
\mu_\nu/T = 6 L - 6 \Delta_p.
\end{align}
The charged current interactions maintain
\begin{align}
\label{eq:chemical eq}
\mu_p + \mu_e = \mu_n + \mu_\nu,
\end{align}
up to some decoupling temperature $T_{d,W}$.
The reaction $p + e \rightarrow n + \nu$ changes the asymmetry of $p$ and $e$ with a rate
\begin{align}
\label{eq:W exchange}
\frac{ \frac{d}{dt} \Delta_p  }{ \Delta_p} = - \sigma v (p + e \rightarrow n+ \nu) \frac{n_p n_e - n_{\bar{p}} n_{\bar{e}}}{\Delta _p},\nonumber \\
\sigma v (p + e \rightarrow n+ \nu) = \frac{1}{8\pi} \frac{(m_p + m_e- m_n)^2}{v^4}
\end{align}
For $T \lsim m_p/25$, $n_{\bar{p}}$  is  smaller than $n_p$, and we obtain
\begin{align}
\label{eq:W exchange}
\frac{ \frac{d}{dt} \Delta_p  }{ \Delta_p} \simeq \sigma v (p + e \rightarrow n +\nu) n_e \simeq -  \sigma v (p + e \rightarrow n +\nu) 2 \left( \frac{m_e}{2\pi T}\right)^{3/2}e^{- m_e/T}.
\end{align}
Here we assume that the symmetric component of $e$ dominates over the asymmetric one.
The decoupling temperature of the process is given by $(d \Delta_p / dt) / \Delta_p (T) =H (T)$. We find
\begin{align}
\label{eq:TdW}
T_{d,W} \simeq \frac{m_e}{18}.
\end{align}
At this temperature the asymmetric component of the mirror electrons is smaller than the symmetric one, as assumed.
Furthermore, since $m_p > m_e $,
it is indeed verified that $n_{\bar{p}}$ is much smaller than $n_p$.

The dominance of  the symmetric component of $e$ implies $|\mu_e/T| \ll 1$.
Eq.~(\ref{eq:L}) shows that $\mu_\nu / T \simeq 6 L$.
On the other hand, at least one of $\mu_n/T$ and $\mu_p/T$ must be much larger than unity to maintain the baryon asymmetry. Thus Eq.~(\ref{eq:chemical eq}) is solved by $\mu_p = \mu_n + 6L T$.
We therefore obtain the relative abundance of $p$ and $n$,
\begin{align}
\frac{\Delta_n}{\Delta_p} = e^{(m_p-m_n)/T} e^{- 6L} \left( \frac{m_n }{ m_p}\right)^{3/2}.
\end{align}
Except for that case with $m_p \simeq m_n$, $|m_p - m_n|/ T_{d,W}$ is much larger than unity.
For $|L| \ll 1$, the baryon asymmetry is stored in the lighter between $p$ and $n$.
If $L=O(1)$ this conclusion may be changed, but we do not pursue this possibility in this paper.

One can repeat the same analysis including all baryons $B'_{uuu}$, $B'_{uud}$, $B'_{udd}$, $B'_{ddd}$, and show that the chemical potentials of those four baryons are the same.  We conclude that the mirror baryon asymmetry is stored in the lightest among $B'_{uuu}$, $B'_{uud}$, $B'_{udd}$ and $B'_{ddd}$, as anticipated in Section IIIC1.

\section{Mirror recombination with electron capture}
\label{sec:rc_EC}
In the following we drop the superscript $'$ for simplicity.
We consider the situation where $m_{p} + m_{e} > m_{n}$, so that the mirror atom is unstable due to the mirror electron capture process, $p + e \rightarrow n+ \nu$. For s-orbit states, the decay rate of a mirror atom is given by
\begin{align}
\Gamma(H(ns) \rightarrow n + \nu) = |\psi(0)|^2 \sigma v (p + e \rightarrow n+ \nu) =  \frac{\left(m_e \alpha \right)^3}{n^5 \pi} \frac{1}{8\pi} \frac{(m_p + m_e- m_n)^2}{v^4} \nonumber \\
\simeq 2\times 10^{-20}~{\rm GeV}  \left(\frac{4}{v'/v}\right)^4 \left(\frac{m_e }{1~{\rm GeV}}\right)^3 \left(\frac{m_p + m_e - m_n}{1~{\rm GeV}}\right)^2 \frac{1}{n^5}.
\end{align}
Around the temperature where mirror recombination occurs, $T\lsim m_e \alpha^2$, the decay rate of the mirror atom is much larger than the Hubble expansion rate, and electron capture is expected to affect the recombination process.

We formulate recombination with  electron capture by modifying the Peebles model~\cite{Peebles:1968ja}.
We consider transitions between the $1s$, $2s$ and $2p$ atomic states as well as the ionized states.
The differential equation governing their fractions, $x_1\equiv n_{1s}/ n_{\rm DM}$, $x_2\equiv (n_{2s} + n_{2p})/ n_{\rm DM}$, $x_e \equiv n_e / n_{\rm DM}$, are given by
\begin{align}
\label{eq:xe}
\dot{x_e} =&
- \left( x_e^2 n_{\rm DM} \alpha_1 -  x_1 \beta_1  \right) P_{1s}
- \left( x_e^2 n_{\rm DM} \alpha_B -  x_2 \beta_B  \right)  \\
\label{eq:x1}
\dot{x_1} = & + \left( x_e^2 n_{\rm DM} \alpha_1 -  x_1 \beta_1  \right) P_{1s} 
+ \frac{3}{4} \Gamma_{2p1s} P_{2s1s}\left( x_2 - 4 x_1 e^{- E_{2s1s}/T}  \right) \nonumber \\
&+ \frac{1}{4} \Gamma_{2s1s} \left( x_2 - 4 x_1 e^{- E_{2s1s}/T}  \right)
- x_1 \Gamma_{1s,ec} \\
\label{eq:x2}
\dot{x_2} = & + \left( x_e^2 n_{\rm DM} \alpha_B -  x_2 \beta_B  \right) 
- \frac{3}{4} \Gamma_{2p1s} P_{2s1s}\left( x_2 - 4 x_1 e^{- E_{2s1s}/T}  \right) \nonumber \\
&- \frac{1}{4} \Gamma_{2s1s} \left( x_2 - 4 x_1 e^{- E_{2s1s}/T}  \right)
-\frac{1}{4} x_2 \Gamma_{2s,ec}
\end{align}
and satisfies the detailed balance relation if electron capture is absent.

\paragraph*{Recombination into the ground state}
The first terms in the r.h.s.~of Eqs.~(\ref{eq:xe}) and (\ref{eq:x1}) are from the process $p + e \leftrightarrow H(1s) + \gamma$.
The coefficient $\alpha_1$ is the thermal average  of the cross section times the velocity of the process $ p + e \rightarrow H(1s) + \gamma$,
which we extract from~\cite{rec_coef} by subtracting the case B coefficient from the case A one.
The coefficient $\beta_1$ is given by
\begin{align}
\beta_1 = \left( \frac{m_e T}{2\pi} \right)^{3/2} e^{- E_{1s}/T} \alpha_1,
\end{align}
where $E_{1s}$ is the binding energy of the 1s state. $P_{1s}$ is the probability that the emitted photon escapes from the capture by the inverse process and is given by the optical depth $\tau_{1s}$ as
\begin{align}
P_{1s} = \frac{1 - e^{-\tau_{1s}}}{\tau_{1s}},~~\tau_{1s} =  \frac{x_1 n_{\rm DM}}{H} \frac{\pi^2 \alpha_1}{E_{1s}^3} \left( \frac{m_e T}{2\pi} \right)^{3/2}.
\end{align}
When electron capture is absent, as recombination proceeds the optical width is so large that the  process $p + e \leftrightarrow H(1s) + \gamma$ does not contribute to recombination. With electron capture, $x_1$ remains very small and the optical depth is almost zero, and we may use the approximation $P_{1s}\simeq1$.

\paragraph*{Recombination into excited states}
The second term in the r.h.s.~of Eq.~(\ref{eq:xe}) is the effect of the process $p + e \leftrightarrow H(n>1) + \gamma$.
The $n>2$ states rapidly cascade down to the $n=2$ states, and we may use the following so-called  case-B coefficient for the evolution of $x_2$,
\begin{align}
\alpha_B \equiv 1.14 \times \sum_{n=2}^\infty \sum_{l=0}^{n-1} \sum_{m=-l}^{l} \vev{\sigma v (p + e\rightarrow H(nlm) + \gamma)}_{\rm thermal}.
\end{align}
The factor of $1.14$ allows the Peebles approximation to agree with a multi-level calculation~\cite{Seager:1999bc}.
The coefficient $\beta_B$ is given by
\begin{align}
\beta_B = \left( \frac{m_e T}{2\pi} \right)^{3/2} e^{- E_{1s}/T} \alpha_B.
\end{align}

\paragraph*{Lyman-$\alpha$ decay 2p$\rightarrow$1s}
The second terms in the r.h.s.~of Eqs.~(\ref{eq:x1}) and (\ref{eq:x2}) are the effect of the process $H(2p) \leftrightarrow H(1s) + \gamma$,
and $\Gamma_{2p1s}$ is the decay width of this process. $E_{2s1s}$ is the difference  of the energy levels of the $n=2$ and 1 states.
$P_{2p1s}$ is the probability that the emitted photon escapes from the capture by the inverse process, and is given by the optical depth $\tau_{2p1s}$ as
\begin{align}
P_{2p1s} = \frac{1 - e^{-\tau_{2p1s}}}{\tau_{2p1s}},~~\tau_{2p1s} =  \frac{x_1 n_{\rm DM}}{H} \frac{3 \pi^2 \Gamma_{2p1s}}{E_{1s}^3}.
\end{align}
As is the case with recombination to the ground state, we may use the approximation $P_{2p1s} \simeq 1$.

\paragraph*{Two-photon decay}
The third terms in the r.h.s.~of Eqs.~(\ref{eq:x1}) and (\ref{eq:x2}) are the effect of the process $H(2s) \leftrightarrow H(1s) + 2\gamma$,
and $\Gamma_{2s1s}$ is the decay width of this process.
Without electron capture, the two-photon decay may dominate over the Lyman-$\alpha$ decay, due to the large optical depth $\tau_{2p1s}$.
With electron capture, the two-photon decay is negligible, and we ignore it.

\paragraph*{Electron capture}
The last terms in the r.h.s.~of Eqs.~(\ref{eq:x1}) and (\ref{eq:x2}) are the effect of the process $H(2s,1s) \rightarrow n + \nu$.
The inverse process is ineffective. This process ensures that $x_{1} \ll 1$, and $P_{1s},P_{2p1s} \simeq 1$

\begin{figure}[t]
\centering
\includegraphics[clip,width=.6\textwidth]{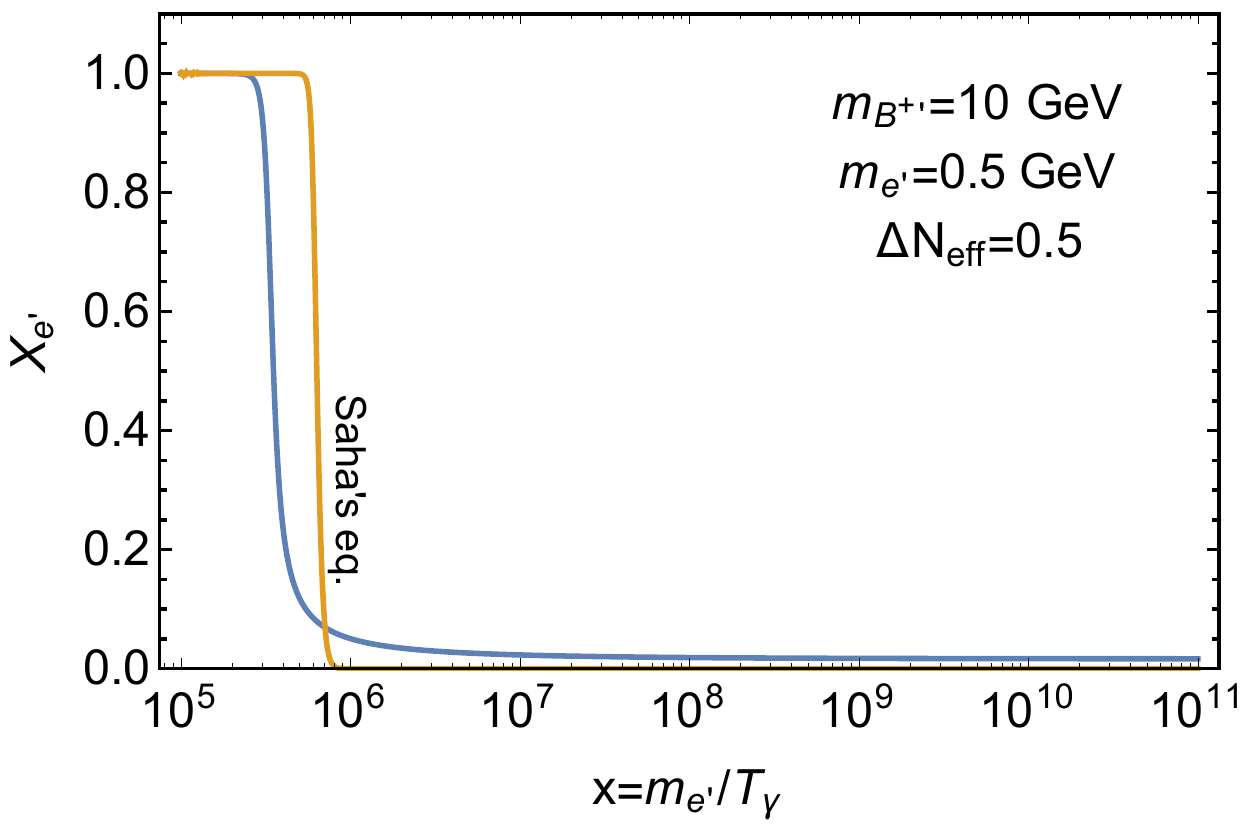}
\caption{
A sample evolution of the ionization fraction when electron capture occurs.
}
\label{fig:n_rc}
\end{figure}

The atomic states are short-lived and we may estimate $x_1$ and $x_2$ by putting $\dot{x_1} = \dot{x_2} =0$, which we call $x_{1,0}$ and $x_{2,0}$.
We find that $x_{2,0}\ll x_{1,0}$ during recombination where $T\ll E_{2s1s}$, and the evolution equation of $x_e$ is given by
\begin{align}
\dot{x_e} = - x_{1,0} \Gamma_{1s,ec}  - x_{2,0} \Gamma_{2s,ec}\simeq -x_{1,0} \Gamma_{1s,ec}.
\end{align}
The full expression for $x_{1,0}$ is not simple, but we can find an approximate solution by adding Eqs.~(\ref{eq:x1}) and (\ref{eq:x2}), and neglecting $x_2$,
\begin{align}
x_{1,0} \simeq \frac{x_e^2 n_{\rm DM} (\alpha_1 + \alpha_B)}{\beta_1 + \Gamma_{1s,ec}}.
\end{align}
The evolution equation of $x_e$ is given by
\begin{align}
\label{eq:xe_app}
\dot{x_e} \simeq - x_e^2 n_{\rm DM} (\alpha_1 + \alpha_B) \frac{\Gamma_{1s,ec}}{\beta_1 + \Gamma_{1s,ec}}.
\end{align}
This equation has a simple interpretation. Once the mirror electron is recombined into atomic states, it rapidly falls into the ground state. The total rate of the formation of the ground state is given by $x_e^2 n_{\rm DM} (\alpha_1 + \alpha_B)$. The ground state mirror electron is again scattered into a free state with a rate $\beta_1$ or is captured by the mirror proton with a rate $\Gamma_{1s,ec}$. 
The latter contributes to recombination, and hence the recombination rate is suppressed by $\Gamma_{1s,ec}/(\beta_1 + \Gamma_{1s,ec})$. 

A sample evolution of the ionization fraction of the mirror electron is shown in Figure~\ref{fig:n_rc}.
Here we use the full expression for $x_{1,0}$. An approximated $x_{1,0}$ gives about a $10$\% larger ionization fraction. 
In the calculation we take $v'/v =4$ and $m_p + m_n-m_e = m_e/2$ to estimate the mirror electron capture rate, but the resultant ionization fraction is insensitive  to these parameters, since during recombination $\beta_1 \ll \Gamma_{1s,ec}$ and the dependence on $\Gamma_{1s,ec}$ drops out  from Eq.~(\ref{eq:xe_app}).

\begingroup
\renewcommand{\addcontentsline}[3]{}
\renewcommand{\section}[2]{}

\endgroup
  
\end{document}